# Cancer Research UK Drug Discovery Process Mining

A study submitted in partial fulfillment

of the requirements for the degree of

MSc Data Science

at

THE UNIVERSITY OF SHEFFIELD

by

Haochao Huang

September 2016

## Abstract

**Background.**

The Drug Discovery Unit (DDU) of Cancer Research UK (CRUK) is using the software Dotmatics for storage and analysis of scientific data during drug discovery process. Whilst the data include event logs, time stamps, activities, and user information are mostly sitting in the database without fully utilising their potential value.

**Aims.**

This dissertation aims at extracting knowledge from event logs data which recorded during drug discovery process, to capture the operational business process of the DDU of Cancer Research UK (CRUK) as it was being executed. It provides the evaluations and methodologies of drawing the process mining panoramic models for the drug discovery process. Thus by enabling the DDU to maximise its efficiency in reviewing its resources and works allocations, patients will benefit from more new treatments faster.

**Methods.**

Two drug discovery process event logs - experiments and registrations datasets have been manipulated with a trial-and-error approach by Excel, Disco, and ProM software. The process mining techniques and algorithms, visualised figures and tables have been applied to the event logs.

**Findings.**

The models, interactive dashboards, and dynamic visualisations have been created,

based on the process mining techniques and algorithms. Only one of the three basic types of process mining, discovery type has been performed successfully in this research. The conformance was failed to check because of not enough data. Lack of information on event classes per case makes the data become isolated, thus lead to a poor performance to apply process mining techniques. Whereas the abundant of event log information can help researchers fully apply the process mining techniques thereby generate more useful discovery.

**Conclusion.**

Management of organisations can be benefit from the process mining methodologies. Disco is excellent for non-experts on management purposes. ProM is great for expert on research purposes. However, the process mining is not once and for all but is a regular operation management process. Indeed, event logs needs to be understand more on the target organisational behaviours and organisational business process. The researchers have to be aware that event logs data are the most important and priority elements in process mining.

## Acknowledgement

I would like to express my deepest gratitude to everyone in the Information School, they offer me the opportunity to study this part-time master programme with full flexibility throughout the past two years. Without their help, I cannot study here while as a full-time worker at the same time.

I would like to appreciate my supervisor particularly, Dr. Gianluca Demartini, for his time and his guidance. Without him, this dissertation would not be possible.

My appreciation also goes to my employer and colleagues, they support me to study and back me up when I off work.

At last but not least, I would like to thank my fellow friends for their support and encouragement throughout the programme.

# Table of contents

# 1. Chapter One Introduction to the study

This dissertation aims at extracting knowledge from event logs data which recorded during drug discovery process, to capture the operational business process of the Drug Discovery Unit (DDU) of Cancer Research UK (CRUK) Manchester Institute as it was being executed. It provides the evaluations and methodologies of drawing the process mining panoramic models for the drug discovery process. Thus by enabling the DDU to maximise its efficiency in reviewing its resources and works allocations, patients will benefit from more new treatments faster.

The term process mining is for the method of refining a structured process description from a set of event logs and real executions from Cancer Research UK in this paper. There are plenty of researches in the mainly health care process mining. The literate review has summarised the current integration of medical field and process mining. However, there is only handful literature directly focusing on the organisational process of drug discovery at the time. It is worth to investigate, explore, and research this topic area in this paper.

There are three types of research questions (Verbeek & Bose, 2010) have been found in the study. Firstly, research questions from a control flow perspective: how the cases were actually being executed; the mean/minimum/maximum throughput time of cases; the paths/routes take too much time on average; the sub-paths/routes of these paths/routes. Secondly, research questions from a performance perspective: the identification of the most frequent flow; the average process time for each activity; the time was spent between any two activities in the process model; the number of activities/cases happen on any time slot. Thirdly, research questions from an organisational perspective: the number of staff were involved in a same activity/case; identification of the most important (busy) staff in the process flow; the staff hand over the work to whom; the staff subcontract the work to whom, the staff work on the same cases.

The implementations are including: to build up the process models based on event logs; to create and analysis the interactive dashboards and dynamic visualisations with process mining techniques and algorithms, to identify the current issues, and to advise potential recommendations.

Chapter one introduces research environment, research organisations, and process mining. In chapter two, a wide range of literature related to process mining and medical area have been introduced. In chapter three, the tools have been introduced, two event logs - experiments and registrations datasets have been manipulated. In chapter four, a trial-and-error approach to implementation has been conducted for the discovery type of process mining research with two event logs. In chapter five, the sections of evaluations, findings, tools, data, guiding principles, and challenges have been discussed. Chapter six is the conclusion and recommendataions for future researchs.

## 1.1 Research Environment:

The Information School of University of Sheffield, which undertook the research topic of drug discovery process mining from Drug Discovery Unit (DDU) of Cancer Research UK (CRUK) Manchester Institute. CRUK was formed in 2002 which is the largest single-disease charity in worldwide and commits over £300 million each year on basic and translational research, aiming to improve the lives of patients with cancer (Jordan, Waddell, & Ogilvie, 2015). DDU was founded in 2009 with a 30 member of staff team working together.

The amount of chemical and biological information has been exploded in recent years. New databases are continually being built and various databases are twice in their size every 1.5 years (Chichester et al., 2010). The research data was provided in the form of an SQLite database, which contains a simplified and anonymised version of a number of Oracle tables from Dotmatics by DDU officially. The student was supervised by Dr. Gianluca Demartini, the professor of Information School, and advised by Adnana Tudose, the DDU’s data scientist.

The fact is that the software Dotmatics which DDU used is primarily designed for the storage and analysis of scientific data, while the data include event logs, time stamps, activities, and user information are mostly sitting in the database without fully utilising their potential value. Therefore, this research will utilise those data by Disco and ProM, in order to give an understanding and improvement of business performance and workflows. Ultimately, by enabling the DDU to maximise its efficiency in reviewing its resources and works allocations, patients will benefit from more new treatments more quickly.

### 1.2 Process mining

In recent years, there have been increasing organisations using information to support the execution of their businesses (Dumas, Van der Aalst, & Ter Hofstede, 2005). The process mining approach, also named workflow mining, relies on an existing process. Obtaining a model that reflects an existing workflow which is a great value to understand and redesign the process. Process mining uses a set of event logs, each representing one event of the process (Blum, Padoy, Feußner, & Navab, 2008). Deriving the model from these logs certainly not replace drug discovery knowledge, but analysis of process can provide an overview for the management of research organisation, benefit from the objective methods.

The information systems supporting operational processes includes Workflow Management Systems (WMS) (Van Der Aalst & Van Hee, 2004), Customer Relationship Management (CRM) systems, Enterprise Resource Planning (ERP) systems (Verbeek & Bose, 2010). Verbeek and Bose (2010) claim that there are three models of processes: "These information systems may contain an explicit model of the processes (for instance, workflow systems like Staware, COSA), may support the tasks involved in the process without necessarily defining an explicit process model (for instance, ERP systems like SAP R/3), or may simply keep track (for auditing purposes) of the tasks that have been performed without providing any support for the actual execution of those tasks (for instance, custom-made information systems in

hospitals)."

All of these information systems usually support logging capabilities which register the information that have been executed in the organisation. These generated logs data typically contain information about cases (i.e. process instances, compound experiments, compound registrations), the activities (i.e. the standard methods used for finding lead compounds as part of the drug discovery process), the times (start and complete time), the resources (typically are the staff or systems) that implemented these tasks, and other kinds of data that have been executed in the organisation. These logs are called event logs and they are fundamental for process mining.

Extraction of information about processes from event logs is the main the task of process mining. Van der Aalst & Weijters assumed that the events are recorded such as (1) each event refers to an activity (i.e., a well-defined step in the process), (2) each event refers to a case (i.e., a process instance), (3) each event can have a performer also referred to as originator (the person executing or initiating the activity), and (4) events have a timestamp and are completely ordered (Van der Aalst & Weijters, 2004). Table 1 indicates an original example of a dataset containing events, activities, and originators. Apart from the information shown in this table, some event logs refer more information on the case itself, i.e., EXPERIMENT_ID referring to more experiment detail and result.

*Table 1. The table of drug discovery unit dataset*

| PROTOCOL_ID | EXPERIMENT_ID | START_DATE | TARGET_NAME | PROGRAM | ANONYMISATION_ISID | Year | PROTOCOL | PROTOCOL_TYPE | COMLETE_DATE |
|---|---|---|---|---|---|---|---|---|---|
| 121 | 138999 | 2011/1/6 16:35 | #N/A | #N/A | staff_27 | 2011 | Chemistry Notebo | NOTEBOOK | 2011/1/13 13:17 |
| 121 | 139000 | 2011/1/6 16:42 | #N/A | #N/A | staff_27 | 2011 | Chemistry Notebo | NOTEBOOK | 2011/1/13 13:17 |
| 121 | 139001 | 2011/1/6 16:59 | #N/A | #N/A | staff_27 | 2011 | Chemistry Notebo | NOTEBOOK | 2011/2/8 10:25 |
| 121 | 139008 | 2011/1/10 9:51 | #N/A | #N/A | staff_22 | 2011 | Chemistry Notebo | NOTEBOOK | 2011/1/17 8:40 |
| 121 | 139009 | 2011/1/10 10:05 | #N/A | #N/A | staff_22 | 2011 | Chemistry Notebo | NOTEBOOK | 2011/1/10 15:45 |
| 121 | 139010 | 2011/1/10 15:47 | #N/A | #N/A | staff_22 | 2011 | Chemistry Notebo | NOTEBOOK | 2011/3/2 8:45 |
| 121 | 139011 | 2011/1/11 13:23 | #N/A | #N/A | staff_22 | 2011 | Chemistry Notebo | NOTEBOOK | NA |
| 121 | 139012 | 2011/1/11 13:26 | #N/A | #N/A | staff_22 | 2011 | Chemistry Notebo | NOTEBOOK | NA |
| 121 | 139013 | 2011/1/12 8:49 | #N/A | #N/A | staff_22 | 2011 | Chemistry Notebo | NOTEBOOK | 2011/1/17 14:17 |
| 121 | 139014 | 2011/1/12 8:58 | #N/A | #N/A | staff_22 | 2011 | Chemistry Notebo | NOTEBOOK | 2011/1/17 14:21 |
| 121 | 139015 | 2011/1/12 11:31 | #N/A | #N/A | staff_22 | 2011 | Chemistry Notebo | NOTEBOOK | 2011/1/19 13:11 |
| 121 | 139016 | 2011/1/12 13:10 | #N/A | #N/A | staff_05 | 2011 | Chemistry Notebo | NOTEBOOK | 2011/5/10 16:08 |
| 121 | 139017 | 2011/1/12 13:23 | #N/A | #N/A | staff_05 | 2011 | Chemistry Notebo | NOTEBOOK | 2011/5/10 16:09 |
| 121 | 139018 | 2011/1/12 17:04 | #N/A | #N/A | staff_05 | 2011 | Chemistry Notebo | NOTEBOOK | 2011/5/10 16:15 |
| 121 | 139019 | 2011/1/13 9:10 | #N/A | #N/A | staff_34 | 2011 | Chemistry Notebo | NOTEBOOK | 2011/1/17 14:19 |
| 121 | 139020 | 2011/1/13 10:02 | #N/A | #N/A | staff_34 | 2011 | Chemistry Notebo | NOTEBOOK | 2011/1/17 14:19 |
| 121 | 139021 | 2011/1/13 10:26 | #N/A | #N/A | staff_27 | 2011 | Chemistry Notebo | NOTEBOOK | 2011/1/20 14:28 |
| 121 | 139022 | 2011/1/13 10:59 | #N/A | #N/A | staff_27 | 2011 | Chemistry Notebo | NOTEBOOK | 2011/7/13 16:20 |
| 121 | 139023 | 2011/1/13 13:05 | #N/A | #N/A | staff_27 | 2011 | Chemistry Notebo | NOTEBOOK | 2011/1/20 14:28 |
| 121 | 139024 | 2011/1/14 11:54 | #N/A | #N/A | staff_34 | 2011 | Chemistry Notebo | NOTEBOOK | 2011/2/8 14:50 |
| 121 | 139028 | 2011/1/17 9:45 | #N/A | #N/A | staff_22 | 2011 | Chemistry Notebo | NOTEBOOK | 2011/1/19 13:03 |
| 121 | 139029 | 2011/1/17 9:53 | #N/A | #N/A | staff_22 | 2011 | Chemistry Notebo | NOTEBOOK | 2011/1/19 13:04 |
| 121 | 139030 | 2011/1/17 10:04 | #N/A | #N/A | staff_27 | 2011 | Chemistry Notebo | NOTEBOOK | 2011/7/13 16:21 |
| 121 | 139031 | 2011/1/17 10:08 | #N/A | #N/A | staff_27 | 2011 | Chemistry Notebo | NOTEBOOK | 2011/7/13 16:21 |

The process mining is defined as a process of discovery of information from event logs. This discovered information can provide a feedback tool that helps in auditing, analysing and improving already enacted business processes or may be used to deploy new systems that support the execution of business processes. The process mining techniques are valuable because the gathered information is objectively compiled according to the event logs that is/was actually happening in the organisation, and not what people *think* that is happening in this organisation (Verbeek & Bose, 2010).

Van der Aalst et al. (2011) have defined several key terms for process mining. The activity is defined as a step in the process. The case is defined as the entity being handled by the process. The event is defined as an action recorded in the log. The event log is defined as the collection of events used as input of process mining.

### 1.3 Drug discovery process

Drug discovery is the process by which new candidate medications are discovered in the fields of medicine, biotechnology, and pharmacology (Lftekhar & Jameel, 2015). In the present time, drug discovery has been developed new technology, such as high-throughput in vitro screening, large compound libraries, combinatorial technology, which defined molecular targets and structure-based design (Lombardino & Lowe, 2004). As Brunning described in Figure 2 which is typically four different stages include Research and Development, Preclinical Studies, Clinical Trials, Review and Approval (Brunning, 2016).

*Figure 2. Understanding the Drug Discovery Process (Brunning, 2016)*

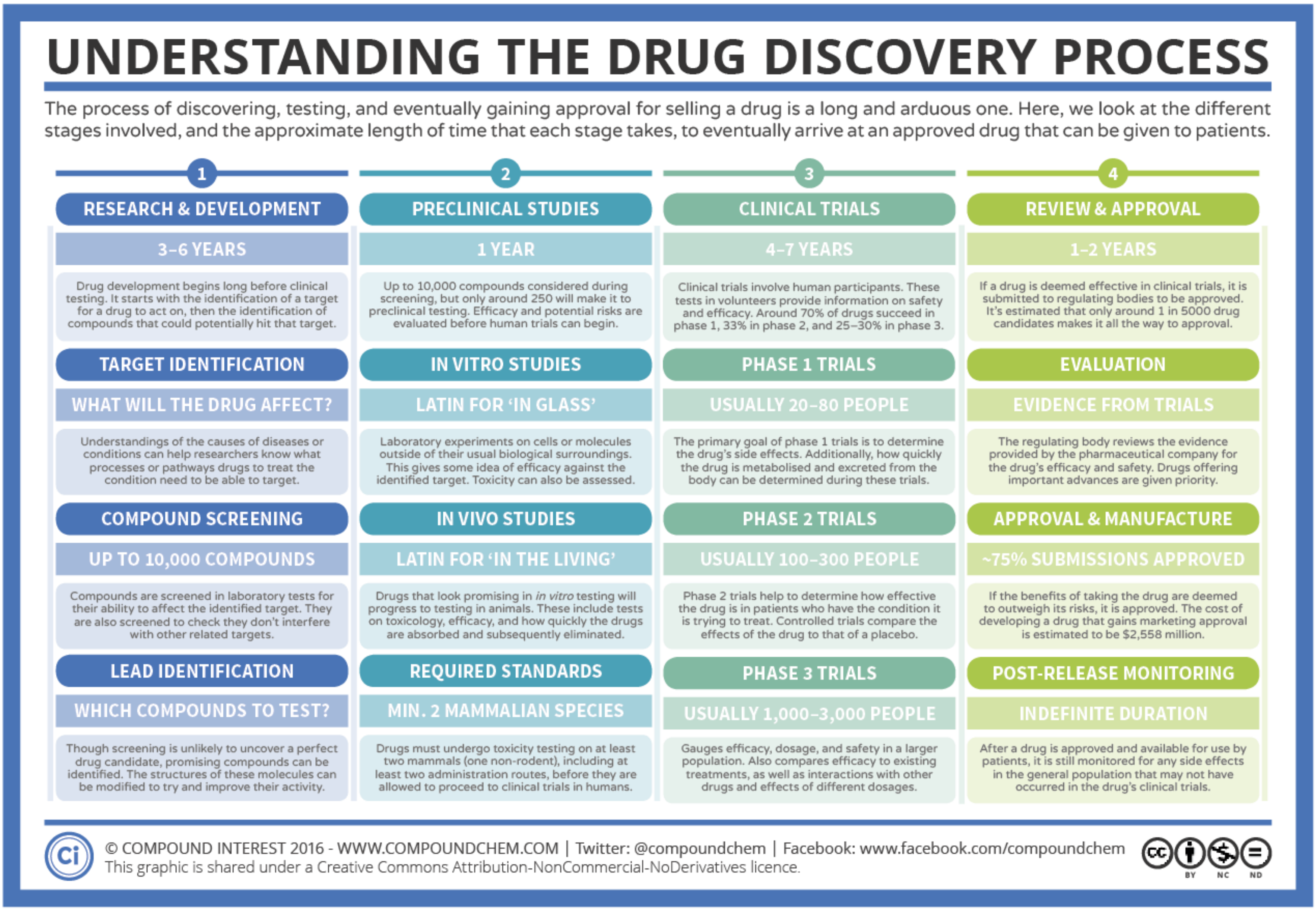

*Figure 3. The drug discovery cycle (Warmuth et al., 2002)*

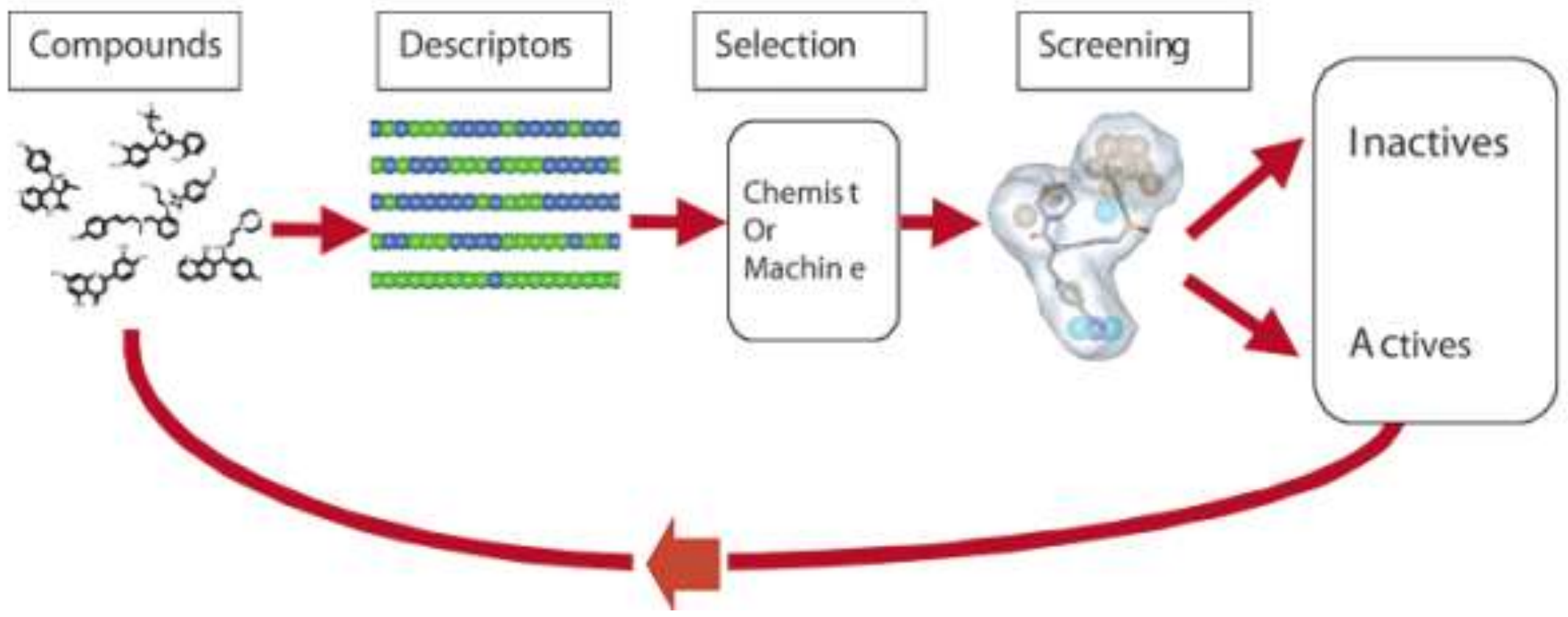

The challenge facing by drug discovery scientist is there are a large number of compounds need to be discovered for finding out rapidly which compounds are active with binding to a particular target. The compounds may be provided by the different suppliers such as vendor catalogs, commercial collections, or chemical and biological partners.

In the Figure 3 drug discovery cycle, it usually starts with some initial set of tested compounds. It is more efficient to test multiple compounds in parallel. But often only a small number of chemical classes can be pursued in parallel. The purpose is to refine the model of activity in each step, based on all tested compounds at the record and to choose the most promising compounds for the next batch. The cycle is repeated until the ultimate goal is attained (Warmuth et al., 2002).

A protocol is a specific study plan developed by researcher or manufacturer. Researchers design clinical trials must follow the protocol to answer research questions about a medical product. Researchers design clinical trials to answer specific research questions related to a medical product. The clinical protocols for studies to be conducted are one of the materials that must be submitted to U.S. Food and Drug Administration (FDA) as an Investigational New Drug Process (IND) application before beginning clinical research. The protocols and safety and efficacy data will be evaluated by statistician and the medical officers from FDA IND review team (U.S. Food and Drug Administration, 2016).

## 1.4 Drug Discovery Unit of Cancer Research UK

After reviewed the four stages of processes of drug discovery, the Drug Discovery Unit is mainly responsible for the stage one research and development process and stage two preclinical studies within Cancer Research UK. Whilst the Christie of Cancer Research UK is the largest single site oncology treatment centre in Europe, the stage three clinical trials are implemented in the organisation (Jordan, Waddell, & Ogilvie, 2015).

New groups as Cancer Research UK is driven by industrial staff are emerging to the delivery of medicines, the Drug Discovery Unit (DDU) at the Manchester Institute was built from the cancer research community into drug discovery programmes (Jordan, Waddell, & Ogilvie, 2015). Not like historically groups grow organically from academic research laboratories, exploiting a particular area of novel biology or new medicines.

However, much of its funding has been dedicated to the fundamental understanding of cancer biology, rather than to presence in drug discovery. Over recent years, industrial drug discovery research has experienced a downturn in headcount. Also, the biotech companies have been forced to downsize research operations and contraction of early-stage pharma research activity. Furthermore, the review acknowledged that the charity had a relatively low presence in small-molecule drug discovery and that this was limiting the potential to exploit the groundbreaking biology emerging from its laboratories (Jordan, Waddell, & Ogilvie, 2015). Therefore, it is necessary that to maximise DDU’s efficiency and improve its business performance and workflows.

# 2. Chapter Two

## 2.1 Literature Review

An iterative procedure of searching compounds that are active against a biological target is the conventional drug discovery process. Every iteration typically refers selecting or synthesising compounds from a number of accessible collections and testing compounds in a biological assay against the target (Warmuth et al., 2002). In order to have a better understanding of event logs and activities, it is necessary to analyse and mining the process data. Therefore the workflow design and allocation can be optimised.

Rojas, Munoz-Gama, Seplveda, and Capurro (2016) claim that the scope of the literature review of the usage of processing mining in healthcare was according to eleven main aspects, including (1) process and (2) data types, (3) frequently posed questions, (4) perspectives, (5) tools (6) techniques, (7) methodologies, (8) implementation and (9) analysis strategies, (10) geographical analysis, and (11) medical areas (Rojas, Munoz-Gama, Seplveda, & Capurro, 2016). This paper follows the scope of eleven aspects to perform the literature review.

### 2.1.1 Process types

Process analysts identified different types of drug discovery processes based on different emphasises. Therefore, process mining techniques and algorithms can be performed properly.

The organisational process of drug discovery focuses on capturing macro collaborative information from workflow and organisational units in order to understand managerial organisational processes (Dumas, Van der Aalst, & Ter Hofstede, 2005). By understanding the managerial organisational processes, management can optimise workflow allocations, assigning projects by shift, and transferring drug information and knowledge between different types of resources. There are plenty of researches in the mainly health care process mining, Rojas et al. (2016) have reviewed more than a hundred relevant health care papers along with case studies in their paper. However, there is only handful literature directly about the organisational process of drug discovery available at the time. It is worth to investigate, explore, and research this topic in this paper.

While another process type is the technical process of drug discovery. It focuses on the process of the specific techniques to enable data-driven drug discovery, integrated resources to support decision–making, and optimise new discoveries efficiently (Chichester et al., 2010). Most of the technical process revolve around the data integration of compounds, targets, pathways and diseases. For example, the research of Warmuth et al. (2002) about data mining problem from computer-aided drug design, the active learning with support vector machine in the drug discovery process. Chandra and Bhaskar (2015) also created a graph mining approach for improving drug discovery process by identifying frequent toxic substructures in chemical compounds.

For the case of CRUK's process type, it mainly includes allocated process of experiments, registrations, protocols, targets, suppliers, projects, programs, and staff.

### 2.1.2 Data types

The types of data determine the process to be analysed and mined. The content of the data is fundamental to construct event logs, then to perform process mining techniques. The event logs typically contain the information of activities, cases, time stamps, and resources which executed through the process (Van Der Aalst, 2012).

The first classification is proposed by Chichester et al. (2010), who supported seven data types in drug discovery flow include: the target information which relate to drug specific for target and protein databank, such as target components; the compound information which are relating to compound properties and annotations, such as molecular weight, toxicity, protein binding; the tissue information which link to other data set, such as medical subject headings; the disease information such as disease class; the pathway information such as pathway organism, the target pharmacology relate to pharmacology compounds, such as target, compound, assay; the compound pharmacology relate to pharmacology compounds and compound–target activity, such as compound, target, activity, assay.

Another classification is according to the data source and its level of abstraction, accuracy, granularity, directness and correctness (Mans, van der Aalst, Vanwersch, & Moleman, 2013). Mans et al. (2013) explained four data sub-types: administrative systems data relate to administration of accounting services, for example, payment administration software; clinical support systems data relate to any information system from a specific unit in a medical organisation, for example, patients management software; logistic systems data relate to all data that support the operational process of the medical organisation, for example, human resources managerial systems of staff shifts; medical devices data, for example medical machines.

For the case of CRUK's data type, it mainly includes data types of protocols data, plates data, experiments data, projects data, and registrations data.

### 2.1.3 Frequently posed questions

The frequently posed questions about the medical process can be responded through process mining techniques and algorithms, therefore it is possible to improve the process. There are five frequently asked medical questions outlined by Mans, van der Aalst, Vanwersch and Moleman (2013), and Rojas, Arias and Sepúlveda (2015): What are the most commonly followed paths and what exceptional paths are followed? What are the differences between care paths followed by different groups? Are internal and external guidelines be complied? What are the bottlenecks in the process? And what are the roles and relations between staff?

In order the compact with the process mining for CRUK drug discovery case, these questions are updated and reconstructed: What happened? It means to identify the activities and process executed. Why did it happen? It means to understand the activities and situation. What will happen? It means to predict when and how an activity will take place. What is the best that can happen? It means to identify the optimisation and improvement of the process (Mans, van der Aalst, & Vanwersch, 2015). Most of these questions are answered and debated in greater detail in the result and discussion sections.

### 2.1.4 Process mining perspectives

Van der Aalst and Weijters (2004) proposed that there are several perspectives in the application of process mining. Four perspectives: control flow; performance; conformance; and organisational are the most commonly used perspectives in the healthcare field. At least one of these perspectives has been applied in more than one hundred case studies which were identified by Rojas et al. (2016).

The control flow perspective (also named process perspective) is to understand the control flow of the drug discovery activities as a whole. Demonstrated by such as Petri net, all kind of paths whatever main and subsidiary flows will be identified and compared for the best optimised characterisation (Van der Aalst & Weijters, 2004). The performance perspectives are to analysis the performance time of activities, to identify the weaknesses, strengths, idle time and synchronisation time (Mans, van der Aalst, & Vanwersch, 2015). The conformance perspective can check the process deviation in terms of designed process model (Mans, van der Aalst, & Vanwersch, 2015). The organisational perspective focus on the originator area, i.e., how many projects, programs, tests are involved by the researchers for a selected time slot. The aim of organisational perspective is to show relations between researchers or to structure the organisation by classifying researchers in terms of roles and units (Van der Aalst & Weijters, 2004).

On the other hand, Van der Aalst and Weijters (2004) suggested that there are three perspectives in term of the different point of views: (1) the process perspective (also named control-flow perspective), (2) the organisational perspective and (3) the case perspective. The process perspective is to understand the control-flow of the drug discovery activities as a whole. Demonstrated by such as Petri net, all kind of paths whatever main and subsidiary flows will be identified and compared for the best optimised characterisation. The organisational perspective focus on the originator area, i.e., how many projects, programs, tests are involved by the researchers for a selected time slot. The aim of organisational perspective is to show relations between researchers or to structure the organisation by classifying researchers in terms of roles and units. The case perspective focuses on properties of cases. The cases may be considered as the protocols, targets, programs, and projects in the drug discovery process. Cases can be characterised by their flow of processes, i.e., how many processes, researchers, activities involved in a case; how long does a case take.

In this paper, the control-flow, organisational, and performance perspectives will be applied.

### 2.1.5 Process mining tools

There are several software have been released online for both business industrial and academic research of process mining applications. They enable techniques and algorithms to be applied to event logs to generate charts, tables, and models.

As a Claes and Poels's (2012) survey shows in Table 4 and Figure 5, the most popular process mining to is ProM. The ProM is an open source tool with a large amount of extensible plug-ins which can implement different techniques and algorithms to analysis (Van Dongen, et al., 2005). Additional software includes ProM Import, Nitro, XESame, Disco, BPMOne.

*Table 4. Process mining tools (Survey: Which process mining tools do you use/know? 119 respondents) (Claes & Poels's, 2012)*

| | *Frequent use* | *Occasional use* | *Tried it once* | *Didn't use but heard about it* | *Never heard about it* |
|---|---|---|---|---|---|
| ARIS Process Performance Manager (Software AG) | 1 | 3 | 12 | 51 | 64 |
| BPMOne (Pallas Athena) | 3 | 8 | 12 | 51 | 57 |
| Disco (Fluxicon) | 10 | 6 | 9 | 29 | 77 |
| Futura Reflect (Futura Technology) | 4 | 5 | 11 | 35 | 76 |
| Interstage Automated Process Discovery (Fujitsu) | 0 | 0 | 3 | 37 | 91 |
| Nitro (Fluxicon) | 11 | 21 | 23 | 24 | 52 |
| ProM (Academic) | 55 | 33 | 9 | 6 | 28 |
| ProM Import (Academic) | 18 | 34 | 23 | 17 | 39 |
| QPR ProcessAnalyzer/Analysis (QPR Software) | 2 | 1 | 3 | 27 | 98 |
| XESame (Academic) | 6 | 12 | 13 | 29 | 71 |

*Figure 5. Process mining tools (Claes & Poels's, 2012)*

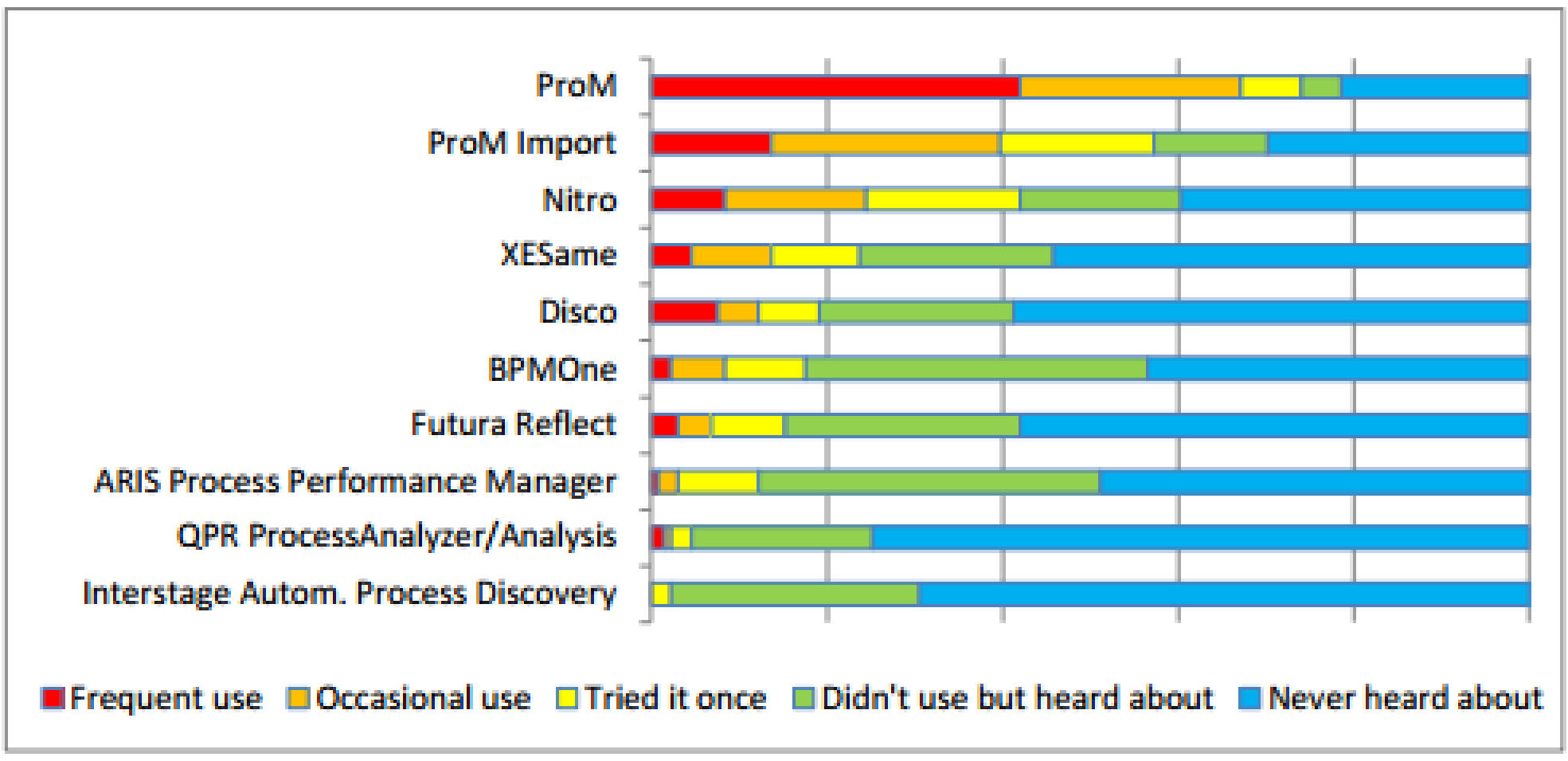

### 2.1.6 Techniques and algorisms

The most commonly used techniques are Heuristics Miner, Fuzzy Miner, and Trace Clustering. Heuristics Miner provides a discovery algorithm which generates process models and it is good at dealing with noise in event logs (Weijters, van Der Aalst, & De Medeiros, 2006). Fuzzy Miner provides a configurable discovery algorithm which utilises its data to generate flexible models at different levels of the process (Günther & Van Der Aalst, 2007). Trace clustering technique allows analysts to distinguish between different process variants within a timeframe, to visualise the process evolvement (Minseok, Günther, & van der Aalst, 2009) (Accorsi & Stocker, 2011).

Other techniques, algorisms and their functions include: Alpha Miner (Van der Aalst, 2011), Genetic Miner (de Medeiros, Weijters, & van der Aalst, 2007) and Inductive Miner (Leemans, Fahland, & van der Aalst, 2013) for identification of process models; Conformance Checker (Rozinat & van der Aalst, 2008) for verification conformance; Performance Sequence Analyser (Verbeek, Buijs, Van Dongen, & van der Aalst, 2010) for execution of performance analysis; Ontologies for including expert knowledge, and Social Miner (Song & Van der Aalst, 2008) for performing organisational analysis. Most of these techniques and algorithms have been plugged-in and ready to use in tools such as ProM.

### 2.1.7 Methodologies

Methodologies are the particular tasks need to be completed along with applying processing mining techniques and algorithms. For instance: collecting and cleaning data; applying techniques and algorithms (Rojas et al., 2016).

The first methodology is the life-cycle model as Figure 6 shows. It is consisted by five stages for a process mining project: plan and justify; extract, build up the control flow model and link to the event logs, integrate process model; provide operational support (Van der Aalst, 2011). Binder et al. (2012), Partington, Suriadi, Ouyang and Karnon (2015) advice that this methodology can guide process mining projects especially for Evidence-Based Medical Compliance Cluster.

*Figure 6. Life-cycle Model (Van der Aalst, 2011)*

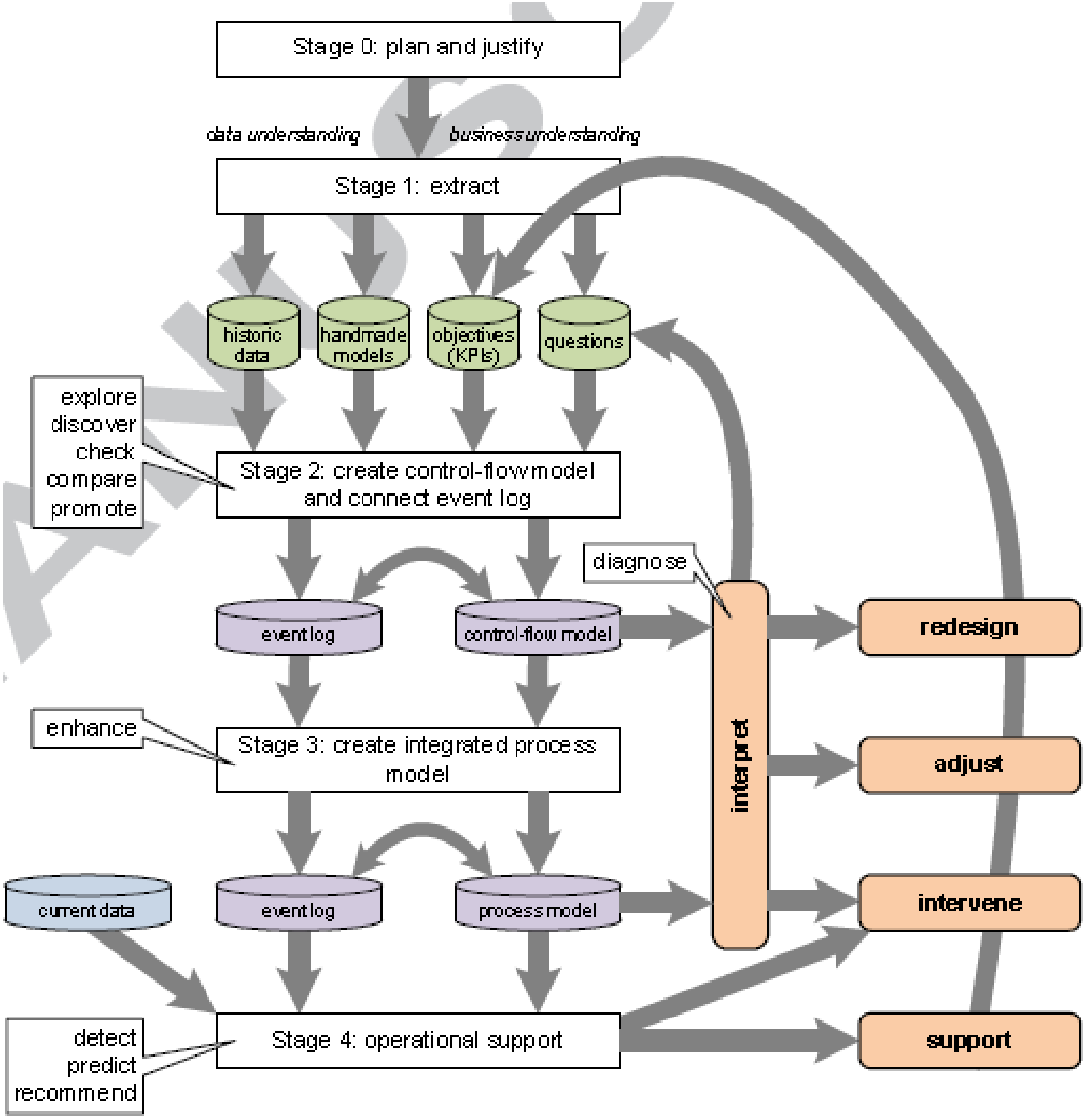

The second methodology uses clustering techniques to conduct analysis for process diagnosis (Bozkaya, Gabriels, & van der Werf, 2009). The original methodology performs a diagnosis process through process mining techniques with five phases: log preparation; log inspection; control flow analysis; performance analysis; and role analysis. Rebuge and Ferreira (2012) developed a new procedure after log inspection for this methodology. This procedure uses Sequence Clustering Analysis (Rebuge & Ferreira, 2012), which relates a series of activities for contending with processes of unstructured or spaghetti type. These improvements utilise activities that helped to identify both typical and infrequent behaviour, which were too complex to detectable previously. Van Doremalen (2012) has improved this particular methodology even more in the later time. Specifically, these relate to the inclusion of tasks which allow information about the activities executed in a process to be added, as well as event log preprocessing to be undertaken, according to the relevant type of analysis being conducted. This improvement based on the tasks: conduction of relevant type of analysis, event log preprocessing, and including information of activities execution in a process. Similar approaches and case studies with this clustering methodology can be found in [(Rebuge and Ferreira, 2012), (Caron, Vanthienen and Baesens, 2013), (De Weerdt, Caron, Vanthienen and Baesens, 2012)].

Additional methodologies can be applied to not only healthcare but also specific medical fields. They are able to help particular these requirements: define particular data reference models for the domain, better data availability from information systems, how to use different tools to answer frequently posed questions [(Caron et al., 2014a), (Lakshmanan, Rozsnyai, & Wang, 2013), (Cho, Song, & Yoo, 2014), (Miclo, Fontanili, Marquès, Bomert, & Lauras, 2015), (Fernandez-Llatas, Lizondo, Monton, Benedi, & Traver, 2015), (Rovani, Maggi, de Leoni, & van der Aalst, 2015), (Rattanavayakorn & Premchaiswadi, 2015)].

### 2.1.8 Implementation strategies

There are three main process mining strategies being identified in the healthcare field after reviewing different case studies in research papers.

The first is manual implementation strategy. This strategy directly collects data from healthcare information systems, which are used to perform process mining techniques and to construct an event log [(Mans, van der Aalst, & Vanwersch, 2015), (Mans, Schonenberg, Song, van der Aalst, & Bakker, 2008), (Mans, Schonenberg, Song, van der Aalst, & Bakker, 2008), (Kim et al., 2013), (Rebuge, Lapão, Freitas, & Cruz-Correia, 2013), (Caron et al., 2014), (Caron, Vanthienen, De Weerdt, & Baesens, 2011b), (Caron, Vanthienen, Weerdt, & Baesens, 2011a), (Delias, Doumpos, Grigoroudis, Manolitzas, & Matsatsinis, 2015), (Rinner et al., 2014), (Dagliati et al., 2014), (Miclo, Fontanili, Marquès, Bomert, & Lauras, 2015), (Antonelli & Bruno, 2015), (Rovani, Maggi, de Leoni, & van der Aalst, 2015), (Forsberg, Rosipko, & Sunshine, 2016), (Kelleher et al., 2014), (Rattanavayakorn & Premchaiswadi, 2015)]. Most of the case studies use this strategy to generate models, figures, table for conducting the analysis. The limitations of this strategy are: the researchers have to manually extract data and construct the correct event logs; it is necessary to understand the tools, techniques, and algorithms before conducting the analysis. The benefits of this strategy are low cost, free to use the open source tools with multiple plug-in packages.

The second is a semi-automated strategy. The data extraction and the event logs construction are the built-in functions of custom-made development tools (Helmering, Harrison, Kabra, & Slette, 2012). These developments link one or more data sources and extract the data required for building the event log through the use of queries. The databases and sources are linked, the queries are ready to be used to generate event logs by this custom-made development tools.

In order to apply the appropriate techniques to conduct the analysis, the knowledge of process mining tools is necessary. But this strategy is defined as an ad hoc manner for the extraction of data, thereby it is limited to be applied to particular tools and environments only.

The third is the fully automated strategy. The implementations are all integrated to a suite, data are extracted, data sources are linked, event logs are constructed, and the process mining techniques are well developed and ready to be applied. This strategy lowers the threshold for researchers who are not experts in data, data sources, process mining techniques, and algorithms. However, this strategy has been developed for particular tools and environment, which may be expensive and not flexible enough as a one size fit all solution.

The studies of examples of these integrated suites are Medtrix Process Mining Studio (Rebuge & Ferreira, 2012), Emotiva Tool (Fernández-Llatas, Benedi, García-Gómez, & Traver, 2013), Careflow Management System Support PM (Quaglini, 2008), Workflow Management Schemata (Neumuth, Liebmann, Wiedemann, & Meixensberger, 2012), Business Process Insight Platform (Lakshmanan, Rozsnyai, & Wang, 2013), Asthma Flow (Prototype Interactive Visual Analytics Tool) (Kumar et al., 2014) and PALIA ILS Suite Web Tool (Fernandez-Llatas et al., 2015).

### 2.1.9 Analysis strategies

The analysis strategies are classified by the ways of how the process mining techniques and algorithms being applied. There are three analysis strategies has been identified from the literature review and case studies.

The first is the basic strategy. This strategy constructs event logs and implements common processing mining techniques and algorithms within available tools, without performing and developing any additional new techniques and algorithms. It does not apply other popular techniques from another field such as statistics. This is relatively the easiest strategy which can be implemented by the researchers in medical and process mining fields who without project development experience. These case studies [(Mans, van der Aalst & Vanwersch, 2015), (Mans, Schonenberg, Song, van der Aalst, & Bakker, 2008), (LANGab, Bürkle, Laumann, & Prokosch, 2008), (Mans, van der Aalst,

Vanwersch, & Moleman, 2013), (Mans et al., 2008), (Caron, Vanthienen, De Weerdt, & Baesens, 2011), (Fei & Meskens, 2008), (Zhou, Wang, & Li, 2014), (Miclo, Fontanili, Marquès, Bomert, & Lauras, 2015), (Antonelli & Bruno, 2015), (Rovani, Maggi, de Leoni, & van der Aalst, 2015), (Rattanavayakorn & Premchaiswadi, 2015), (Helm & Paster, 2015)] have applied this strategy.

The second is the additional strategy. This strategy not only applies existing but also new process mining techniques and algorithms, which are for specific tasks and fields with existing available tools. The goal of this strategy is to find the new implementation of techniques to handle big and multi-sources datasets, and to deal with various, unstructured and complicated processes. This needs additional analysis and time to develop new techniques, which maybe involve a lot of manual works and become cost and time consuming. Two case studies have applied this strategy [(Bose & van der Aalst, 2011), (Fernandez-Llatas, Lizondo, Monton, Benedi, & Traver, 2015)].

The third is the extra strategy. This strategy integrated analysis of other fields apart from process mining techniques and algorithms, which includes statistical analysis [(Bose & van der Aalst, 2011), (Rebuge, Lapão, Freitas, & Cruz-Correia, 2013), (Neumuth, Liebmann, Wiedemann, & Meixensberger, 2012), (Forsberg, Rosipko, & Sunshine, 2016), (Kelleher et al., 2014)], roll-up or drill-down analysis (De Weerdt, Caron, Vanthienen, & Baesens, 2012), data mining (McGregor, Catley, & James, 2011), ontologies [(Grando, Van Der Aalst, & Mans, 2011), (Grando, Schonenberg, & van der Aalst, 2011)], and simulation models (Zhou, Wang, & Li, 2014). This strategy has the potential to create new knowledge through the combination of techniques across different fields. It may become more challenging for a project team because it involves more resources and researchers that are expert in each field.

### 2.1.10 Geographical analysis

Claes and Poels (2012) have conducted a survey to get insights in process mining community as Table 7 and Figure 8 show. It implied that the Europe, especially Netherlands, Belgium, and Germany were having the largest amount of people work in process mining area in 2012. There are no records of people work in Africa in this area.

*Table 7. Country where respondents work (Claes & Poels, 2012)*

| | |
|---|---|
| Netherlands | 27 |
| Belgium | 16 |
| Germany | 7 |
| Israel | 5 |
| United States | 5 |
| Austria | 3 |
| China | 2 |
| Australia | 2 |
| United Kingdom | 2 |
| Chile | 2 |
| Portugal | 2 |
| Switzerland | 2 |
| Spain | 1 |
| India | 1 |
| Poland | 1 |
| France | 1 |
| Peru | 1 |
| Japan | 1 |
| Ireland | 1 |
| Finland | 1 |
| Slovakia | 1 |
| Turkey | 1 |
| Norway | 1 |
| Sweden | 1 |
| Italy | 1 |
| Greece | 1 |

*Figure 8. Country where respondents work (Claes & Poels, 2012)*

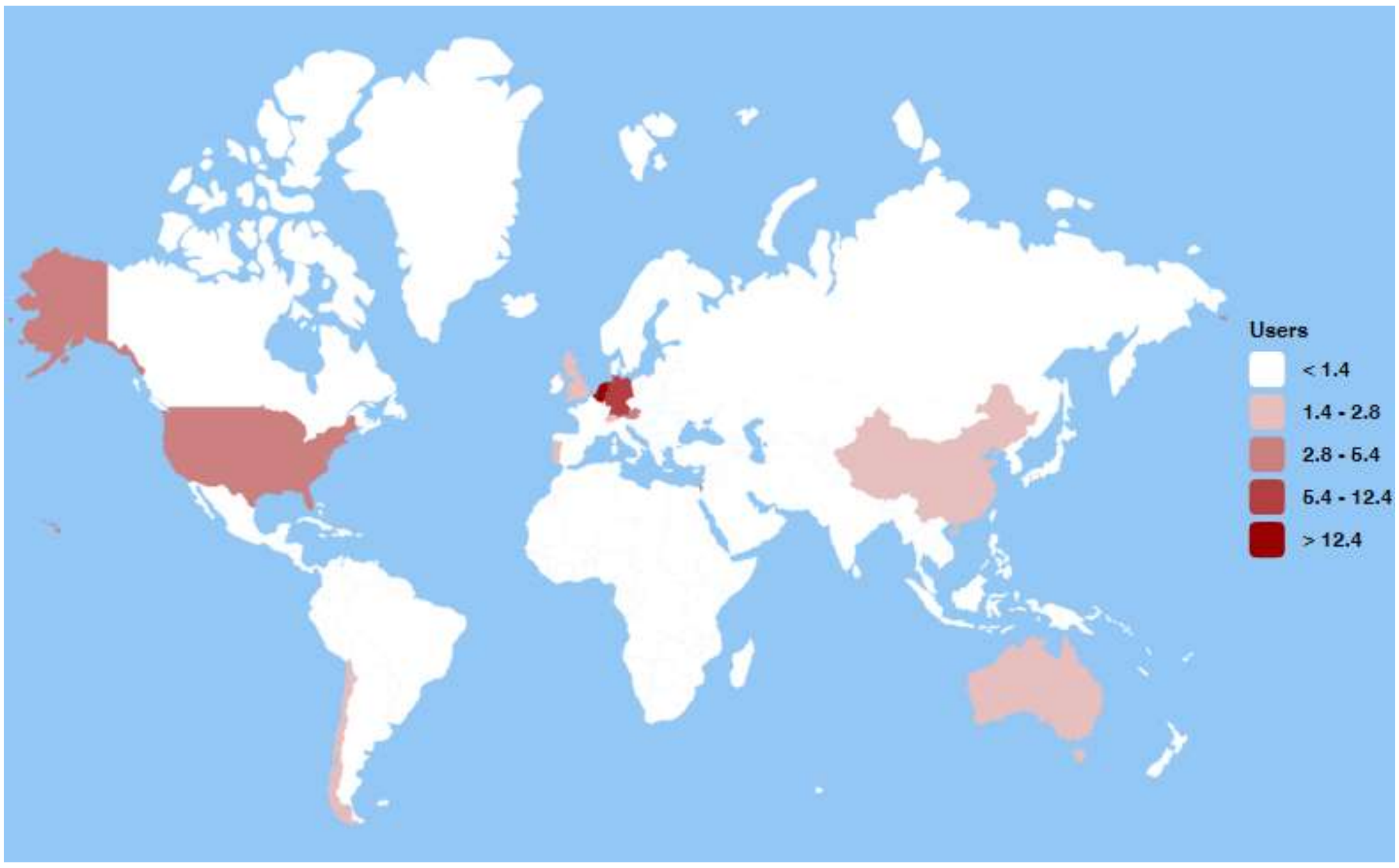

### 2.1.11 Medical fields

Rojas et al. (2016) identified and classified 22 different medical fields in his research of case studies as Figure 9 indicate. The case studies of oncology and surgery are the main medical fields being focused and researched for the application of process mining. In relating to drug discovery, only one field of medication case study being identified. It implies that there are still great potential application and integration to drug discovery with process mining.

*Figure 9. Case studies by medical fields (Rojas et al., 2016)*

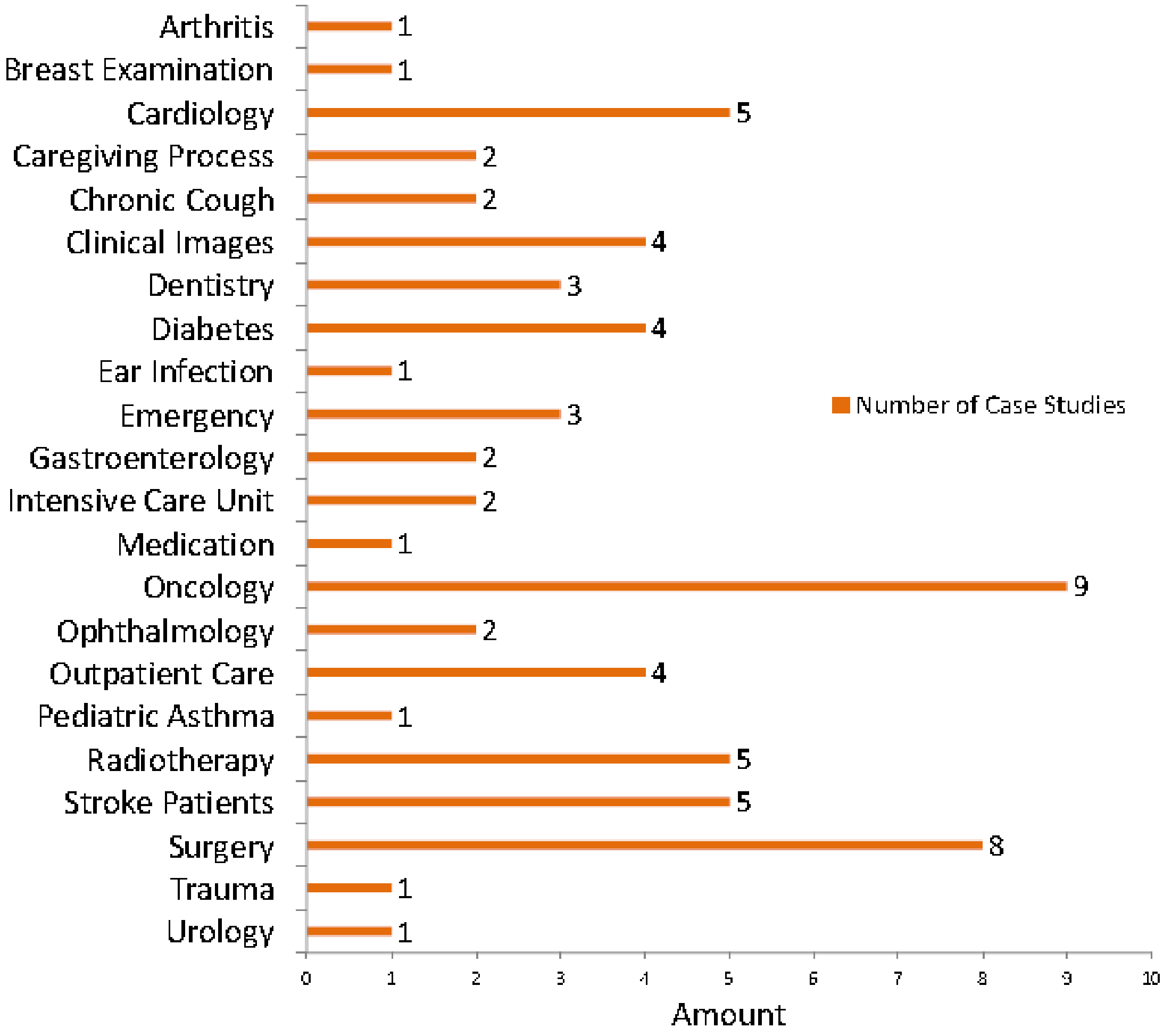

# 3. Chapter Three Methodology

## 3.1 Merging, Cleaning, Manipulating data

First of all, once the data has been distributed, it has to be adjusted before apply any process mining methodology, techniques, and algorithms. It can be sorted out manually by spreadsheet, such as Excel, or it can be dealing with R with command. In this study, the pre-process procedures have been demonstrated by Excel.

### 3.1.1 Dealing with Missing Data

It is very common that the dataset received are having the missing value. The reason behind it has to be checked if they are missing completely at random, missing at random, or missing not at random. We can substitute those missing value only if they are missing completely at random.

Specifically for the ProM software, these settings of configuring additional conversion were performed when importing dataset: the error handling was set to omit trace on error; the empty cells (missing value) were excluded (sparse) in the conversion; the XFactory implementation was set to standard/naive for the memory usage; the attributes were added to both start and complete event. The sparse/dense log affects how empty cells in the data are handled. It might be more efficient or even required to only add attributes to events if the attributes actually contain data.

In order to apply process mining technique to the event logs, the data cleaning step is necessary to consider only the data that are interested in. For instance, the cases that have completed is the only information may be interested in mining process; the cases are still running and without an archiving task as last may not be considered, the name of each executed staff has been anonymised to such as staff_01 to staff_50.

The dataset provided by Drug Discovery Unit of CRUK are containing data records from 2010 to 2016. However, the date of 2010 only started from in March, and data of 2016 ended in February 2016. So both the data of the year 2010 and 2016 have been removed because they are not complete when applying process mining techniques and algorithms. Therefore, only the dataset from 2011 to 2015 will be processed in this paper.

### 3.1.2 Combine different dataset into a large dataset

It is common that there are different database which generate different sheets. Only the data in one same data sheet we can apply methodology. In order to combine this different data sheet into one large sheet, the common point of the dataset such as foreign key and primary key need to be found. This must be done after having a good understanding of the dataset. To add rows of data to an existing dataset, to add more columns of data to an existing dataset, to searching for duplicated data and removing them.

The original dataset provided by DDU of CRUK are containing 7 tables: 1_reg_batches_sanitised, 2_reg_data_sanitised, 3_reg_projects_sanitised, 4_tm_exp_metadata_sanitised, 5_tm_experiments_sanitised, 6_tm_plate_results, and 7_tm_protocols. They are classified as two main groups of data based on their prefix of naming: “reg” –registration of compound and “tm” – experiment of the compound.
The first group of dataset “reg” – registration of compound are shown as table 10, 11. 12 Based on the foreign key of REG_ID and PROJECT_ID, these three tables were merged together as one event log - Table 13 event_log_reg. This log contains information: cases - REG_ID, tasks/activities/events – PROJECT_NAME/ BATCH_NUMBER, start date and time stamp – CREATION_DATE, and the originators/resources - ANONYMINSTRATION_ISID. The other attributes such as SUPPLIER can also help the identification and description of cases and activities.

*Table 10. Sample table of 1_reg_batches_sanitised*

| REG_ID | BATCH_ID | CREATION_DATE | BATCH_NUMBER | PROJECT_ID | SUPPLIER | SUPPLIER_REF | ANONYMISATION_ISID | YYYYMMDD |
|---|---|---|---|---|---|---|---|---|
| 1 | 200384 | 2010/3/19 9:41 | 1 | NA | NA | NA | staff_34 | 20100319 |
| 2 | 200385 | 2010/3/19 9:57 | 1 | NA | NA | NA | staff_34 | 20100319 |
| 3 | 200386 | 2010/3/19 10:16 | 1 | NA | NA | NA | staff_34 | 20100319 |
| 4 | 200387 | 2010/3/19 10:23 | 1 | NA | NA | NA | staff_34 | 20100319 |
| 5 | 200388 | 2010/3/19 10:31 | 1 | NA | NA | NA | staff_34 | 20100319 |
| 6 | 200389 | 2010/3/19 10:44 | 1 | NA | NA | NA | staff_34 | 20100319 |
| 7 | 200390 | 2010/3/19 10:52 | 1 | NA | NA | NA | staff_34 | 20100319 |
| 8 | 200391 | 2010/3/19 10:58 | 1 | NA | NA | NA | staff_34 | 20100319 |
| 9 | 200392 | 2010/3/19 11:05 | 1 | NA | NA | NA | staff_34 | 20100319 |
| 10 | 200393 | 2010/3/19 11:18 | 1 | NA | NA | NA | staff_34 | 20100319 |
| 11 | 200394 | 2010/3/19 11:30 | 1 | NA | NA | NA | staff_34 | 20100319 |
| 12 | 200395 | 2010/3/19 11:36 | 1 | NA | NA | NA | staff_34 | 20100319 |
| 13 | 200396 | 2010/3/19 11:42 | 1 | NA | NA | NA | staff_34 | 20100319 |
| 14 | 200397 | 2010/3/19 11:48 | 1 | NA | NA | NA | staff_34 | 20100319 |
| 15 | 200398 | 2010/3/19 11:55 | 1 | NA | NA | NA | staff_34 | 20100319 |
| 16 | 200399 | 2010/3/19 12:02 | 1 | NA | NA | NA | staff_34 | 20100319 |
| 17 | 200400 | 2010/3/19 12:07 | 1 | NA | NA | NA | staff_34 | 20100319 |
| 18 | 200401 | 2010/3/19 12:16 | 1 | NA | NA | NA | staff_34 | 20100319 |
| 19 | 200402 | 2010/3/19 12:23 | 1 | NA | NA | NA | staff_34 | 20100319 |
| 20 | 200403 | 2010/3/19 13:06 | 1 | NA | NA | NA | staff_34 | 20100319 |
| 21 | 200404 | 2010/3/19 13:12 | 1 | NA | NA | NA | staff_34 | 20100319 |

*Table 11. Sample table of 2_reg_data_sanitised*

| REG_ID | PROJECT_ID | REG_DATE | SUPPLIER | SUPPLIER_REF | TYPE | ANONYMISATION_ISID | YYYYMMDD | COMBINE_STAFF_DATE |
|---|---|---|---|---|---|---|---|---|
| 7484 | 328 | 2012/1/3 15:53 | CRT_Kinase | CRT0051086 | PARENT | staff_39 | 20120103 | staff_3920120103 |
| 7485 | 328 | 2012/1/3 15:53 | CRT_Kinase | CRT0051087 | PARENT | staff_39 | 20120103 | staff_3920120103 |
| 7486 | 328 | 2012/1/3 15:53 | CRT_Kinase | CRT0051088 | PARENT | staff_39 | 20120103 | staff_3920120103 |
| 7487 | 328 | 2012/1/3 15:53 | CRT_Kinase | CRT0051089 | PARENT | staff_39 | 20120103 | staff_3920120103 |
| 7488 | 328 | 2012/1/3 15:53 | CRT_Kinase | CRT0051090 | PARENT | staff_39 | 20120103 | staff_3920120103 |
| 7489 | 328 | 2012/1/3 15:53 | CRT_Kinase | CRT0051091 | PARENT | staff_39 | 20120103 | staff_3920120103 |
| 7490 | 328 | 2012/1/3 15:53 | CRT_Kinase | CRT0051092 | PARENT | staff_39 | 20120103 | staff_3920120103 |
| 7491 | 328 | 2012/1/3 15:53 | CRT_Kinase | CRT0051093 | PARENT | staff_39 | 20120103 | staff_3920120103 |
| 7492 | 328 | 2012/1/3 15:53 | CRT_Kinase | CRT0051094 | PARENT | staff_39 | 20120103 | staff_3920120103 |
| 7493 | 328 | 2012/1/3 15:53 | CRT_Kinase | CRT0051095 | PARENT | staff_39 | 20120103 | staff_3920120103 |
| 7498 | 328 | 2012/1/3 15:53 | CRT_Kinase | CRT0051100 | PARENT | staff_39 | 20120103 | staff_3920120103 |
| 7499 | 328 | 2012/1/3 15:53 | CRT_Kinase | CRT0051101 | PARENT | staff_39 | 20120103 | staff_3920120103 |
| 7500 | 328 | 2012/1/3 15:53 | CRT_Kinase | CRT0051102 | PARENT | staff_39 | 20120103 | staff_3920120103 |
| 7494 | 328 | 2012/1/3 15:53 | CRT_Kinase | CRT0051096 | PARENT | staff_39 | 20120103 | staff_3920120103 |
| 7495 | 328 | 2012/1/3 15:53 | CRT_Kinase | CRT0051097 | PARENT | staff_39 | 20120103 | staff_3920120103 |
| 7496 | 328 | 2012/1/3 15:53 | CRT_Kinase | CRT0051098 | PARENT | staff_39 | 20120103 | staff_3920120103 |
| 7497 | 328 | 2012/1/3 15:53 | CRT_Kinase | CRT0051099 | PARENT | staff_39 | 20120103 | staff_3920120103 |
| 7501 | 328 | 2012/1/3 15:53 | CRT_Kinase | CRT0051103 | PARENT | staff_39 | 20120103 | staff_3920120103 |
| 7502 | 328 | 2012/1/3 15:53 | CRT_Kinase | CRT0051104 | PARENT | staff_39 | 20120103 | staff_3920120103 |
| 7503 | 328 | 2012/1/3 15:53 | CRT_Kinase | CRT0051105 | PARENT | staff_39 | 20120103 | staff_3920120103 |
| 7504 | 328 | 2012/1/3 15:53 | CRT_Kinase | CRT0051106 | PARENT | staff_39 | 20120103 | staff_3920120103 |
| 7505 | 328 | 2012/1/3 15:53 | CRT_Kinase | CRT0051107 | PARENT | staff_39 | 20120103 | staff_3920120103 |
| 7506 | 328 | 2012/1/3 15:53 | CRT_Kinase | CRT0051108 | PARENT | staff_39 | 20120103 | staff_3920120103 |
| 7507 | 328 | 2012/1/3 15:53 | CRT_Kinase | CRT0051109 | PARENT | staff_39 | 20120103 | staff_3920120103 |
| 7508 | 328 | 2012/1/3 15:53 | CRT_Kinase | CRT0051110 | PARENT | staff_39 | 20120103 | staff_3920120103 |

*Table 12. Sample table of 3_reg_projects_sanitised*

| ID | PROJECT_NAME | PARENT | CREATION_DATE |
|---|---|---|---|
| 246 | Tdp2 | 0 | 00:00.0 |
| 327 | LSD1 | 0 | NA |
| 346 | MLK | 0 | NA |
| 347 | LZK | 0 | NA |
| 240 | Fragments | 0 | NA |
| 266 | Tool_compounds | 0 | NA |
| 328 | TARGET1 | 0 | NA |
| 241 | Dehydrogenase | 0 | 00:00.0 |
| 244 | Autotaxin | 0 | 00:00.0 |
| 268 | TARGET2G | 0 | NA |

*Table 13. Sample table of event_log_reg*

| REG_ID | PROJECT_ID | CREATION_DATE | BATCH_NUMBER | SUPPLIER | ANONYMISA | Year | PROJECT_NAME |
|---|---|---|---|---|---|---|---|
| 4143 | 268 | 2011/11/1 | 1 | Internal | staff_18 | 2011 | TARGET2G |
| 4144 | 241 | 2011/11/1 | 1 | Internal | staff_18 | 2011 | Dehydrogenase |
| 4145 | 241 | 2011/11/1 | 1 | Internal | staff_18 | 2011 | Dehydrogenase |
| 4146 | 241 | 2011/11/1 | 1 | Internal | staff_18 | 2011 | Dehydrogenase |
| 4147 | 246 | 2011/11/2 | 1 | Internal | staff_11 | 2011 | Tdp2 |
| 4148 | 241 | 2011/11/3 | 1 | Internal | staff_27 | 2011 | Dehydrogenase |
| 4149 | 241 | 2011/11/3 | 1 | Internal | staff_27 | 2011 | Dehydrogenase |
| 4150 | 241 | 2011/11/4 | 1 | Internal | staff_34 | 2011 | Dehydrogenase |
| 4151 | 241 | 2011/11/4 | 1 | Internal | staff_34 | 2011 | Dehydrogenase |
| 4152 | 241 | 2011/11/4 | 1 | Internal | staff_34 | 2011 | Dehydrogenase |
| 4153 | 241 | 2011/11/4 | 1 | Internal | staff_18 | 2011 | Dehydrogenase |
| 4154 | 241 | 2011/11/4 | 1 | Internal | staff_34 | 2011 | Dehydrogenase |
| 4155 | 241 | 2011/11/4 | 1 | Internal | staff_27 | 2011 | Dehydrogenase |
| 4156 | 246 | 2011/11/4 | 1 | Internal | staff_31 | 2011 | Tdp2 |
| 4157 | 246 | 2011/11/4 | 1 | Internal | staff_31 | 2011 | Tdp2 |
| 4158 | 241 | 2011/11/4 | 1 | Internal | staff_18 | 2011 | Dehydrogenase |
| 4159 | 241 | 2011/11/4 | 1 | Internal | staff_18 | 2011 | Dehydrogenase |
| 4160 | 246 | 2011/11/4 | 1 | Internal | staff_13 | 2011 | Tdp2 |
| 4161 | 246 | 2011/11/4 | 1 | Internal | staff_13 | 2011 | Tdp2 |
| 4162 | 246 | 2011/11/4 | 1 | Internal | staff_19 | 2011 | Tdp2 |
| 4163 | 241 | 2011/11/4 | 1 | Internal | staff_18 | 2011 | Dehydrogenase |
| 4164 | 246 | 2011/11/4 | 1 | Internal | staff_19 | 2011 | Tdp2 |
| 4165 | 246 | 2011/11/4 | 1 | Internal | staff_19 | 2011 | Tdp2 |

The second group of dataset "tm" – experiment of the compound are shown as table 14, 15, 16, 17. Based on the foreign key of EXPERIMENT_ID and PROTOCOL_ID, these four tables were merged together as one event log - Table 18 event_log_tm. This experiment event log contains information: cases - EXPERIMENT_ID, tasks/activities/events – PROTOCOL, start date and time stamp – CREATED_DATE, complete date and time stamp - COMPLETED_DATE and the originators/resources -

ANONYMINSTRATION_ISID. The other attributes such as PROTOCOL_TYPE/ PROTOCOL_ID/ TARGET_NAME/ PROGRAM can also help the identification and description of cases and activities.

*Table 14. Sample table of 4_tm_exp_metadata_sanitised*

| PROTOCOL_ID | EXPERIMENT_ID | TARGET_NAME | PROGRAM | ASSAY_FORMAT | ASSAY_FORMAT_CELL | ASSAY_FORMAT_TOX | CELL_CONDITIONS | CELL_CONDITIONS_ | CELL_LINE |
|---|---|---|---|---|---|---|---|---|---|
| 341 | 150000 | TARGET1 | KINASE | NA | NA | NA | None | NA | LC-2/ad |
| 146 | 138600 | IDH(R132H) | DEHYDROGENASE | Amplite | NA | NA | NA | NA | NA |
| 222 | 140700 | NA | NA | NA | NA | NA | NA | NA | NA |
| 222 | 141300 | NA | TDP2 | NA | NA | NA | NA | NA | NA |
| 201 | 141600 | wtIDH | DEHYDROGENASE | Amplite | NA | NA | NA | NA | NA |
| 145 | 141700 | G6PD | DEHYDROGENASE | Amplite | NA | NA | NA | NA | NA |
| 222 | 141800 | NA | TARGET2 | NA | NA | NA | NA | NA | NA |
| 222 | 141900 | NA | DEHYDROGENASE | NA | NA | NA | NA | NA | NA |
| 222 | 142300 | NA | DEHYDROGENASE | NA | NA | NA | NA | NA | NA |
| 145 | 143300 | G6PD | DEHYDROGENASE | Amplite | NA | NA | NA | NA | NA |
| 222 | 147000 | NA | TARGET2 | NA | NA | NA | NA | NA | NA |
| 145 | 147200 | TARGET1 | KINASE | HTRF | NA | NA | NA | NA | NA |
| 145 | 147400 | TARGET1 | KINASE | HTRF | NA | NA | NA | NA | NA |
| 261 | 148800 | TARGET2G | TARGET2 | HTRF | NA | NA | NA | NA | NA |
| 222 | 149600 | NA | LSD1 | NA | NA | NA | NA | NA | NA |
| 222 | 149700 | NA | TARGET VALIDATION | NA | NA | NA | NA | NA | NA |
| 341 | 149900 | KDR | KINASE | NA | ELISA (phospho endp | NA | #NAME? | NA | MZ_CRC-1 |
| 401 | 150500 | NA | NA | NA | NA | NA | NA | NA | NA |
| 341 | 150600 | TARGET1 | KINASE | NA | ELISA (phospho endp | NA | None | NA | LC-2/ad |
| 421 | 152000 | TARGET1-M918T | KINASE | NA | Proliferation | NA | None | NA | MZ_CRC-1 |
| 145 | 152400 | TARGET1 | KINASE | HTRF | NA | NA | NA | NA | NA |
| 145 | 152500 | TARGET1 | KINASE | HTRF | NA | NA | NA | NA | NA |
| 341 | 153000 | KDR | KINASE | NA | ELISA (phospho endp | NA | #NAME? | NA | MZ_CRC-1 |
| 421 | 154100 | TARGET1-M918T | KINASE | NA | Proliferation | NA | None | NA | MZ_CRC-1 |
| 421 | 155800 | ALK | KINASE | NA | Proliferation | NA | None | NA | H2228 |

*Table 15. Sample table of 5_tm_experiments_sanitised*

| PROTOCOL_ID | EXPERIMENT_ID | CREATED_DATE | COMPLETED_DATE | PLATE_COUNT | MODIFIED_DATE | DELETED | COUNTERSIGNED_DATE | PDF_COMPLETE | ANONYMISATION_ISID | COUNTERSIGNEI | ANONYMISATION_COMPLETED_ISID |
|---|---|---|---|---|---|---|---|---|---|---|---|
| 146 | 138535 | 2010/9/13 9:45 | 2010/11/24 16:29 | 1 | ############ | NA | NA | NA | staff_49 | NA | staff_49 |
| 146 | 138537 | 2010/9/13 9:57 | 2010/11/24 16:30 | 1 | ############ | NA | NA | NA | staff_49 | NA | staff_49 |
| 146 | 138538 | 2010/9/13 10:09 | 2010/11/24 16:30 | 1 | ############ | NA | NA | NA | staff_49 | NA | staff_49 |
| 146 | 138539 | 2010/9/13 10:13 | 2010/11/24 16:30 | 1 | ############ | NA | NA | NA | staff_49 | NA | staff_49 |
| 146 | 138540 | 2010/9/13 10:24 | 2010/11/24 16:30 | 1 | ############ | NA | NA | NA | staff_49 | NA | staff_49 |
| 146 | 138541 | 2010/9/13 10:35 | 2010/11/24 16:30 | 1 | ############ | NA | NA | NA | staff_49 | NA | staff_49 |
| 146 | 138542 | 2010/9/13 10:41 | 2010/11/24 16:30 | 1 | ############ | NA | NA | NA | staff_49 | NA | staff_49 |
| 146 | 138543 | 2010/9/13 10:56 | 2010/11/24 16:30 | 1 | ############ | NA | NA | NA | staff_49 | NA | staff_49 |
| 146 | 138544 | 2010/9/13 11:12 | 2010/11/24 16:30 | 1 | ############ | NA | NA | NA | staff_49 | NA | staff_49 |
| 146 | 138545 | 2010/9/13 11:21 | 2010/11/24 16:31 | 1 | ############ | NA | NA | NA | staff_49 | NA | staff_49 |
| 146 | 138546 | 2010/9/13 11:22 | 2010/11/24 16:31 | 1 | ############ | NA | NA | NA | staff_49 | NA | staff_49 |
| 146 | 138547 | 2010/9/13 11:36 | 2010/11/24 16:31 | 1 | ############ | NA | NA | NA | staff_49 | NA | staff_49 |
| 146 | 138548 | 2010/9/13 11:37 | 2010/11/24 16:32 | 1 | ############ | NA | NA | NA | staff_49 | NA | staff_49 |
| 146 | 138549 | 2010/9/13 11:44 | 2010/11/24 16:32 | 1 | ############ | NA | NA | NA | staff_49 | NA | staff_49 |
| 146 | 138550 | 2010/9/13 11:52 | 2010/11/24 16:32 | 1 | ############ | NA | NA | NA | staff_49 | NA | staff_49 |
| 146 | 138551 | 2010/9/13 11:58 | 2010/11/24 16:32 | 1 | ############ | NA | NA | NA | staff_49 | NA | staff_49 |
| 145 | 138553 | 2010/9/14 15:34 | 2010/11/24 16:22 | 1 | ############ | NA | NA | NA | staff_49 | NA | staff_49 |
| 145 | 138554 | 2010/9/14 15:47 | 2010/11/24 16:22 | 1 | ############ | NA | NA | NA | staff_49 | NA | staff_49 |
| 145 | 138555 | 2010/9/14 15:51 | 2010/11/24 16:22 | 1 | ############ | NA | NA | NA | staff_49 | NA | staff_49 |

*Table 16. Sample table of 6_tm_plate_results*

| PROTOCOL_ | EXPERIMENT_ID | CREATED_DATE | PLATE_ID | SAMPLE_PLATE_ID | PLATE_DATE | PLATE_ZPRIME | PLATE_AVG_LOW | PLATE_AVG_HIGH | PLATE_STD_LOW | PLATE_STD_HIGH | PLATE_ROBUST_ZPRIME | PLATE_AVG_CONT | PLATE_STD_CONT |
|---|---|---|---|---|---|---|---|---|---|---|---|---|---|
| 101 | 138443 | 2010/6/8 | 1 | 1 | NA | -3.09358 | 820.7143 | 10602.7143 | 652.3365 | 12695.4762 | -28.7 | NA | NA |
| 101 | 138444 | 2010/6/8 | 1 | 1 | NA | 0.58665 | 1143.75 | 23628.5 | 758.3576 | 2339.6832 | 0.95916 | NA | NA |
| 81 | 138445 | 2010/6/8 | 1 | 1 | NA | 0.58665 | 1143.75 | 23628.5 | 758.3576 | 2339.6832 | 0.95916 | NA | NA |
| 101 | 138446 | 2010/6/8 | 1 | 1 | NA | 0.58665 | 1143.75 | 23628.5 | 758.3576 | 2339.6832 | 0.95916 | NA | NA |
| 102 | 138447 | 2010/6/9 | 1 | 1 | NA | -2.48613 | 822.25 | 12063.125 | 603.9619 | 12458.4306 | -3.61013 | NA | NA |
| 102 | 138448 | 2010/6/9 | 1 | 1 | NA | -2.48613 | 822.25 | 12063.125 | 603.9619 | 12458.4306 | -3.61013 | NA | NA |
| 141 | 138491 | 2010/8/18 | 1 | 1 | NA | NA | NA | NA | NA | NA | NA | NA | NA |
| 141 | 138492 | 2010/8/18 | 1 | 1 | NA | NA | NA | NA | NA | NA | NA | NA | NA |
| 142 | 138493 | 2010/8/20 | 1 | 1 | NA | -1.29085 | 55907.25 | 115035.3125 | 14912.1796 | 30238.9832 | 0.70942 | NA | NA |
| 143 | 138499 | 2010/8/23 | 1 | 1 | NA | 0.75746 | 59367.8438 | 122499.9063 | 2921.5443 | 2182.4033 | 0.70942 | NA | NA |
| 144 | 138501 | 2010/8/23 | 1 | 1 | NA | 0.75746 | 59367.8438 | 122499.9063 | 2921.5443 | 2182.4033 | 0.70942 | NA | NA |
| 144 | 138502 | 2010/8/23 | 1 | 1 | NA | 0.7566 | 25075.9688 | 113602.3438 | 4510.2112 | 2672.2875 | 0.79652 | NA | NA |
| 144 | 138503 | 2010/8/23 | 1 | 1 | NA | 0.74055 | 60771.8438 | 119780.1875 | 3005.0323 | 2098.2045 | 0.67596 | NA | NA |
| 144 | 138505 | 2010/8/23 | 1 | 1 | NA | 0.80104 | 28672.625 | 112535.7813 | 3301.0331 | 2260.7274 | 0.7587 | NA | NA |
| 141 | 138506 | 2010/8/24 | 1 | 1 | NA | 0.79861 | 11009.9063 | 78455.8125 | 631.7104 | 3895.9545 | 0.85108 | NA | NA |
| 144 | 138507 | 2010/8/24 | 1 | 1 | NA | 0.72293 | 61778.3125 | 119143.5938 | 3094.8317 | 2203.191 | 0.63513 | NA | NA |
| 144 | 138508 | 2010/8/24 | 1 | 1 | NA | 0.79859 | 30693 | 111513.9688 | 3301.4065 | 2124.6144 | 0.78266 | NA | NA |
| 144 | 138512 | 2010/9/7 | 1 | 1 | NA | 0.75746 | 59367.8438 | 122499.9063 | 2921.5443 | 2182.4033 | 0.70942 | NA | NA |
| 144 | 138512 | 2010/9/7 | 2 | 2 | NA | 0.74055 | 60771.8438 | 119780.1875 | 3005.0323 | 2098.2045 | 0.67596 | NA | NA |
| 144 | 138512 | 2010/9/7 | 3 | 3 | NA | 0.72293 | 61778.3125 | 119143.5938 | 3094.8317 | 2203.191 | 0.63513 | NA | NA |
| 145 | 138514 | 2010/9/8 | 1 | 1 | NA | 0.79861 | 11009.9063 | 78455.8125 | 631.7104 | 3895.9545 | 0.85108 | NA | NA |
| 146 | 138515 | 2010/9/8 | 1 | 1 | NA | 0.75746 | 59367.8438 | 122499.9063 | 2921.5443 | 2182.4033 | 0.70942 | NA | NA |
| 146 | 138530 | 2010/9/10 | 1 | 1 | NA | 0.75746 | 59367.8438 | 122499.9063 | 2921.5443 | 2182.4033 | 0.70942 | NA | NA |

*Table 17. Sample table of 7_tm_protocols*

| PROTOCOL_ID | PROTOCOL | CREATED_DATE | DESCR | PROTOCOL_TYPE |
|---|---|---|---|---|
| 81 | BeatsonTest | 2010/2/10 14:45 | Sample from some Beatson files | SCREENING |
| 101 | PICR test | 2010/6/8 14:14 | Testing dotmatics GT 08-06-10 | SCREENING |
| 102 | IC50 test | 2010/6/9 10:19 | IC50 | SCREENING |
| 121 | Chemistry Notebook | 2010/6/16 10:47 | medchem notebook | NOTEBOOK |
| 141 | IC50 | 2010/8/18 13:51 | 10 point IC50 determination | SCREENING |
| 142 | CRT IC50 | 2010/8/20 13:49 | 10 POINT IC50 PERFORMED BY CRT | SCREENING |
| 143 | 384 IC50 manual | 2010/8/23 10:35 | 16 compounds / plate | SCREENING |
| 144 | EC50 | 2010/8/23 13:56 | 10 point EC50 determination | SCREENING |
| 145 | IC50 ENZYME | 2010/9/8 9:34 | 10 point IC50 determination | SCREENING |
| 146 | EC50 ENZYME | 2010/9/8 11:10 | 10 point EC50 determination | SCREENING |
| 161 | Chemistry Ideas | 2010/9/10 11:25 | Chemical ideas capture notebook | NOTEBOOK |
| 181 | PRIMARY SCREEN (loss of signal) | 2010/11/4 14:11 | Primary screen - loss of signal assay | SCREENING |
| 201 | PRIMARY SCREEN (gain of signal) | 2010/11/9 9:10 | Primary screen - gain of signal assay | SCREENING |
| 222 | Biology Notebook | 2011/1/31 11:57 | Biology Electronic Lab Notebbok | NOTEBOOK |
| 241 | IC50 test protocol | 2012/4/16 15:32 | IC50 test protocol | SCREENING |
| 261 | EC50 ENZYME PARTIAL INHIBITION | 2012/6/8 0:00 | 10 point EC50 determination | SCREENING |
| 281 | THERMAL SHIFT | 2012/10/12 9:36 | Thermal shift assay | HCS |
| 301 | ELN - Cyprotex Upload | 2012/11/6 9:56 | This is used to upload Cyprotex DMPK data from Excel reports | NOTEBOOK |
| 321 | EC50 CELL | 2012/12/19 10:55 | EC50 dose response cell biology | SCREENING |
| 341 | IC50 CELL | 2013/1/21 0:00 | 8 point dose response | SCREENING |
| 361 | ELN - Thermal Shift Upload | 2013/4/19 0:00 | This is used to upload the output from the ThermalShift R package to Dotmatics | NOTEBOOK |

*Table 18. Sample table of event_log_tm*

| PROTOCOL_ID | EXPERIMENT_ID | START_DATE | TARGET_NAME | PROGRAM | ANONYMISATION_ISID | Year | PROTOCOL | PROTOCOL_TYPE | COMLETE_DATE |
|---|---|---|---|---|---|---|---|---|---|
| 121 | 138999 | 2011/1/6 16:35 | #N/A | #N/A | staff_27 | 2011 | Chemistry Notebo | NOTEBOOK | 2011/1/13 13:17 |
| 121 | 139000 | 2011/1/6 16:42 | #N/A | #N/A | staff_27 | 2011 | Chemistry Notebo | NOTEBOOK | 2011/1/13 13:17 |
| 121 | 139001 | 2011/1/6 16:59 | #N/A | #N/A | staff_27 | 2011 | Chemistry Notebo | NOTEBOOK | 2011/2/8 10:25 |
| 121 | 139008 | 2011/1/10 9:51 | #N/A | #N/A | staff_22 | 2011 | Chemistry Notebo | NOTEBOOK | 2011/1/17 8:40 |
| 121 | 139009 | 2011/1/10 10:05 | #N/A | #N/A | staff_22 | 2011 | Chemistry Notebo | NOTEBOOK | 2011/1/10 15:45 |
| 121 | 139010 | 2011/1/10 15:47 | #N/A | #N/A | staff_22 | 2011 | Chemistry Notebo | NOTEBOOK | 2011/3/2 8:45 |
| 121 | 139011 | 2011/1/11 13:23 | #N/A | #N/A | staff_22 | 2011 | Chemistry Notebo | NOTEBOOK | NA |
| 121 | 139012 | 2011/1/11 13:26 | #N/A | #N/A | staff_22 | 2011 | Chemistry Notebo | NOTEBOOK | NA |
| 121 | 139013 | 2011/1/12 8:49 | #N/A | #N/A | staff_22 | 2011 | Chemistry Notebo | NOTEBOOK | 2011/1/17 14:17 |
| 121 | 139014 | 2011/1/12 8:58 | #N/A | #N/A | staff_22 | 2011 | Chemistry Notebo | NOTEBOOK | 2011/1/17 14:21 |
| 121 | 139015 | 2011/1/12 11:31 | #N/A | #N/A | staff_22 | 2011 | Chemistry Notebo | NOTEBOOK | 2011/1/19 13:11 |
| 121 | 139016 | 2011/1/12 13:10 | #N/A | #N/A | staff_05 | 2011 | Chemistry Notebo | NOTEBOOK | 2011/5/10 16:08 |
| 121 | 139017 | 2011/1/12 13:23 | #N/A | #N/A | staff_05 | 2011 | Chemistry Notebo | NOTEBOOK | 2011/5/10 16:09 |
| 121 | 139018 | 2011/1/12 17:04 | #N/A | #N/A | staff_05 | 2011 | Chemistry Notebo | NOTEBOOK | 2011/5/10 16:15 |
| 121 | 139019 | 2011/1/13 9:10 | #N/A | #N/A | staff_34 | 2011 | Chemistry Notebo | NOTEBOOK | 2011/1/17 14:19 |
| 121 | 139020 | 2011/1/13 10:02 | #N/A | #N/A | staff_34 | 2011 | Chemistry Notebo | NOTEBOOK | 2011/1/17 14:19 |
| 121 | 139021 | 2011/1/13 10:26 | #N/A | #N/A | staff_27 | 2011 | Chemistry Notebo | NOTEBOOK | 2011/1/20 14:28 |
| 121 | 139022 | 2011/1/13 10:59 | #N/A | #N/A | staff_27 | 2011 | Chemistry Notebo | NOTEBOOK | 2011/7/13 16:20 |
| 121 | 139023 | 2011/1/13 13:05 | #N/A | #N/A | staff_27 | 2011 | Chemistry Notebo | NOTEBOOK | 2011/1/20 14:28 |
| 121 | 139024 | 2011/1/14 11:54 | #N/A | #N/A | staff_34 | 2011 | Chemistry Notebo | NOTEBOOK | 2011/2/8 14:50 |
| 121 | 139028 | 2011/1/17 9:45 | #N/A | #N/A | staff_22 | 2011 | Chemistry Notebo | NOTEBOOK | 2011/1/19 13:03 |
| 121 | 139029 | 2011/1/17 9:53 | #N/A | #N/A | staff_22 | 2011 | Chemistry Notebo | NOTEBOOK | 2011/1/19 13:04 |
| 121 | 139030 | 2011/1/17 10:04 | #N/A | #N/A | staff_27 | 2011 | Chemistry Notebo | NOTEBOOK | 2011/7/13 16:21 |
| 121 | 139031 | 2011/1/17 10:08 | #N/A | #N/A | staff_27 | 2011 | Chemistry Notebo | NOTEBOOK | 2011/7/13 16:21 |

### 3.2 Process mining tools and techniques

This section is an overview of the process mining techniques and algorithm which have been used in the implementation. In this study, the pre-process procedures have been performed by Excel, implementations of process mining techniques and algorithms have been executed by Disco and ProM.

As a closed source software, the Disco has embedded the process mining techniques and algorithm into its functions. The functions of discovery (construct a model for a given event log), analysis, interactive dashboard, and filter have been all embedded in Disco originally. The illustrations will come along when these functions are being performed. The process maps of the models are based on Christian's Fuzzy miner (Günther & Van Der Aalst, 2007). The CSV file can be read by Disco without conversion. All of the functions of Disco have been applied in this research.

The ProM is an open source software come with plenty of plugins. The functions of plugins have been categorised as discovery, conformance checking, enhancement, filtering, analytics, toggle interactive filters, and toggle batch filter plugins. The plugin action information is available on the interface for illustration. These plugins below have been used in this research.

Conversion plug-ins implement conversions between different data formats, for instance from CSV to XES format. The XES is an Extensible Markup Language (XML) based standard. The XES has become the standard and adopted by the IEEE (Institute of Electrical and Electronic Engineers, Inc.) Task Force on Process Mining as the default interchange format for event logs (Van der Aalst et al., 2011).

Mining plugins implement some mining algorithms (Alpha miner, Inductive Miner-infrequent (IMf) miner, Activity Cluster, Causal activity, Graphviz Petri Net Visualisation, Inductive Miner, Fuzzy Model Miner) that construct Petri net based on event logs (Van Dongen, de Medeiros, Verbeek, Weijters, & Van Der Aalst, 2005) Analysis plugins implement some property analysis on some mining result such as Dotted chart.

Social network miner uses the event log file to generate a social network map of people (Van der Aalst, 2004). It requires the event log to contain the originator/ resource element.

### 3.3 Ethical concerns

The ethical concerns of this research have been classified as “low risk”. The potential risk of participating are the same as those experienced in everyday life. The ethics application has been approved by Information School which has been attached in the appendix.

The research data will be stored on my laptop and the University’s Google Drive which is password protected. They can be accessed by myself, my supervisor, the school’s examination officer. ProM, Disco, and Excel have been used to analysis the data. The personal data about the participants are in anonymised format to ensure appropriate protection. Personal data is only present in the staff ID, user ID and will be associated with an arbitrary code number. All of the research data will be deleted 3 months after the dissertation being completed and marked.

The informed consent has been obtained from a member of staff - Adnana Tudose in the CRUK, which has been attached in the appendix, but this study has not use any information of interview. There are no human research participants and human tissue involves to the research activities. The potential participant is the data provided by Drug Discovery Unit of Cancer Research UK. CRUK has agreed to provide their scientific processes data which include such as batch, project, supplier, type, program, protocol, experiment, assay, staff ID and user ID.

# 4. Chapter Four Result

## 4.1 Case study one - process log of experiments

### 4.1.1 Descriptive statistics

There was 44 staff to execute the number of 26240 experiments in the year 2011-2015 as Figure 22 shows, including 494 experiments were executed by unknown staff. It means there were missing data between activities and staff. The total average number was 583, the median was 408 for experiments being executed in 2011-2015 by each staff. Staff_40 executed the largest number of experiments with 2610.

Figure 19, 20 and 21 shows the number of experiments being executed by staff in total and annually 2011-2015. There was 26 staff have executed 3370 experiments in the year 2011; 30 staff have executed 5048 experiments in the year 2012; 44 staff have executed 5698 experiments in the year 2013; 44 staff have executed 5095 experiments in the year 2014; 30 staff have executed 7029 experiments in the year 2015.

The average number of experiments being executed in 2011 was 125; in 2012 was 163; in 2013 was 196; in 2014 was 182, in 2015 was 227.

Staff_35 executed the largest number of experiments 441 in 2011; Staff_20 executed 812 in 2012; Staff_20 executed 613 in 2013; Staff_40 executed 819 in 2014; Staff_47 executed 1007 in 2015.

*Figure 19. The sum of the number of experiments executed by staff, 2011-2015*

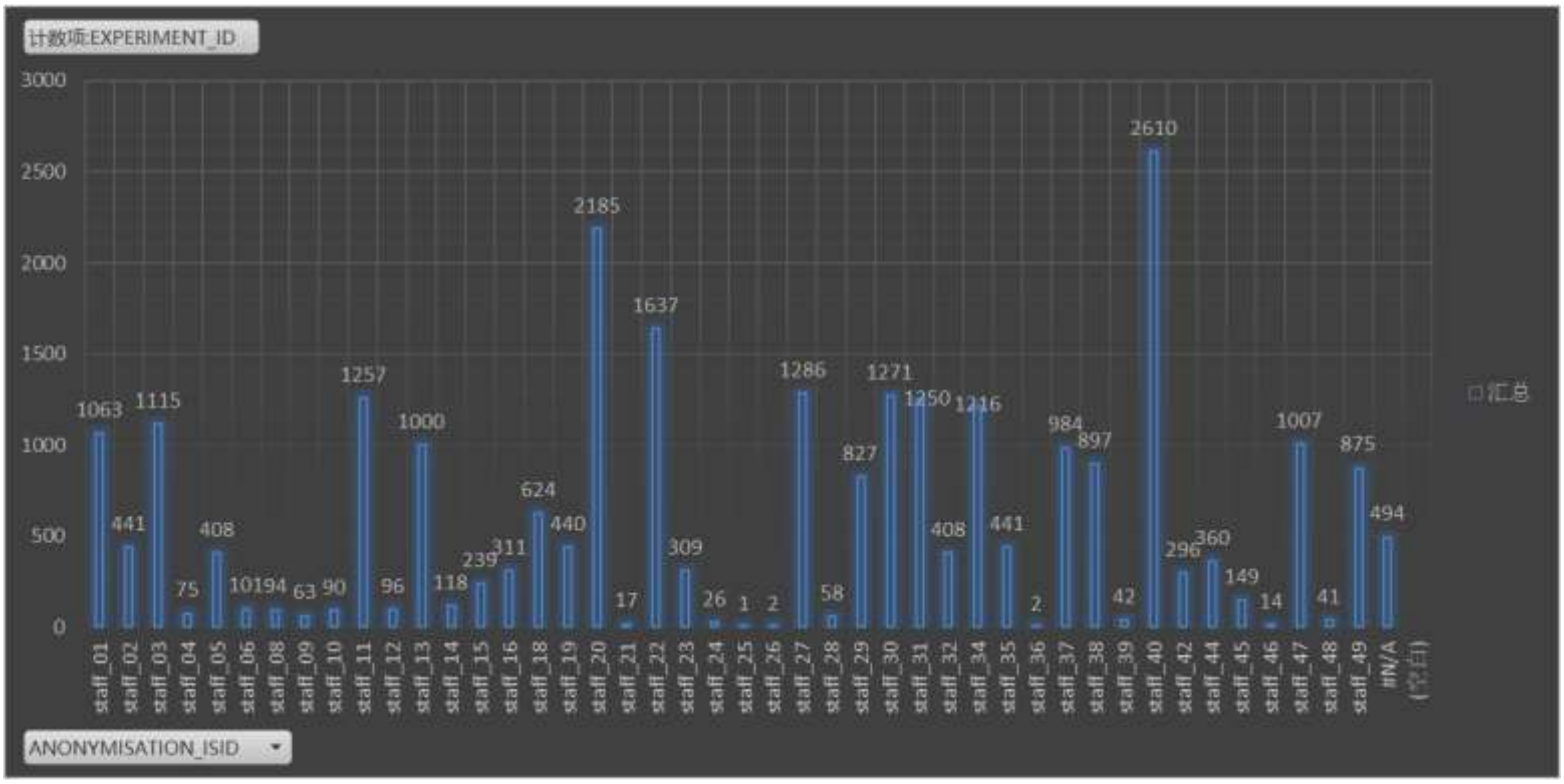

*Figure 20. The sum and average of the number of experiments annually 2011-2015*

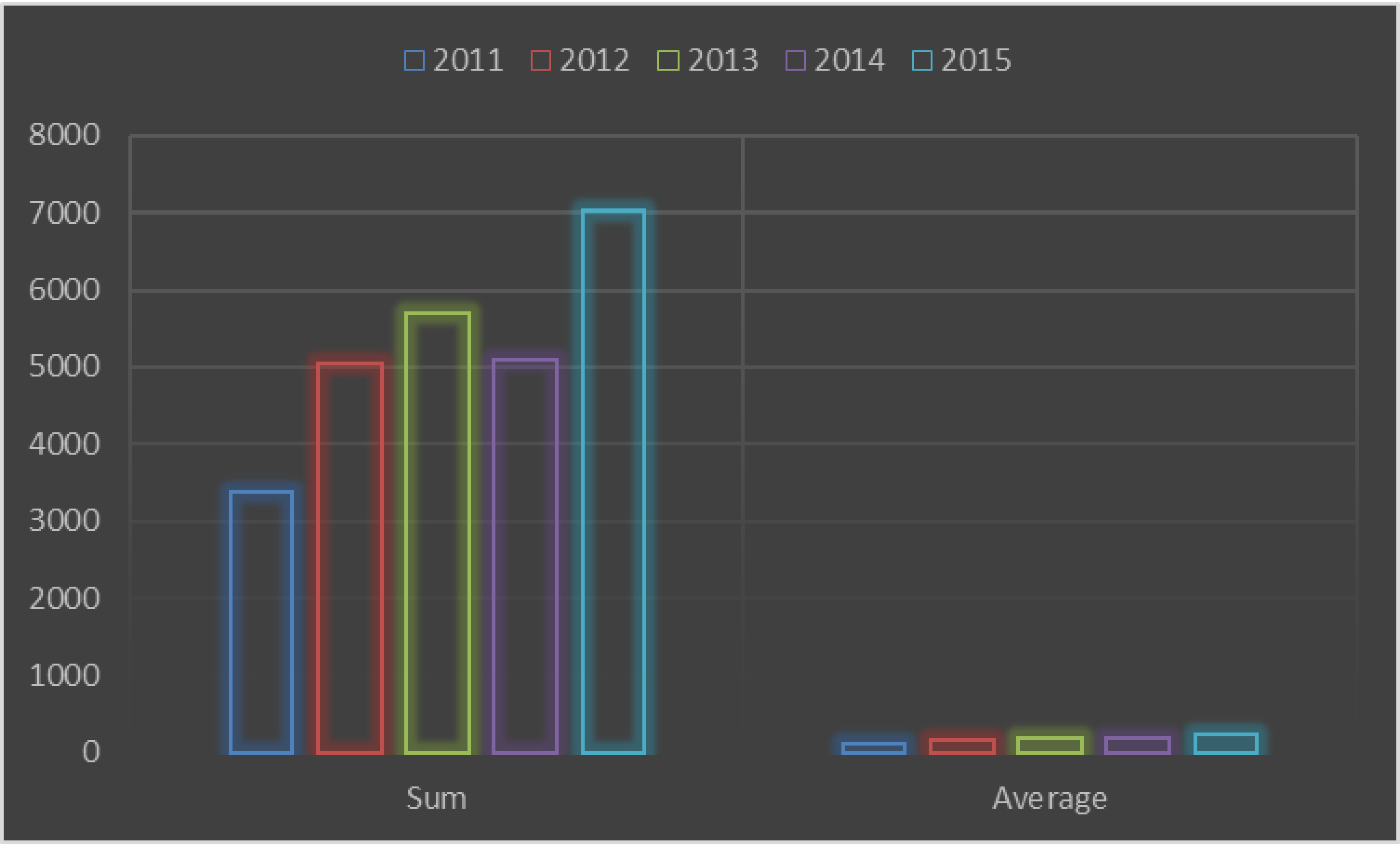

*Figure 21. The number of experiments executed by staff annually 2011-2015*

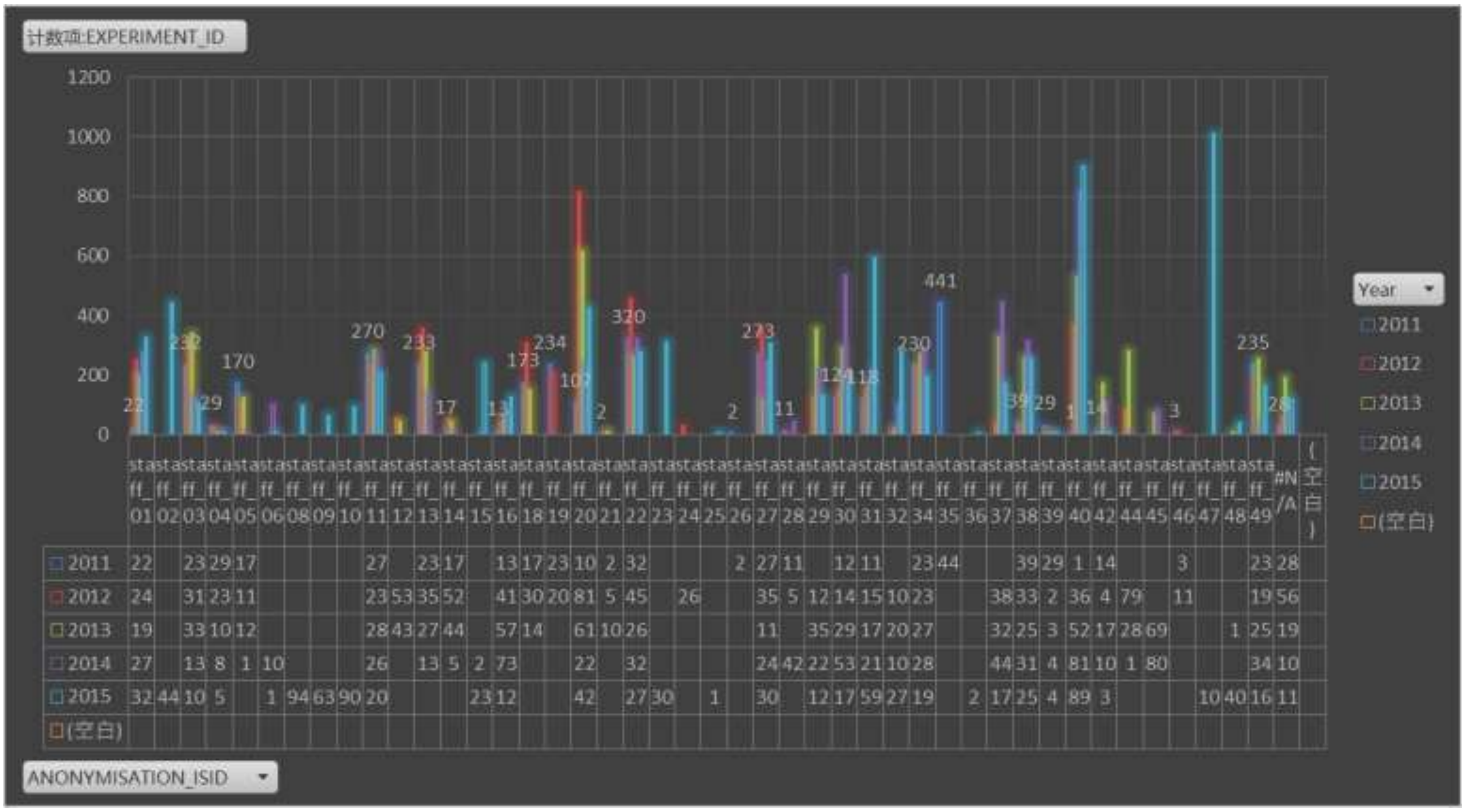

*Figure 22. The descriptive statistics overview for experiments 2011-2015 of Disco software*

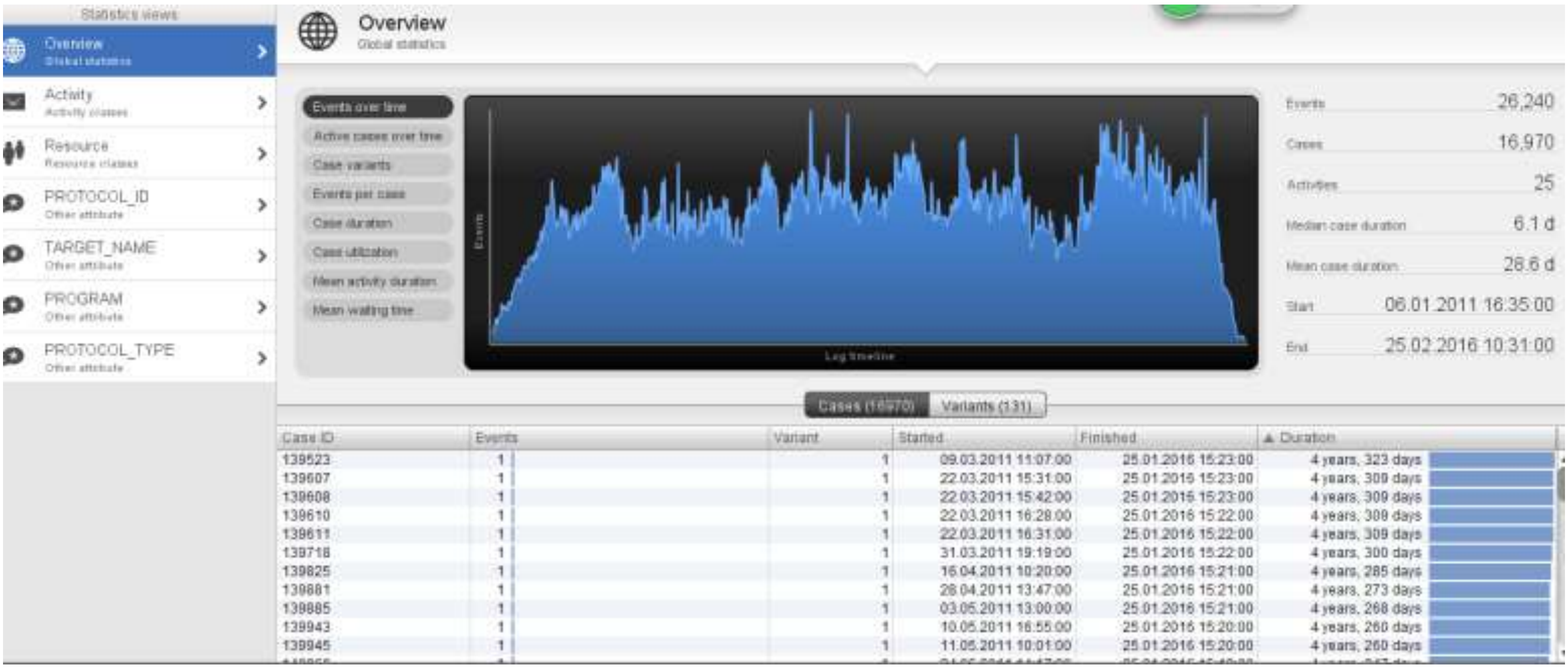

*Figure 23. The descriptive statistics of activities for experiments 2011-2015 of Disco software*

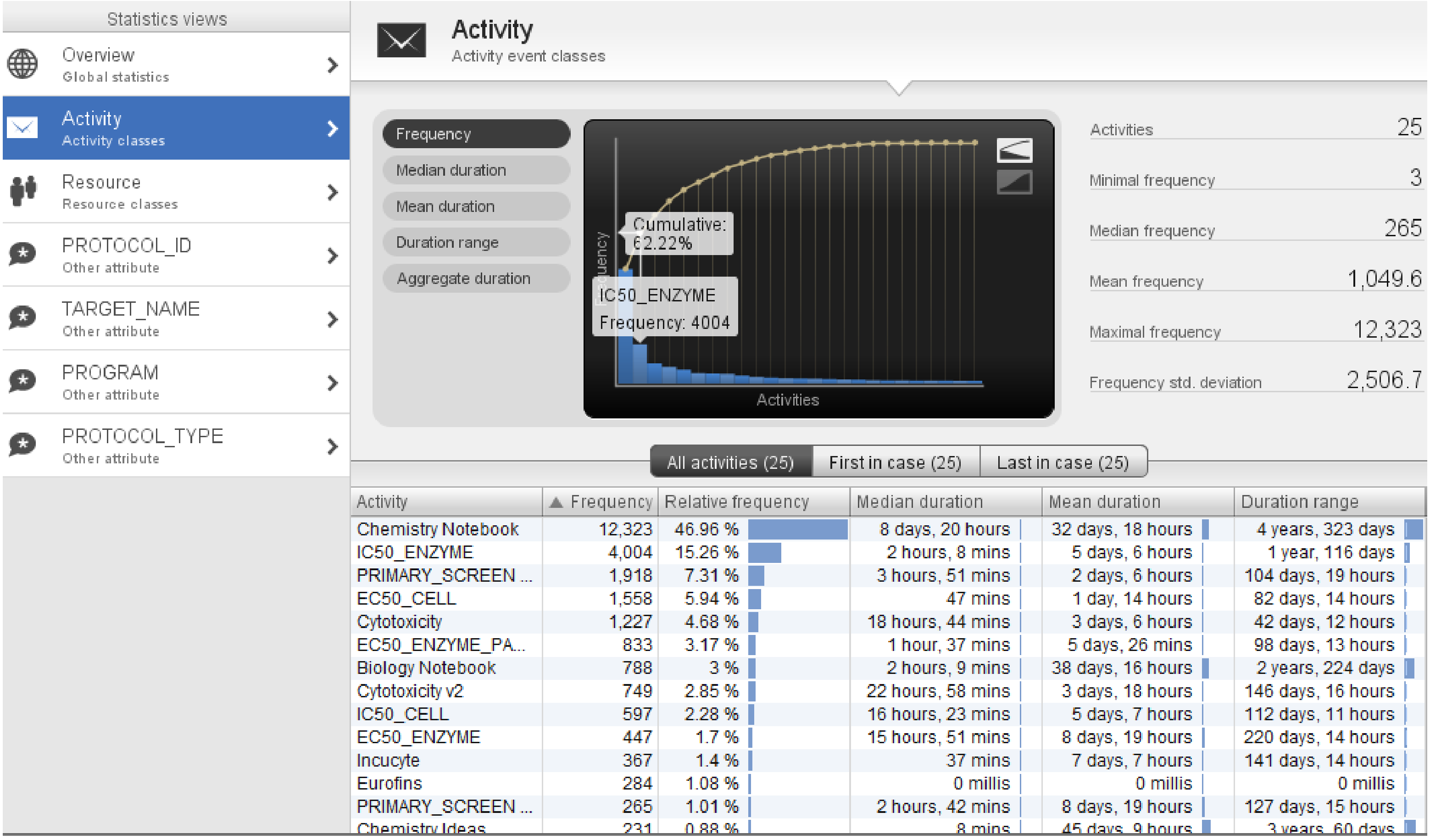

*Figure 24. The descriptive statistics of resources for experiments 2011-2015 of Disco software*

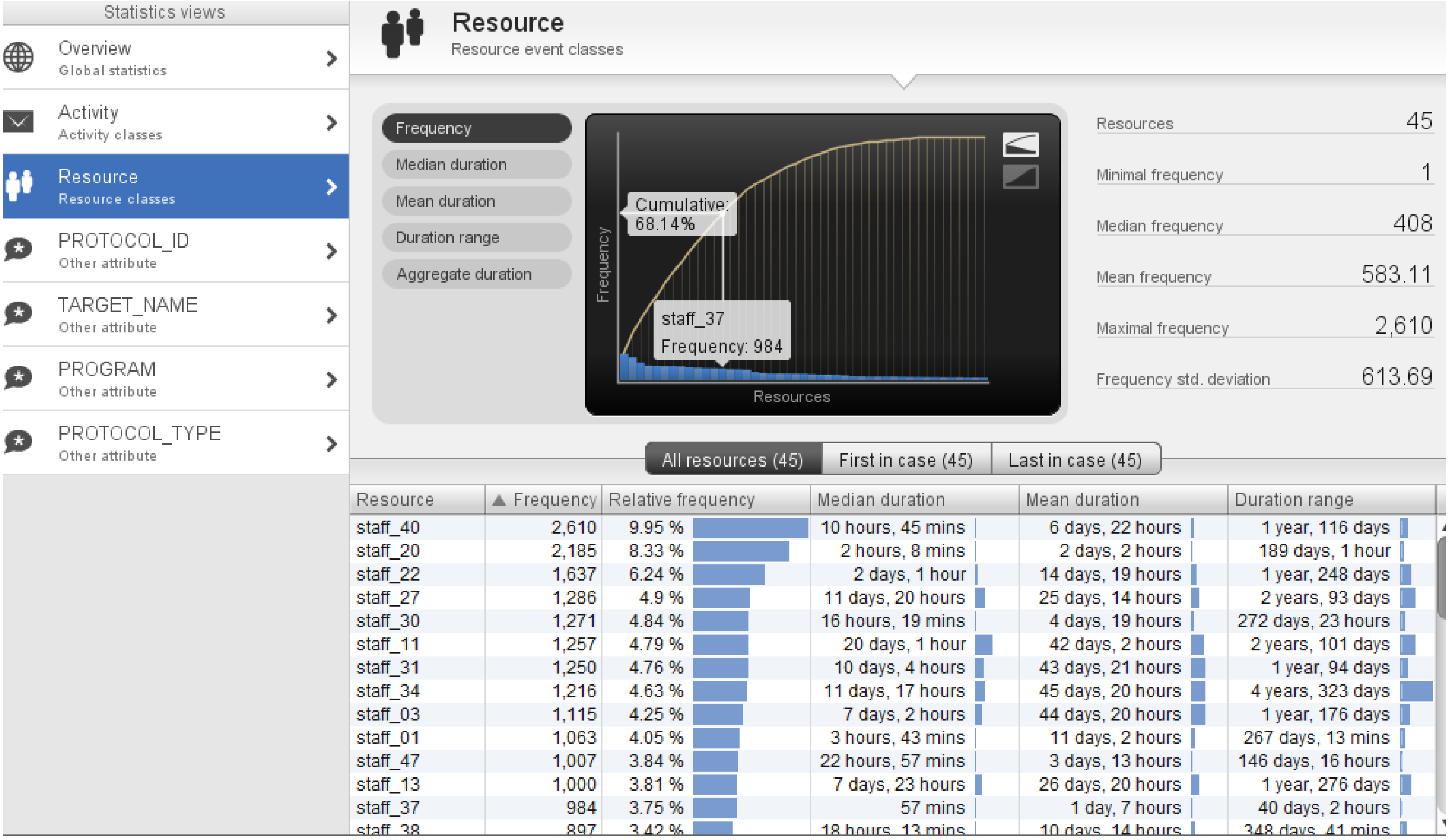

As Figure 22, 23, 24 shows, there are 26240 events, 16970 cases, 25 activities, 45 resources with missing value as "NA" in the log. The mean case duration is 28.6 days, the median case duration is 6.1 days, the minimum case duration is 0 millisecond, and the maximum is 4 years 323 days.

### 4.1.2 Interactive dashboards and dynamic visualisations

#### *4.1.2.1 Control flow perspective*

The process maps Figure 25 were generated by Disco which was based on Christian's Fuzzy miner (Günther & Van Der Aalst, 2007) with Disco further development. The start of the process was represented by the triangle symbol at the top, the end of the process was represented by the stop symbol at the bottom of the map. Arrows visualise the process flows between activities, boxes represent the activities. Activities occurred from the very beginning to the very end of the process were pointed by dashed arrows (Rozinat, n.d.).

As Figure 26 show, the thickness of paths and colouring of activities indicate main paths and highlight frequent activities of the process flows. The activity Chemistry Notebook was having the biggest amount of cases – 12323 which were highlighted in dark blue. The activity IC50_ENZYME was the second biggest with 4004 cases which were highlighted in light blue. The activity Cytotoxicity has 172 loops by itself.

*Figure 25. Full process map of experiment event log 2011-2015 of ProM software*

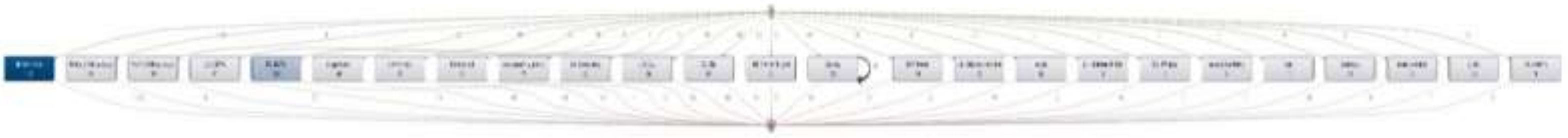

*Figure 26. Enlarge of full process map of experiment event log 2011-2015 of ProM software*

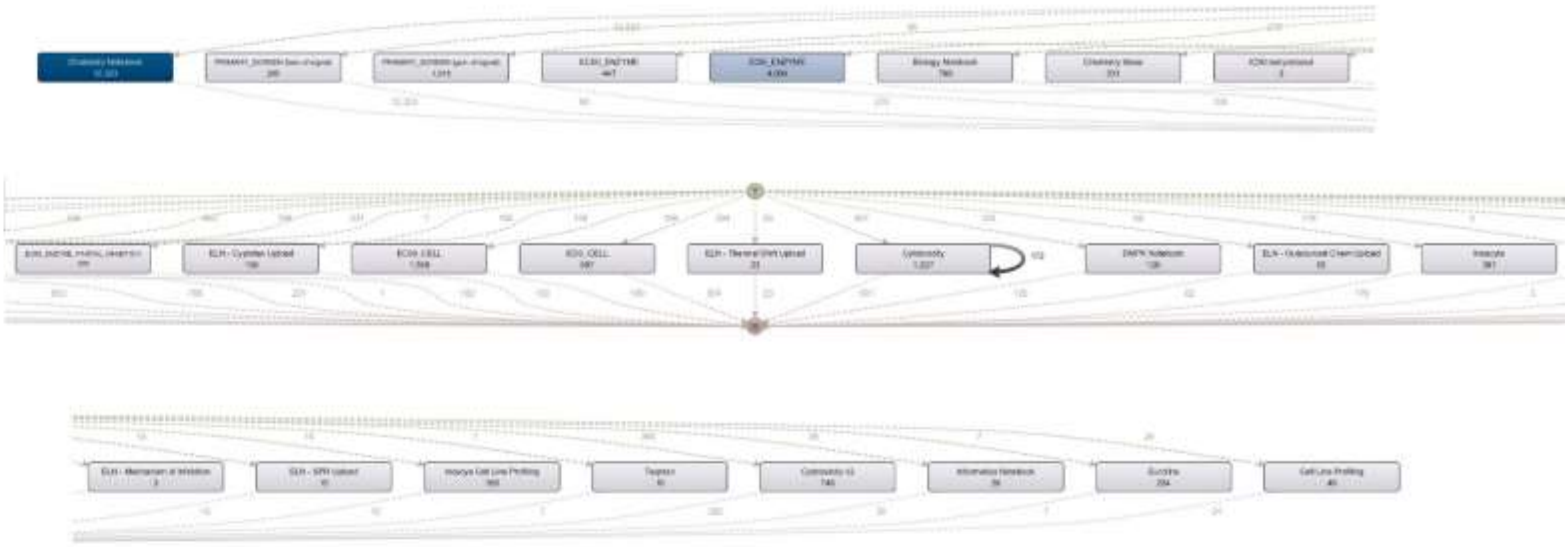

However, the Figure 27 generated by ProM with Inductive Miner-infrequent (IMf) miner (Leemans, Fahland, & van der Aalst, 2016) shows there were 18 activities it has more loop activities, which is different from Disco’s. It is because for ProM software, the empty cells (missing value) of complete date rows were set to exclude (sparse) in the conversion. This rule was applying all below figures generated by ProM.

*Figure 27. Animation of full process map of experiment event log 2011-2015 of ProM software*

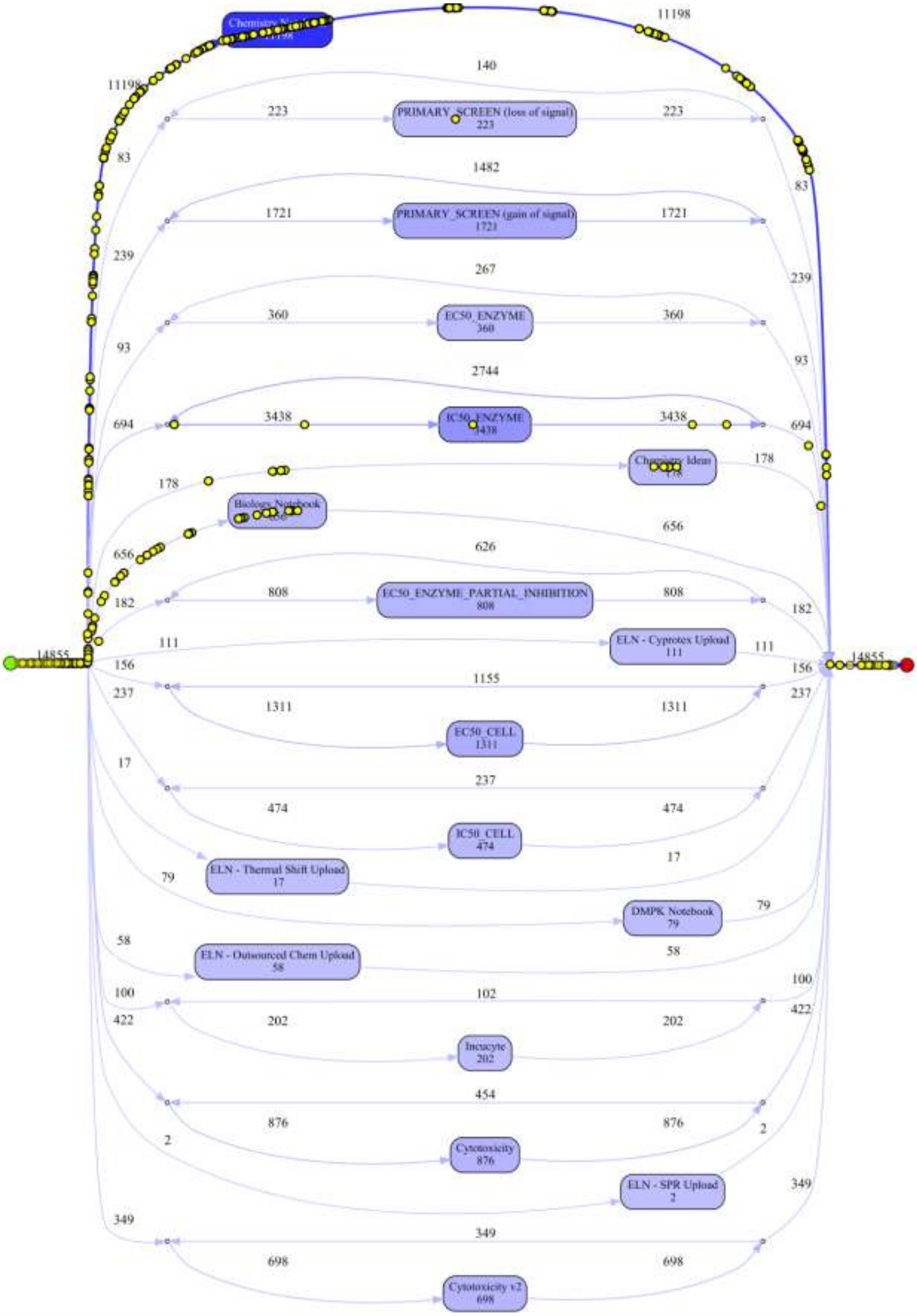

The activity cluster array plugin for ProM has the purpose of mining higher-level, repetitive clusters which do not consider the ordering of low-level events (Process mining research tools application, 2014). There were 21 clusters being clustered in Figure 28.

*Figure 28. Activity Cluster array of experiment event log 2011-2015 of ProM software*

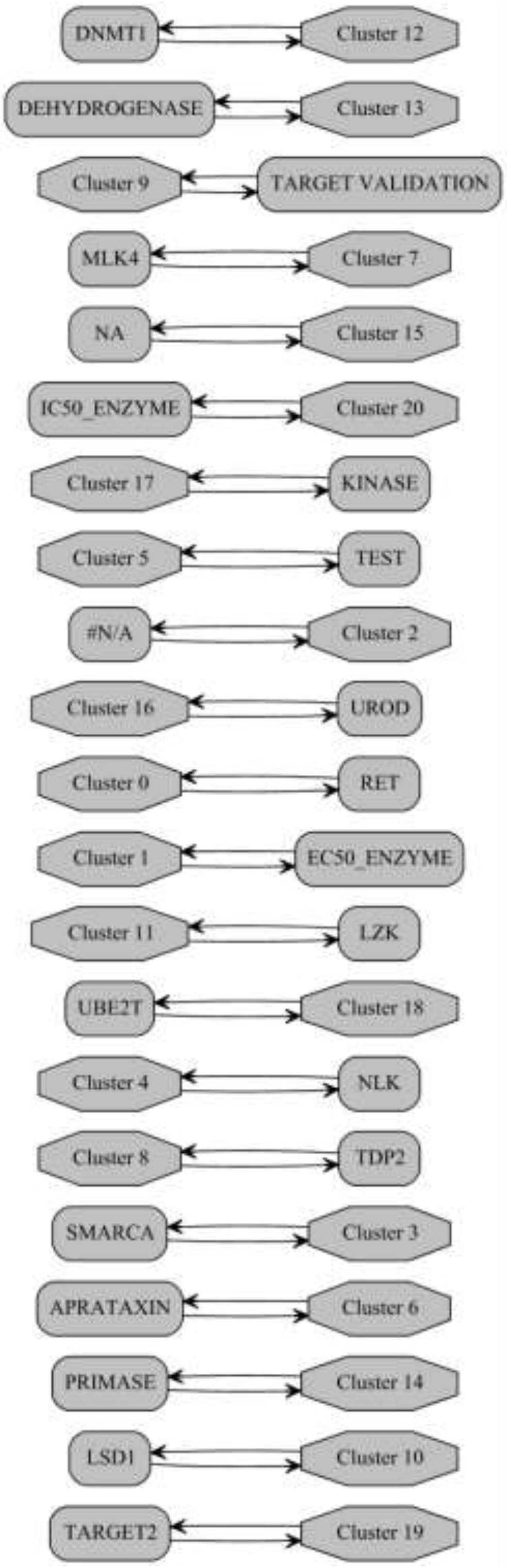

### *4.1.2.2 Performance perspective*

The biggest repetition was PRIMARY_SCREEN (gain of Signal) with 125 cases as Figure 29 shows.

*Figure 29. Repetition frequency process map of experiment event log 2011-2015 of Disco software*

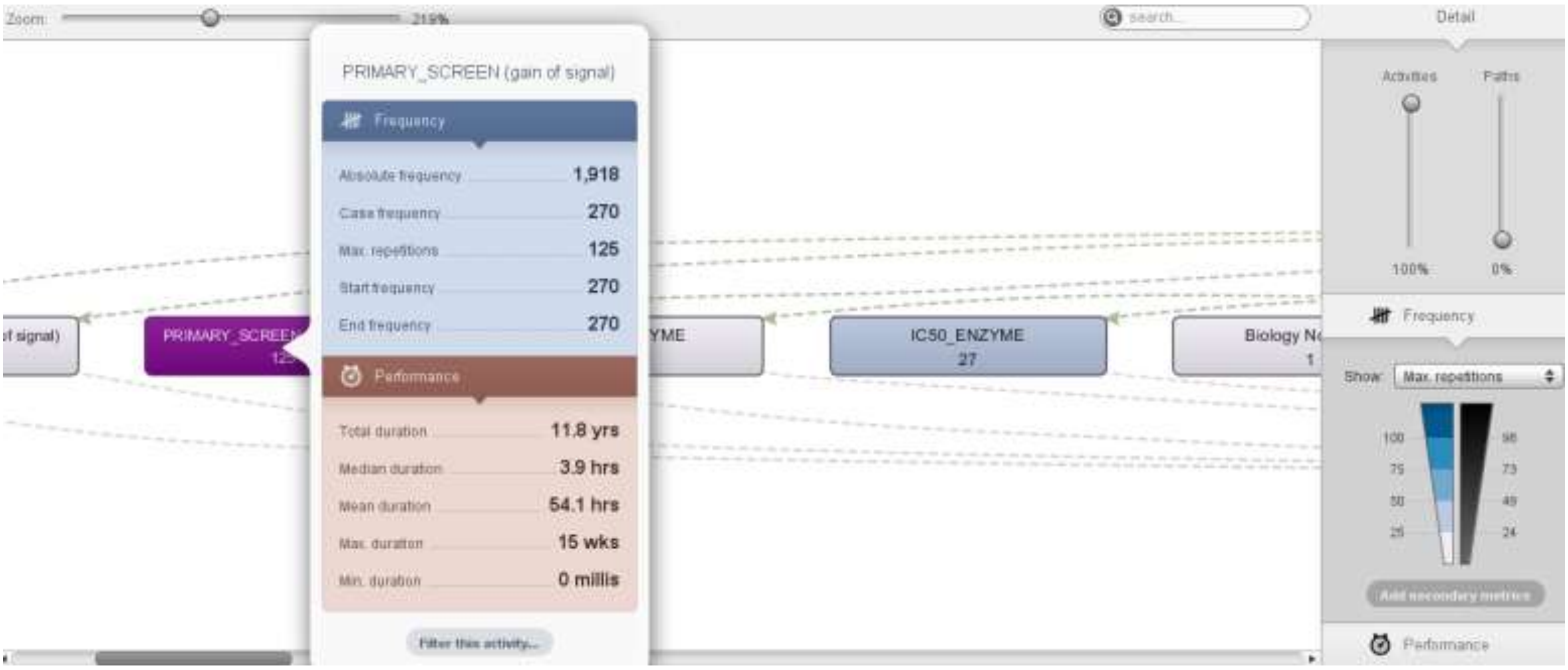

The longest mean duration performance was ELN – Thermal Shift Upload with 22.5 weeks as Figure 30 shows. The second longest mean duration performance was ELN – Cyprotex Upload with 14.1 weeks.

*Figure 30. Mean duration performance process map of experiment event log 2011-2015 of Disco software*

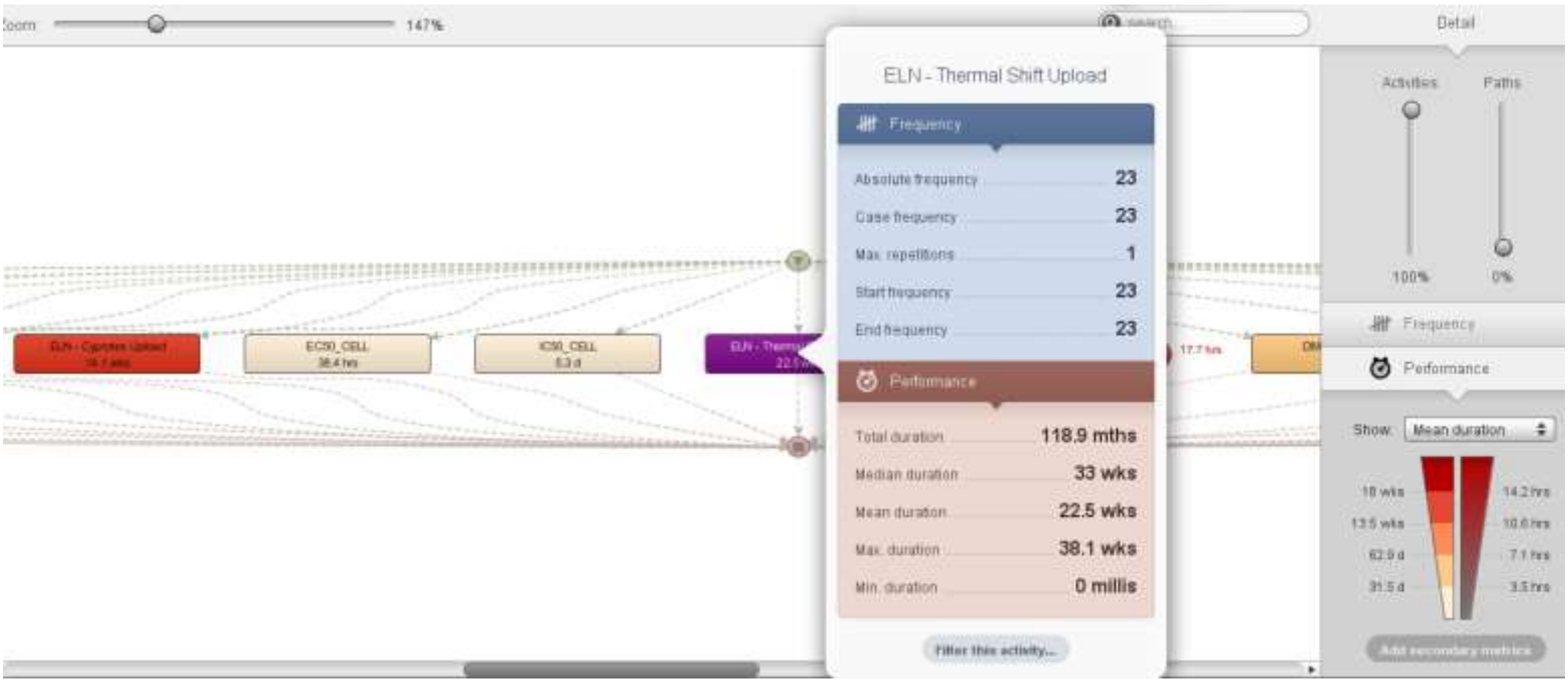

Over the years from 2011 to 2015, the number of events and cases were all having fluctuation. The year 2011, 2012 reached the highest point at the second half of the year. While the year 2014 and 2015 reached the highest point in the first half of the year as Figure 31 and 32 show.

*Figure 31. The number of events over time 2011-2015 of Disco software*

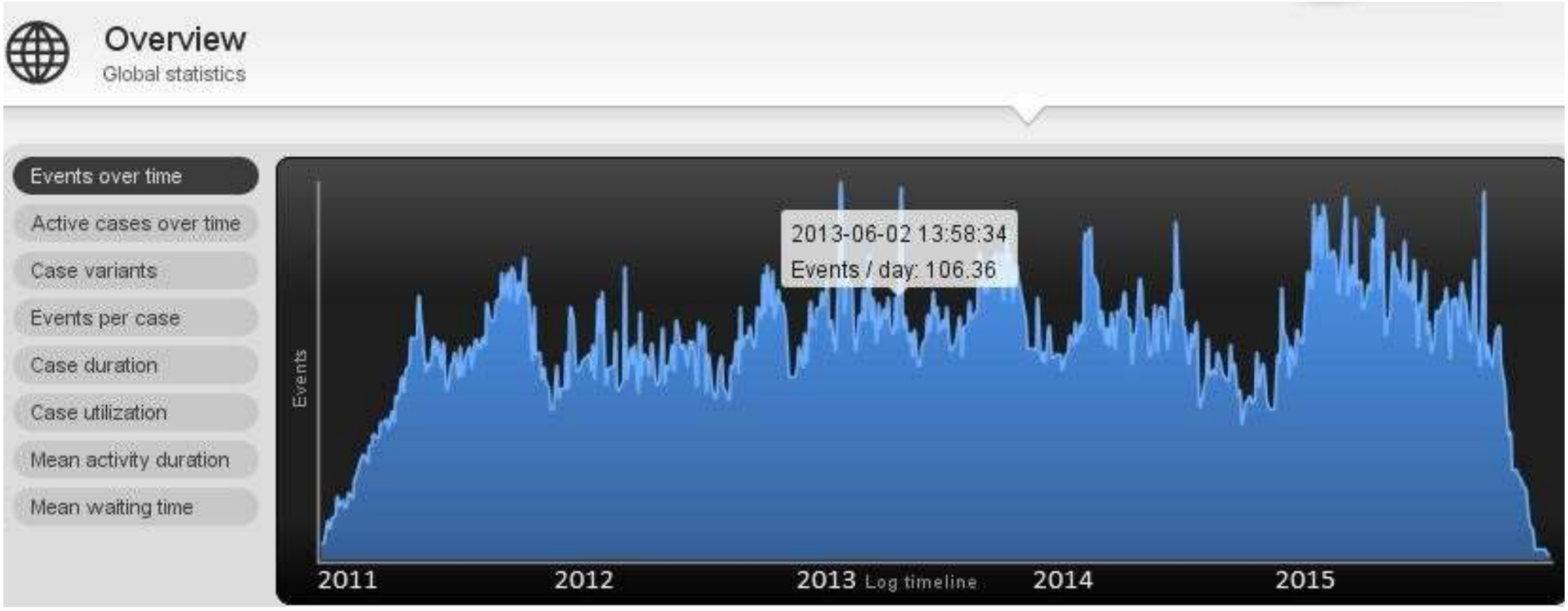

*Figure 32. The number of cases over time 2011-2015 of Disco software*

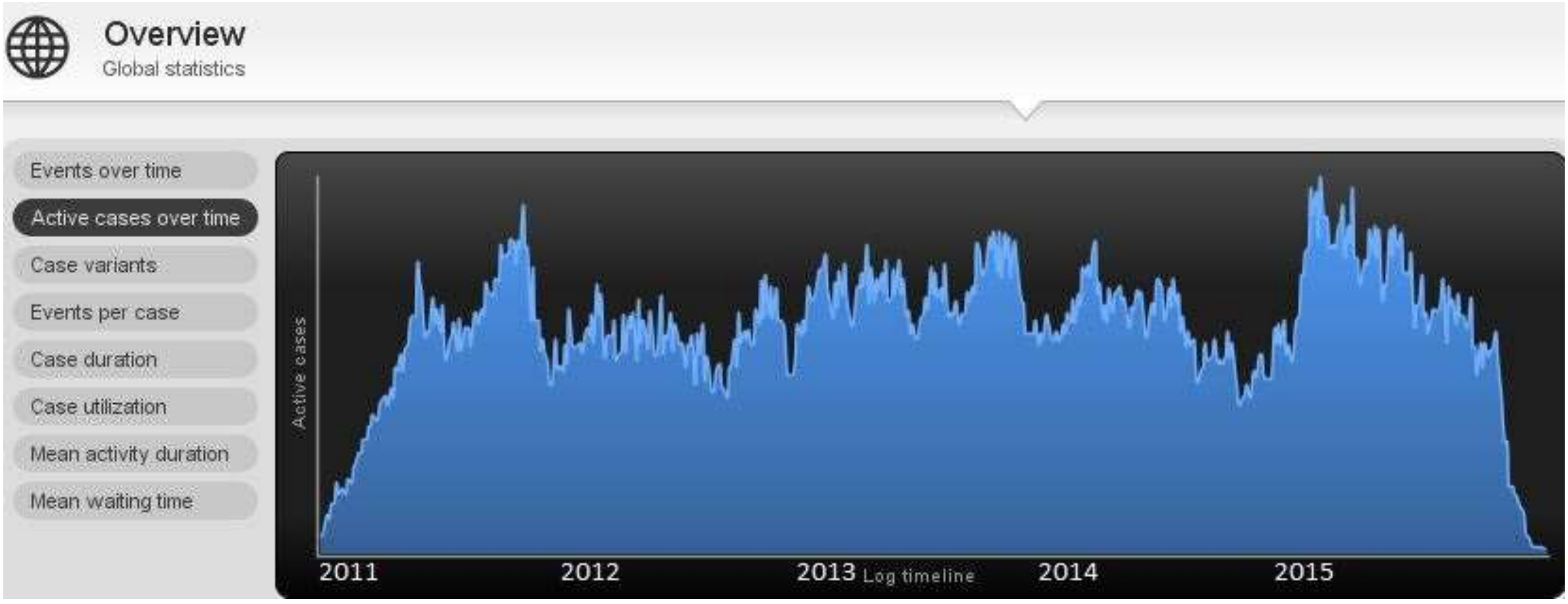

Figure 33 shows the animated dynamic visualised process map of experiment event log at the time 04/11/2013. The animation can be played along with the timeline at the bottom of the dashboard. The spots move along with dashed arrow represented the cases were being performed. The blue boxes represented the activities were being executed, while grey boxes represented the activities was not yet be executed or has ceased.

*Figure 33. Animation process map of experiment event log 2011-2015 of Disco software*

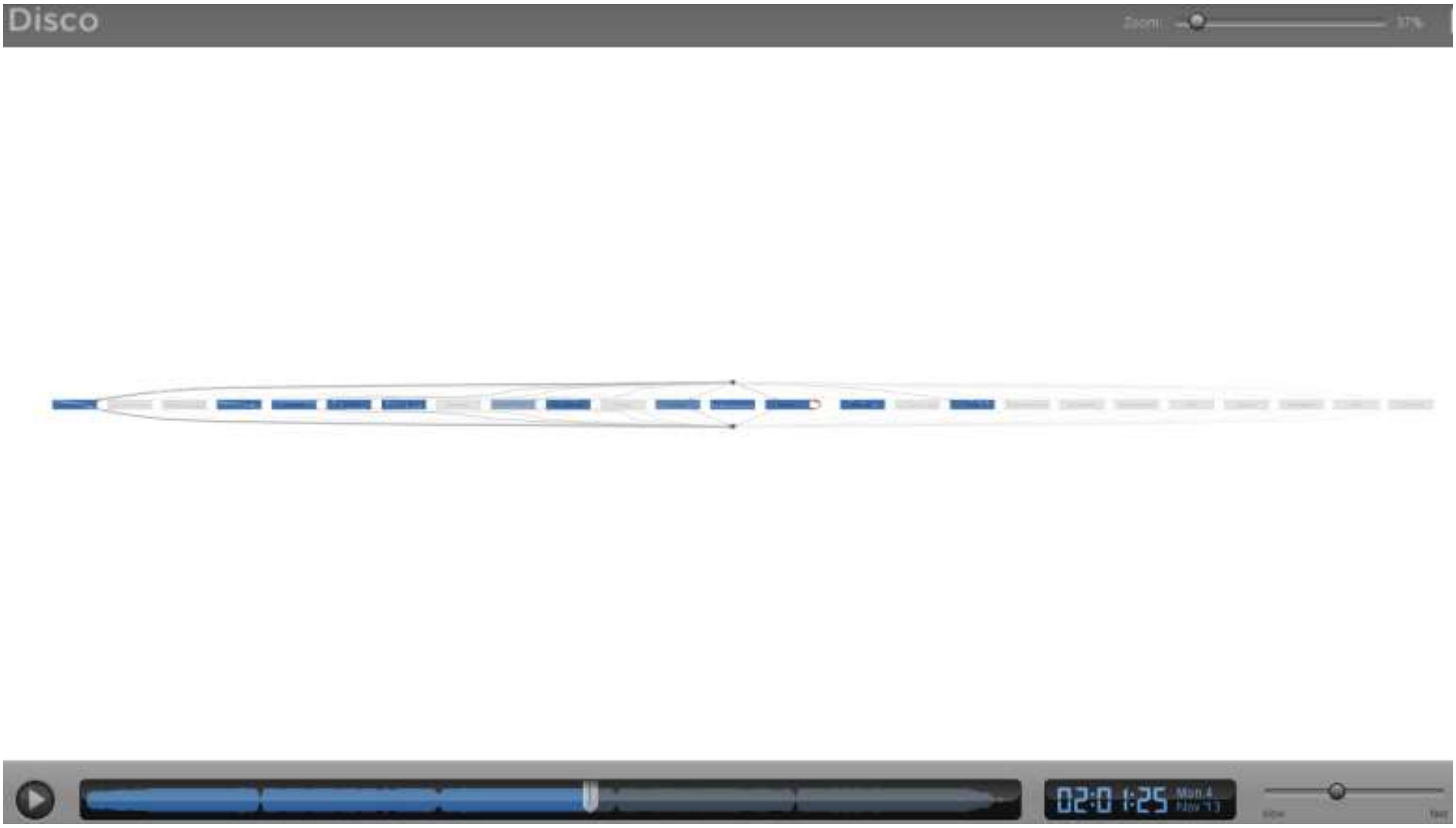

*Figure 34. Legend of dots (applied to Figures 35 to 40)*

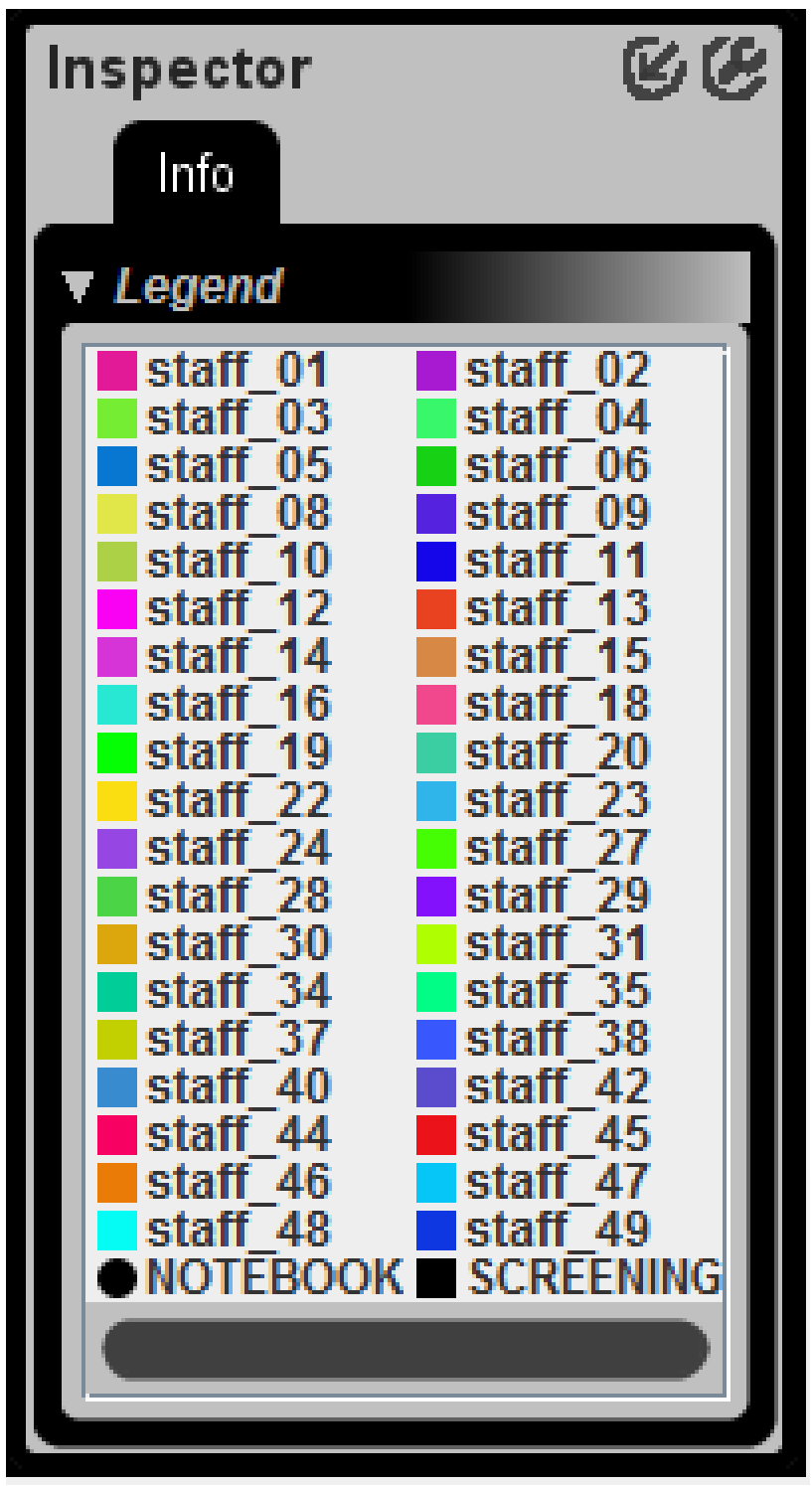

Figure 35 shows the activities change overtime. The activities Biology Notebook, Chemistry Notebook, EC50_ENZYME, IC50_ENZYME, PRIMARY_SCREEN (gain of signal) were executed all the time from 2011 to 2015. The activity Chemistry Ideas was executed from 2011 to 2014, then it stopped in the second half of 2014. The activities Cytotoxicity, DMPK Notebook, EC50_CELL, ELN-SPR Upload, ELN-Outsourced Chem Upload, ELN-Thermal Shift Upload, Incucyte only started to be executed in 2013.

*Figure 35. Dotted chart of activities being executed overtime 2011-2015 of ProM software*

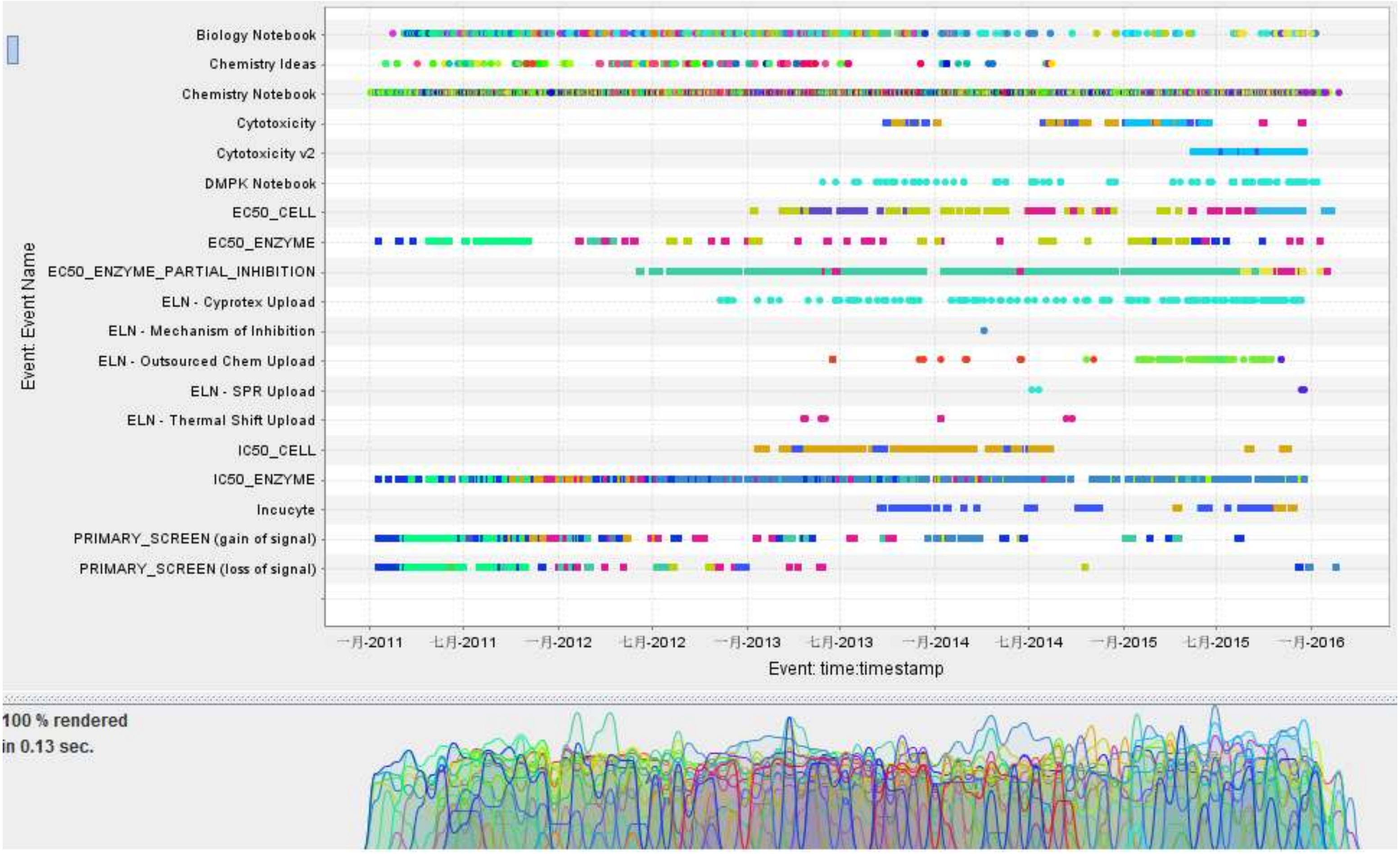

Figure 36 shows the cases increase overtime, which imply that there were sharply increasing cases in 2013 and 2015 including started and completed cases. Both Disco and ProM can also filter the data as detail as months, weeks, days, and minutes.

*Figure 36. Dotted chart of cases overtime 2011-2015 of ProM software*

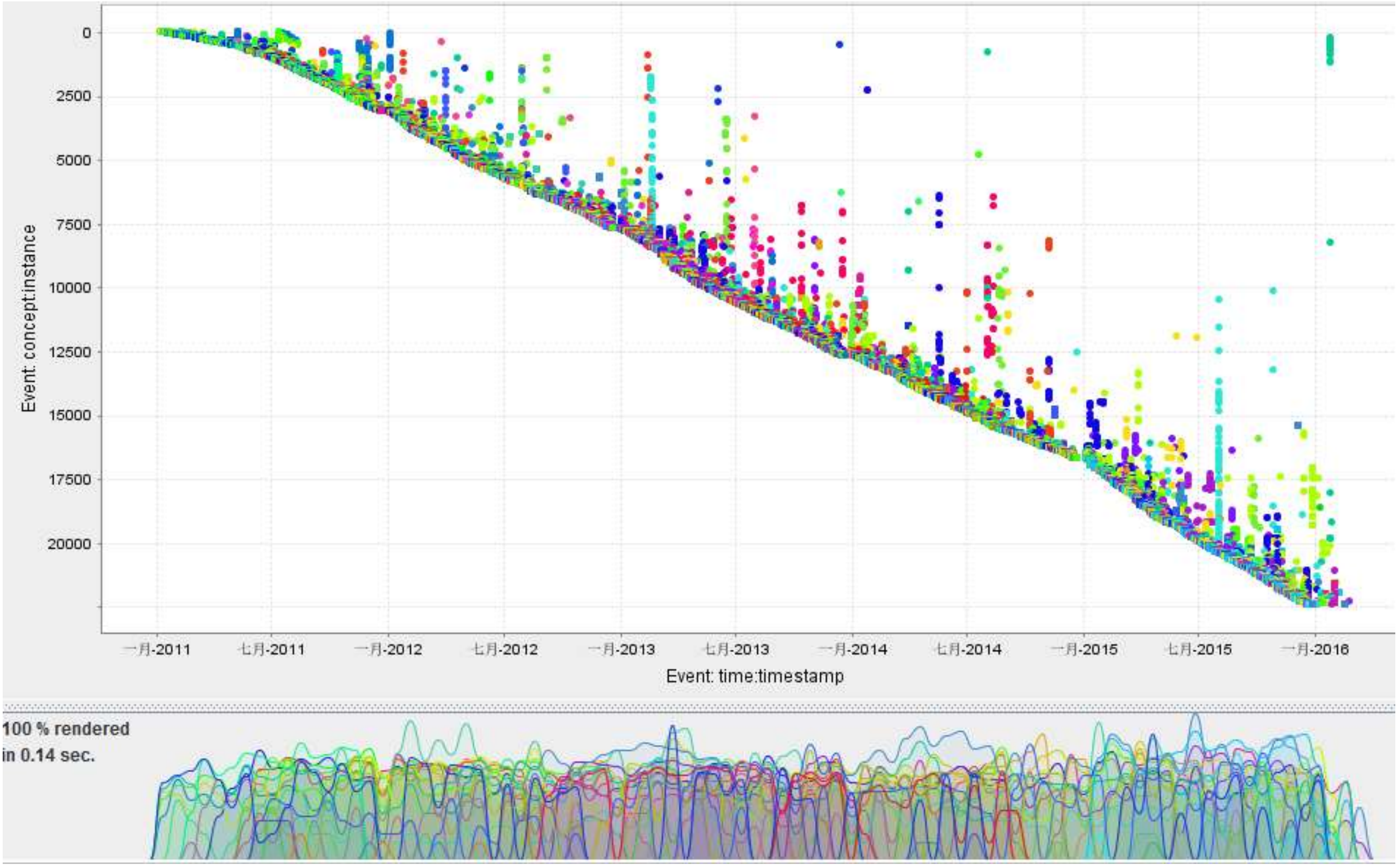

#### *4.1.2.3 Organisational perspective*

The figure 37 implies that which activities were mainly executed by whom; who work on the same activities. The circle and square dot represent protocol types were Notebook and Screening relatively. The activities ELN-Cyprotex Upload, ELN-Mechanism of Inhibition, ELN-Thermal Shirt Upload were executed by staff_16, staff_40, and staff_01 representatively only.

The Figure 37 demonstrates that there were mainly two types of staff based on their activities. The first type of staff responsible for most of the notebook activities except for Biology Notebook, they did not execute any screening activities. The second type of staff responsible for all screening activities and one Biology Notebook activity only. The exceptional researcher was staff_40, he/she executed one extra ELN -Mechanism of Inhibition notebook activity for a short period of time in 2014.

*Figure 37. Dotted chart of activities executed by staff 2011-2015 of ProM software*

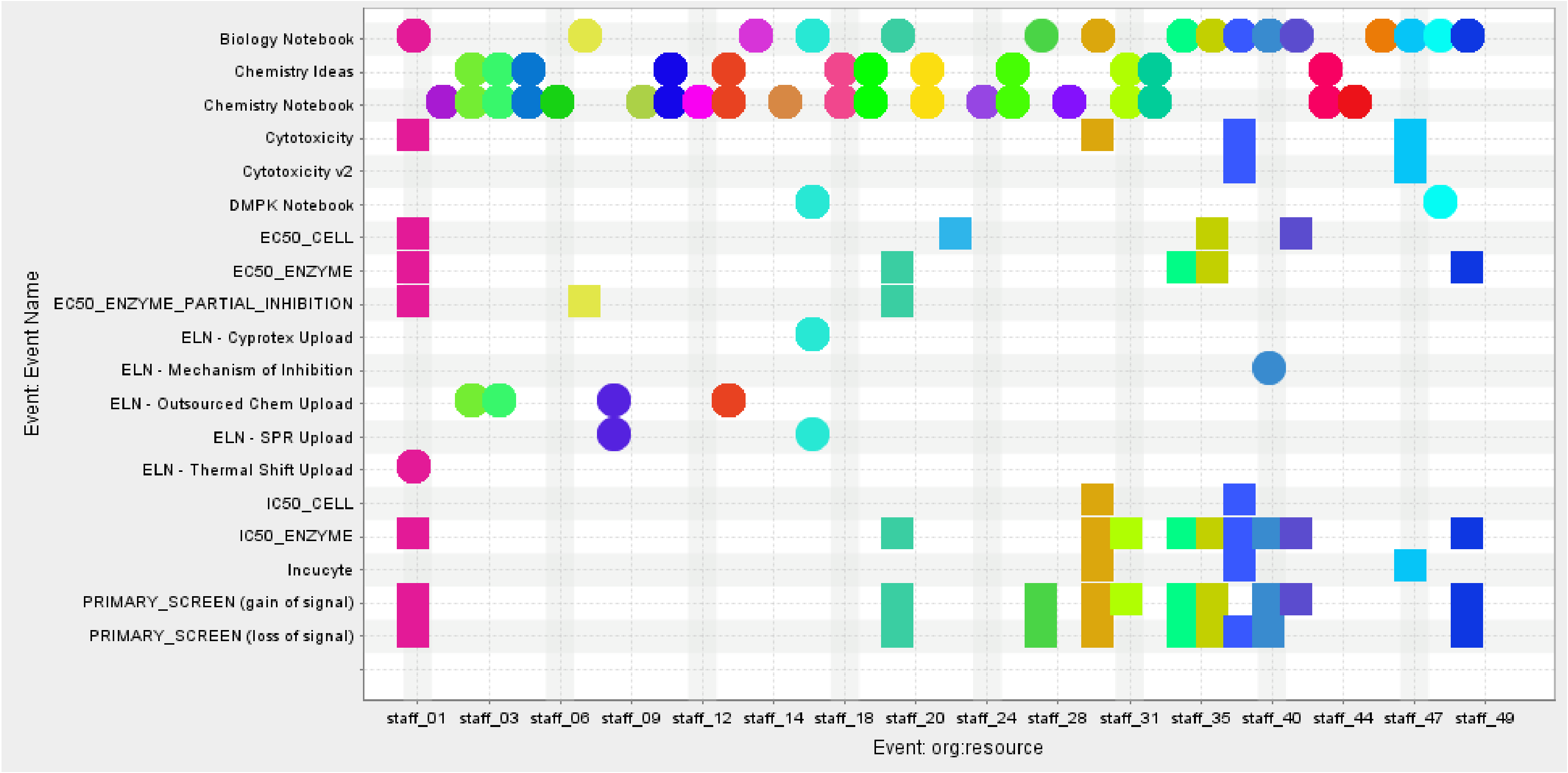

Figure 38, 39 and 40 indicate one similar story of the data quality. The activities “#N/A” and “NA” were generated due to missing value, which was because the executors did not input the relevant data field particularly at the time of implementation. These missing value were generated by the staff who only responsible for notebook activities mainly but not any screening activities. The exceptional researchers were staff_20 and staff_35, they generate missing value, while executing both notebook and screening activities. Another exceptional researcher staff_48 only executes notebook activities but without generating missing value.

*Figure 38. Dotted chart of programs overtime 2011-2015 of ProM software*

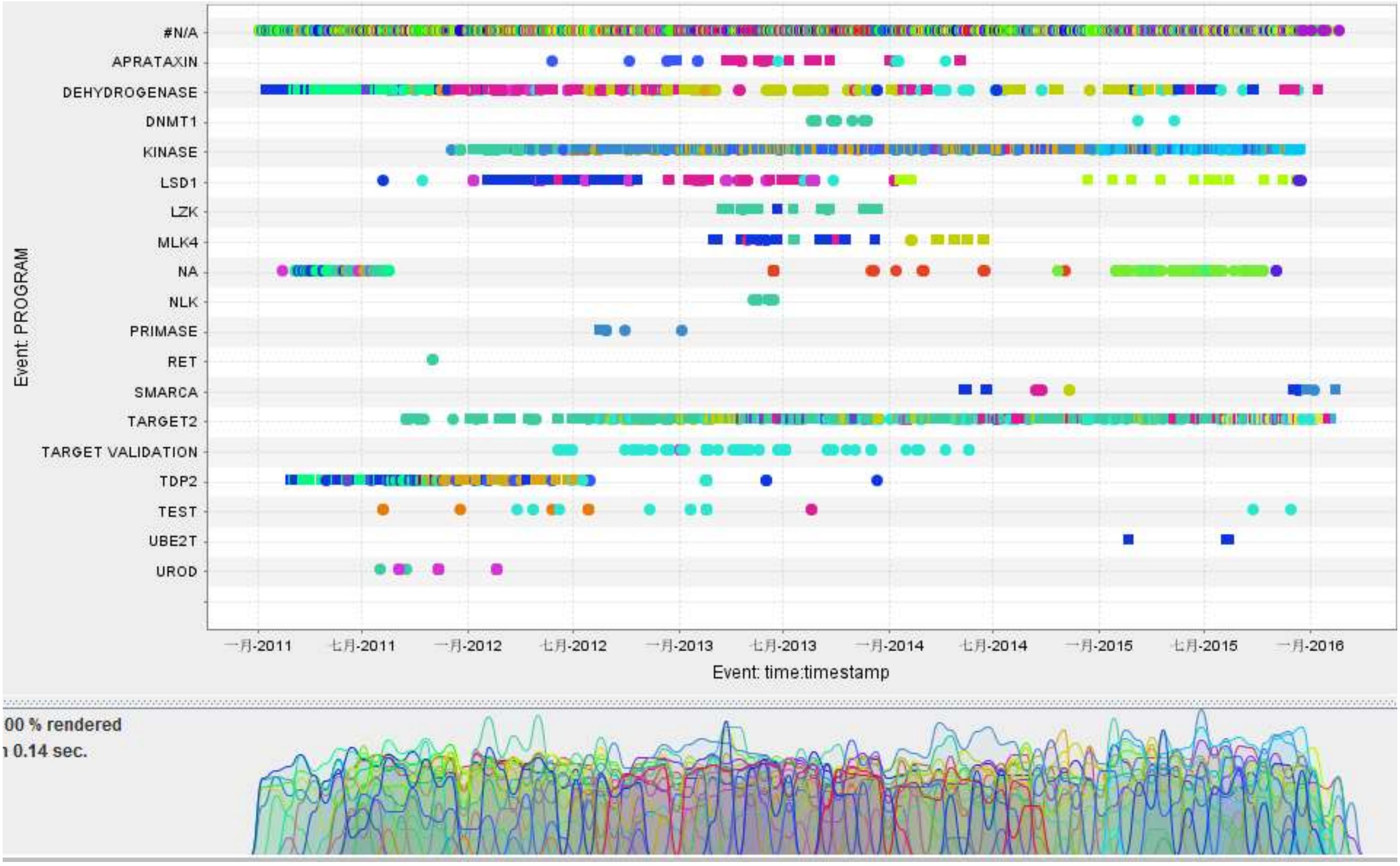

*Figure 39. Dotted chart of programs executed by staff 2011-2015 of ProM software*

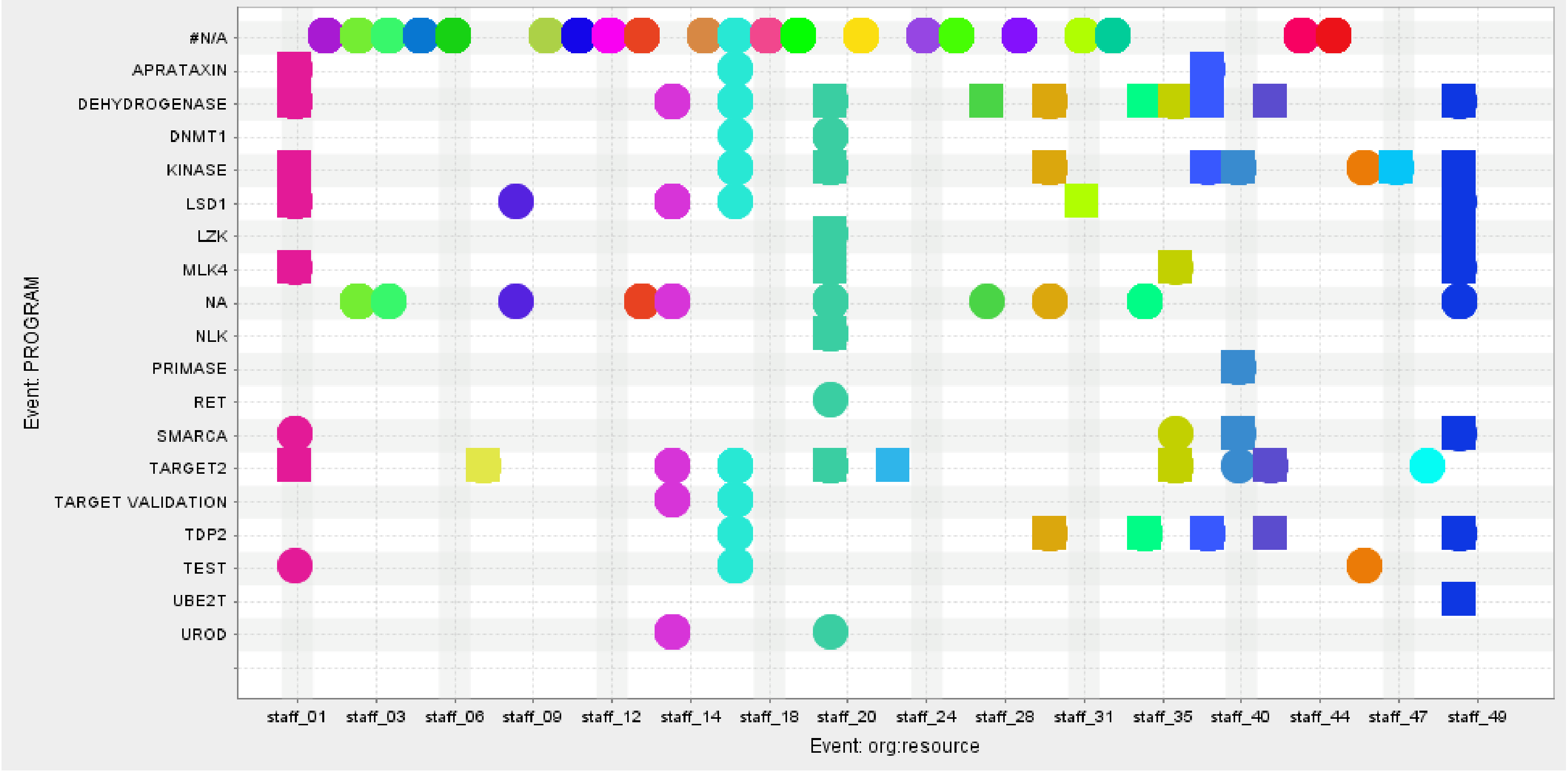

*Figure 40. Dotted chart of targets executed by staff 2011-2015 of ProM software*

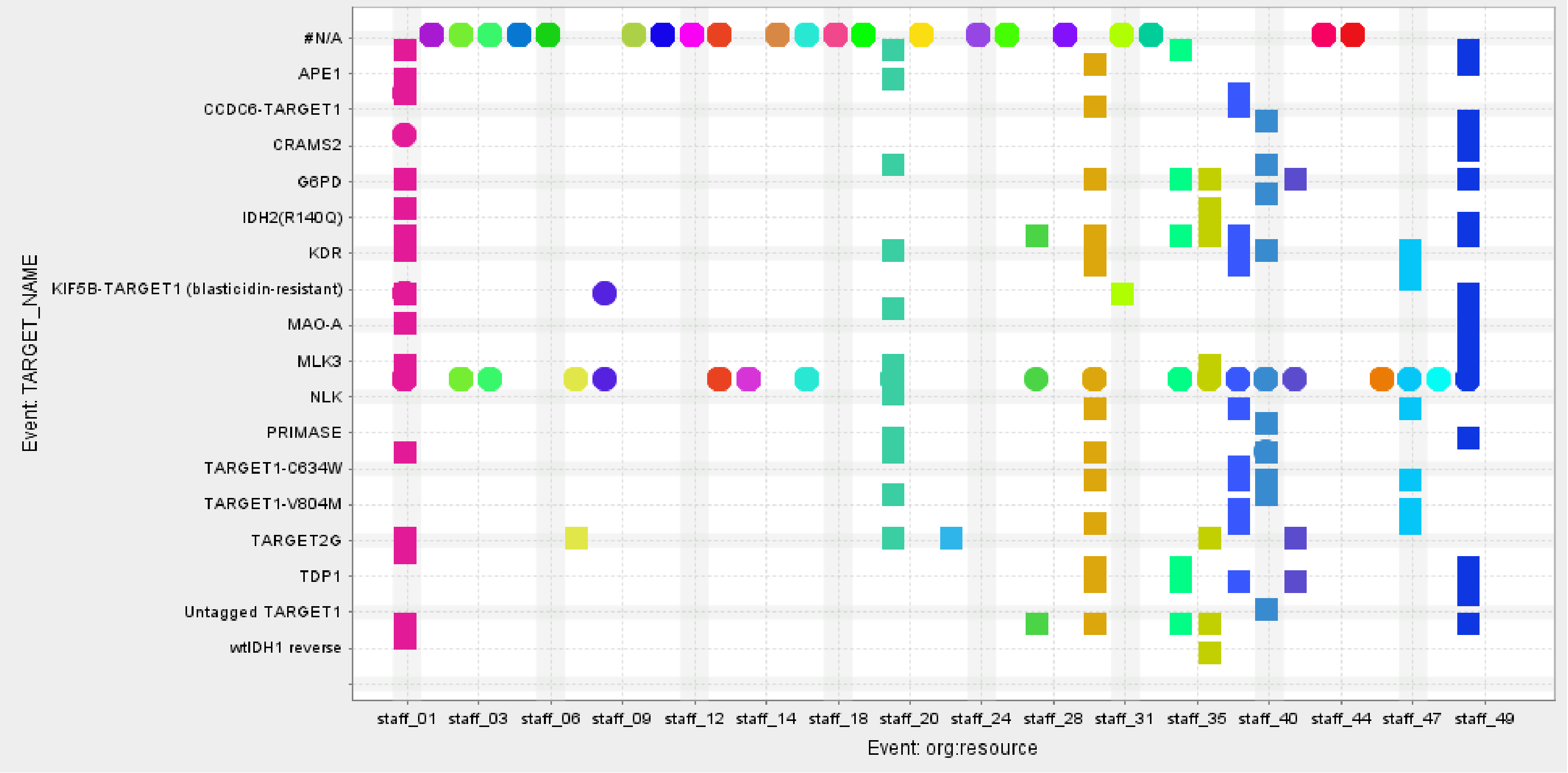

## 4.2 Case study two - Process log of registrations

### 4.2.1 Descriptive statistics

There was 23 staff to execute the number of 32215 registration activities in the year 2011-2015 as Figure 44 shows. Its 23 staff were far less than the number of 44 staff who execute experiment activities. It is believed that because of registration were mainly screening activities, experiments were including both screening and notebook activities.

There were no registration activities were executed by unknown staff, which means there was no missing data between activities and staff. The total average number was 1401, the median was 149 for registrations being executed in 2011-2015 by each staff. Staff_39 executed the largest number of experiments with 23919.

Figure 41, 42 and 43 shows the number of registrations being executed by staff in total and annually 2011-2015. There were 12 staff have executed 2768 registrations in the year 2011; 17 staff have executed 12323 experiments in the year 2012; 15 staff have executed 1674 experiments in the year 2013; 13 staff have executed 986 experiments in the year 2014; 13 staff have executed 14464 experiments in the year 2015.

Staff_39 executed the largest number of experiments 1015 in 2011; Staff_39 executed 1041 in 2012; Staff_13 executed 550 in 2013; Staff_34 executed 267 in 2014; Staff_39 executed 1248 in 2015.

*Figure 41. The sum of the number of registrations executed by staff, 2011-2015*

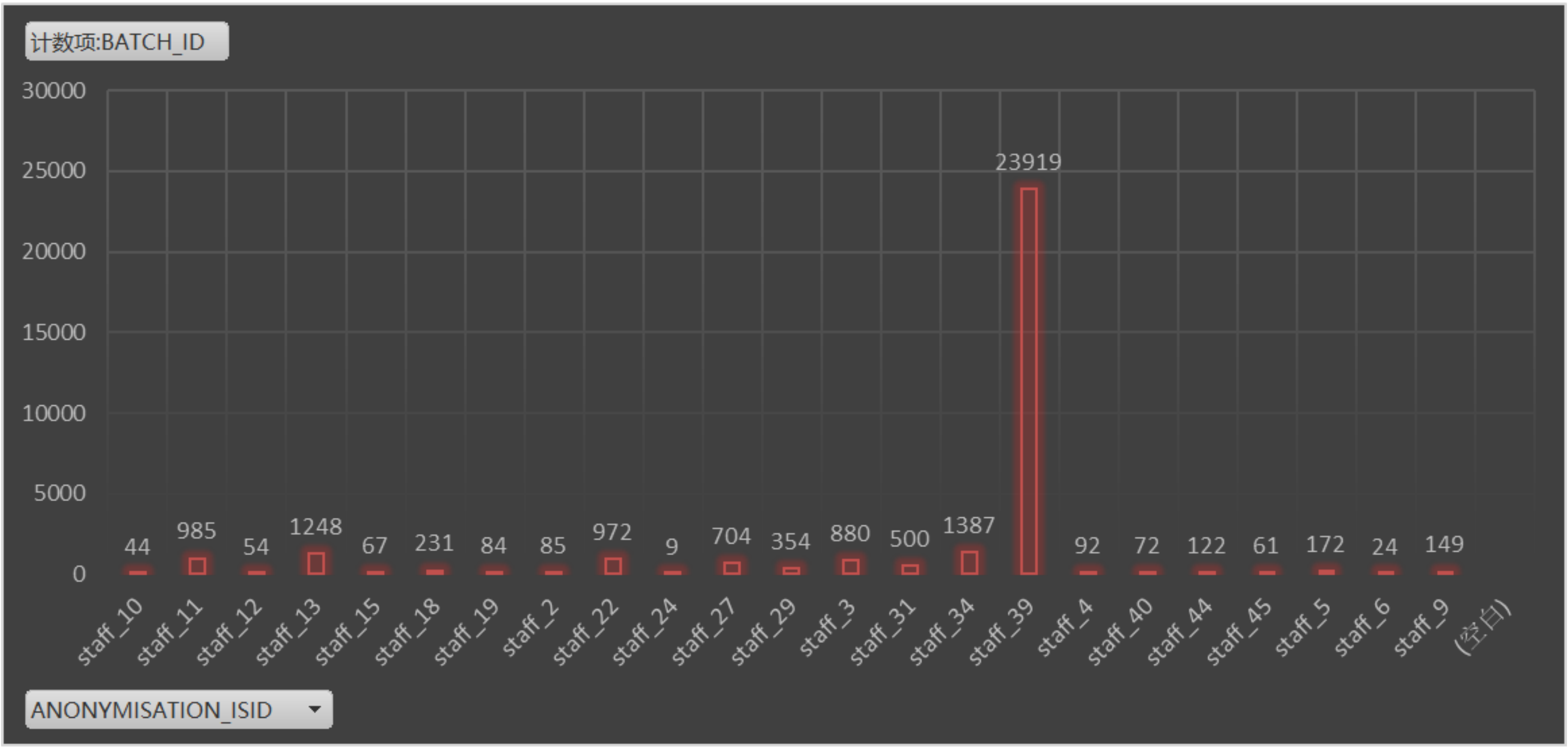

*Figure 42. The sum and average of the number of registrations annually 2011-2015*

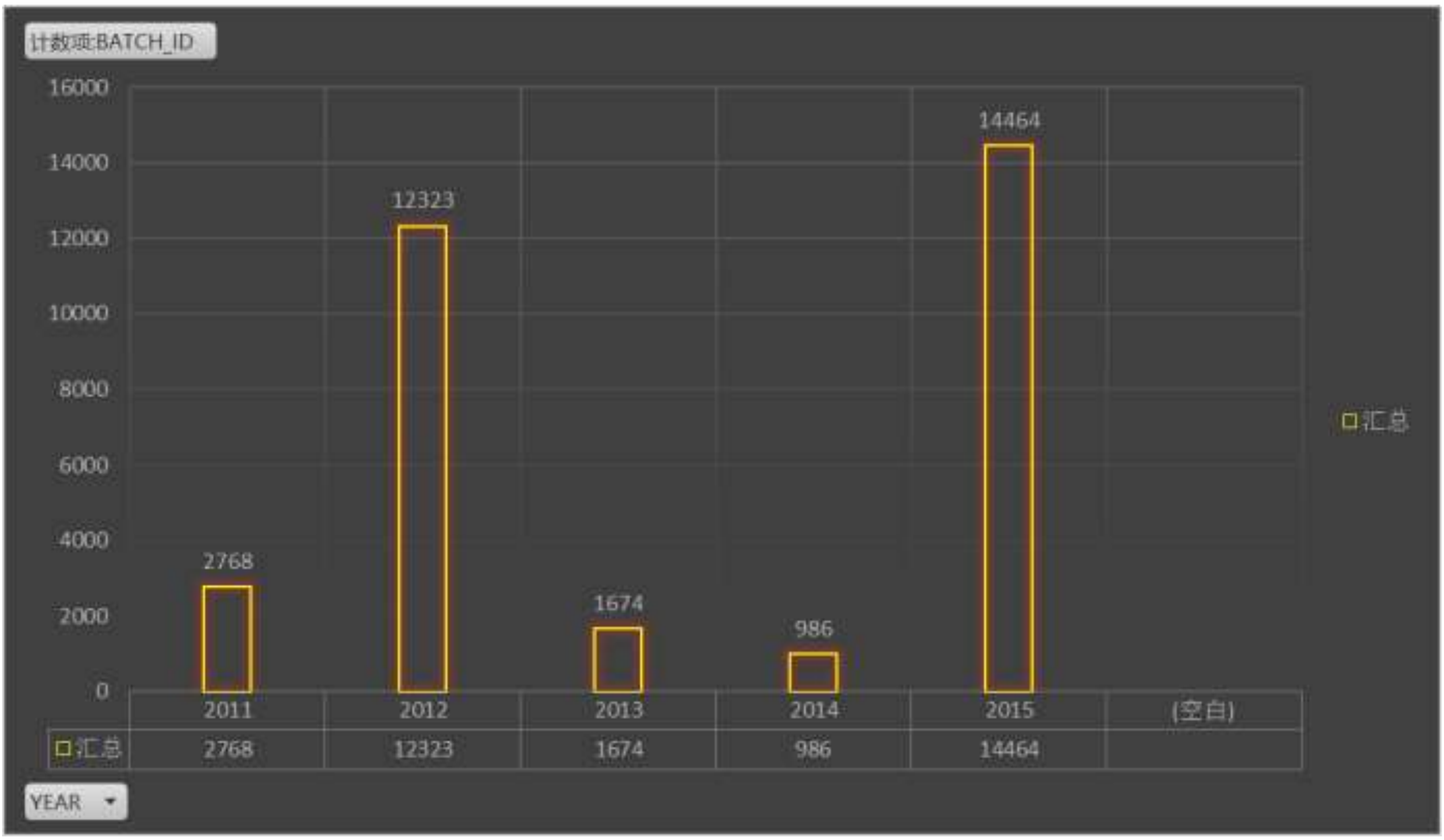

| | 2011 | 2012 | 2013 | 2014 | 2015 | (空白) |
|---|---|---|---|---|---|---|
| 汇总 | 2768 | 12323 | 1674 | 986 | 14464 | |

*Figure 43. The number of registrations executed by staff annually 2011-2015*

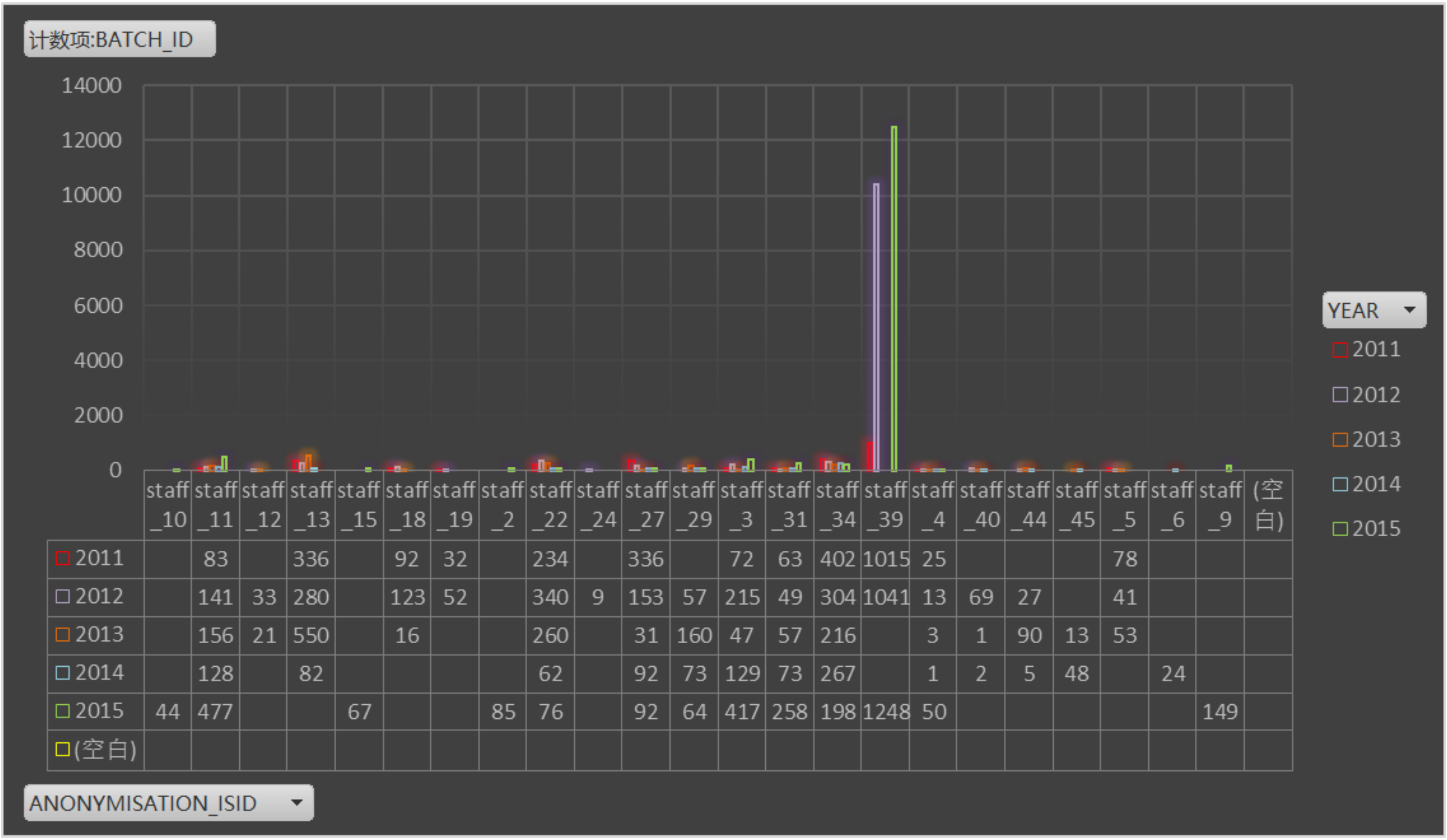

| | staff_10 | staff_11 | staff_12 | staff_13 | staff_15 | staff_18 | staff_19 | staff_2 | staff_22 | staff_24 | staff_27 | staff_29 | staff_3 | staff_31 | staff_34 | staff_39 | staff_4 | staff_40 | staff_44 | staff_45 | staff_5 | staff_6 | staff_9 | (空白) |
|---|---|---|---|---|---|---|---|---|---|---|---|---|---|---|---|---|---|---|---|---|---|---|---|---|
| 2011 | | 83 | | 336 | | 92 | 32 | | 234 | | 336 | | 72 | 63 | 402 | 1015 | 25 | | | | 78 | | | |
| 2012 | | 141 | 33 | 280 | | 123 | 52 | | 340 | 9 | 153 | 57 | 215 | 49 | 304 | 1041 | 13 | 69 | 27 | | 41 | | | |
| 2013 | | 156 | 21 | 550 | | 16 | | | 260 | | 31 | 160 | 47 | 57 | 216 | | 3 | 1 | 90 | 13 | 53 | | | |
| 2014 | | 128 | | 82 | | | | | 62 | | 92 | 73 | 129 | 73 | 267 | | 1 | 2 | 5 | 48 | | 24 | | |
| 2015 | 44 | 477 | | | 67 | | | 85 | 76 | | 92 | 64 | 417 | 258 | 198 | 1248 | 50 | | | | | | 149 | |
| (空白) | | | | | | | | | | | | | | | | | | | | | | | | |

*Figure 44. The descriptive statistics overview for registrations 2011-2015 of Disco software*

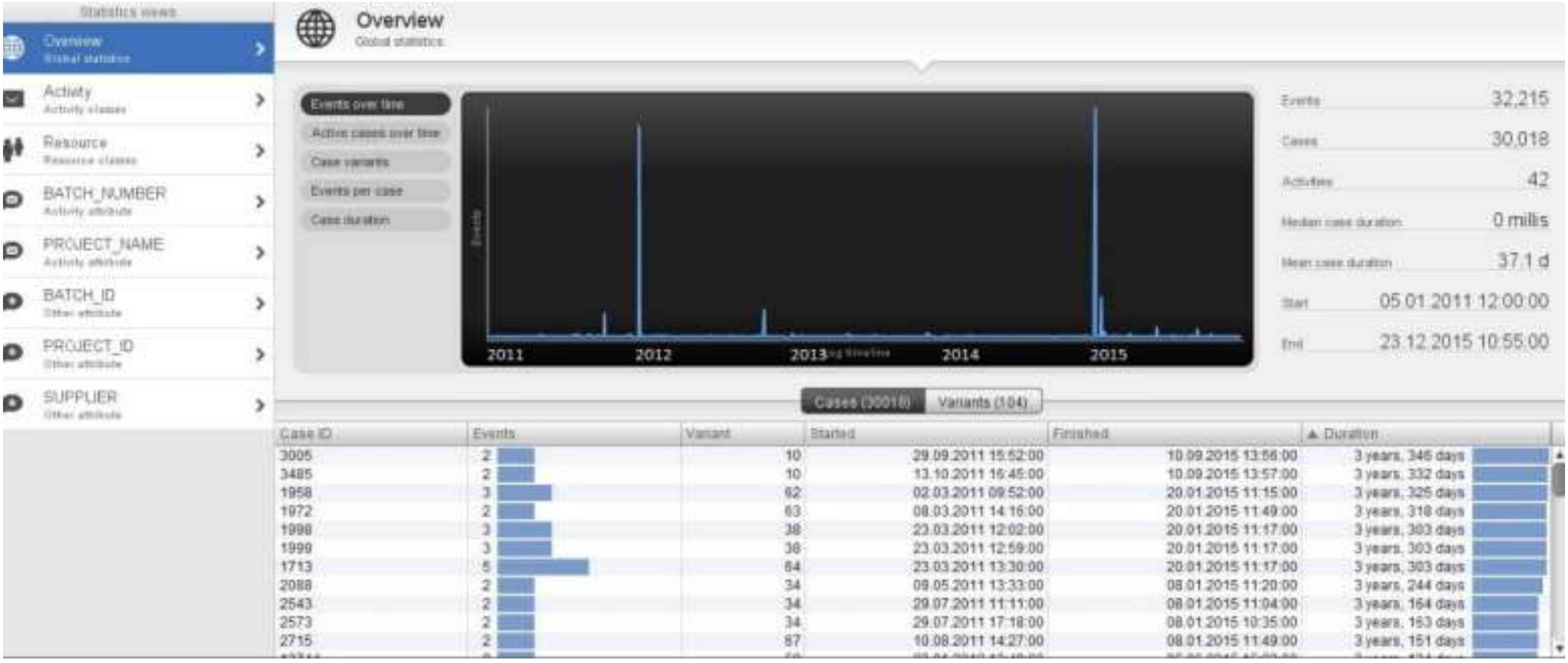

*Figure 45. The descriptive statistics of activities for* registrations *2011-2015 of Disco software*

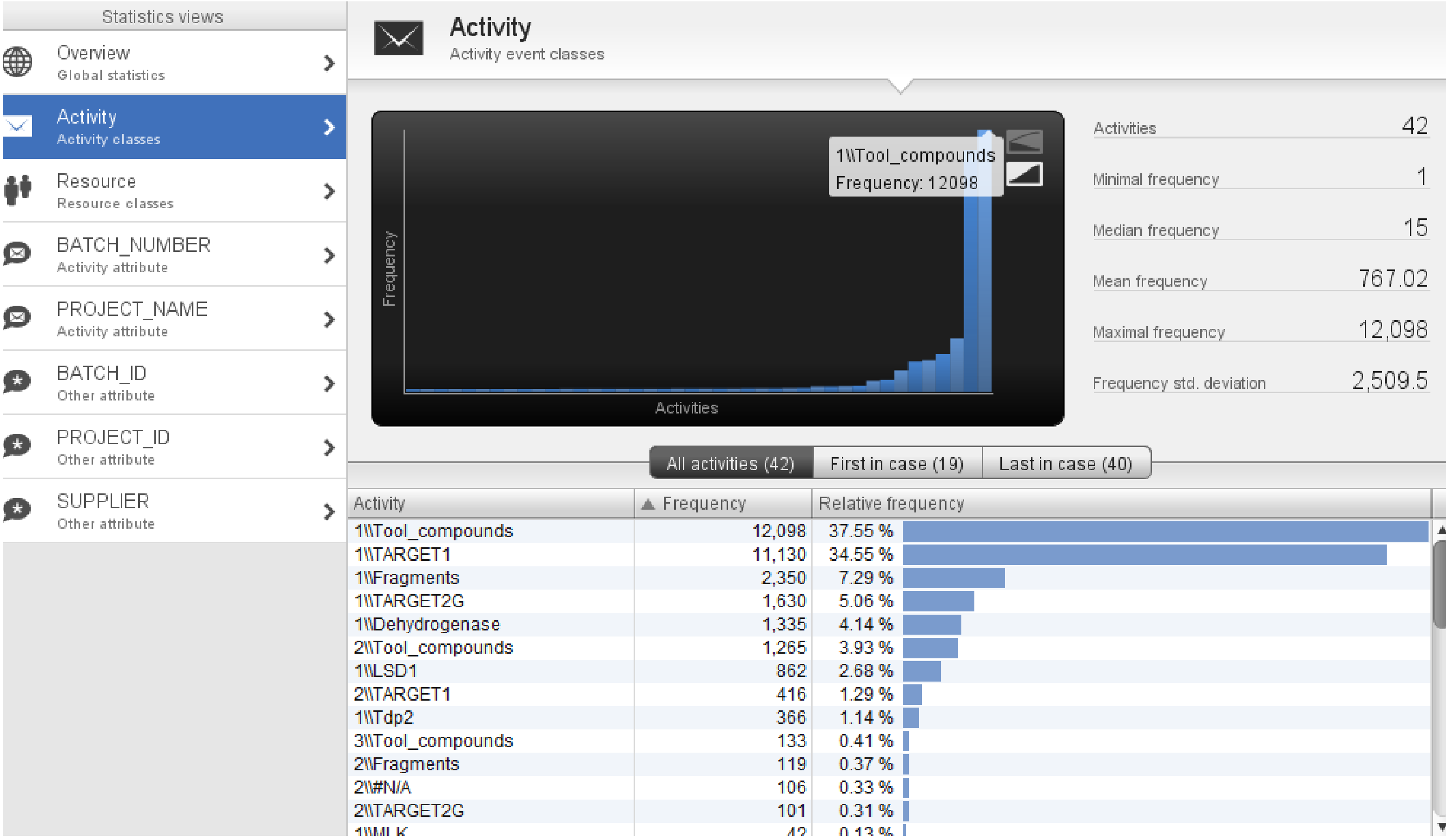

*Figure 46. The descriptive statistics of resources for* registrations *2011-2015 of Disco Software*

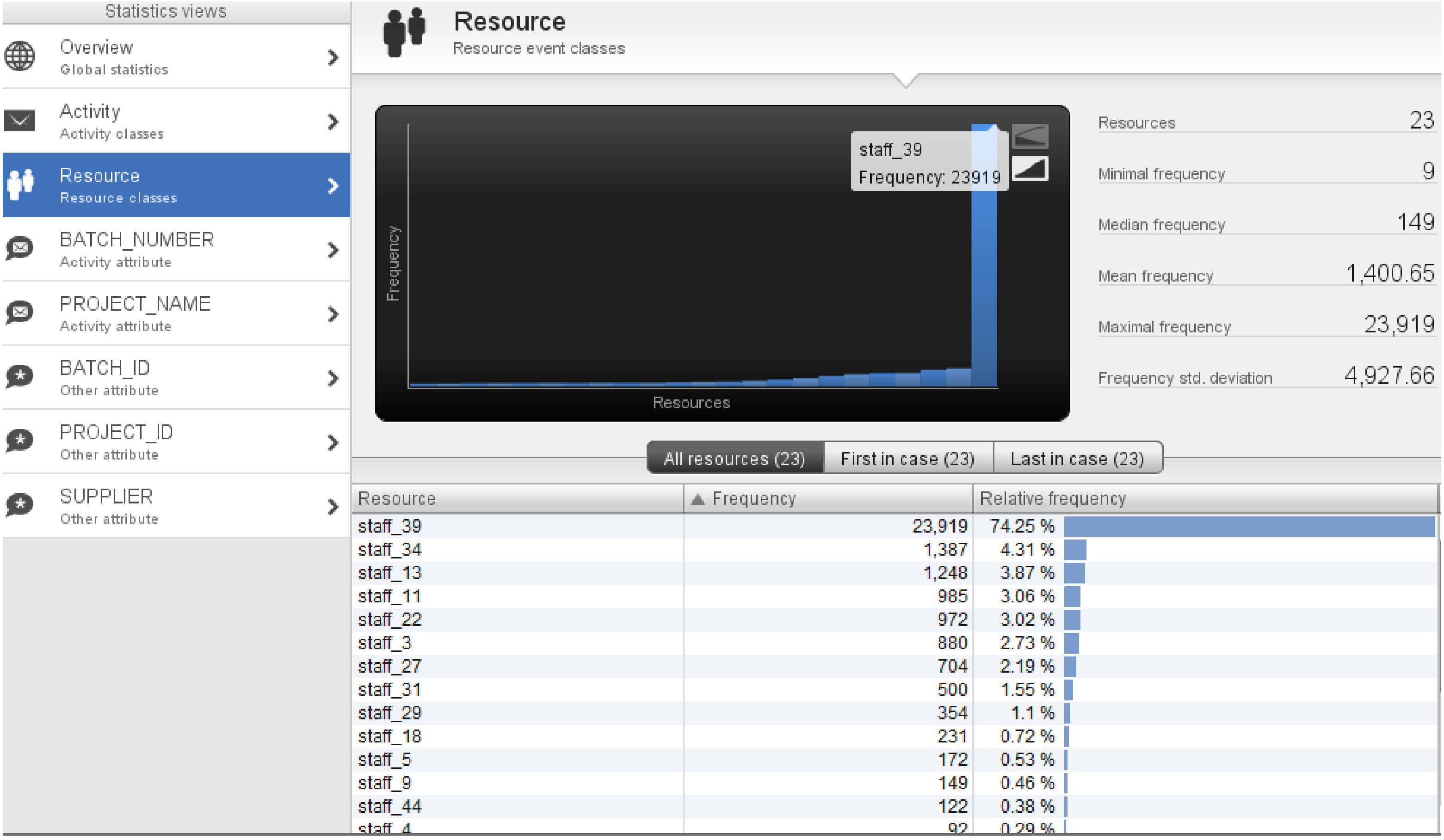

As Figure 43, 45, 46 shows, there are 32215 events, 30018 cases, 42 activities, 23 resources involve in the log.

The mean case duration is 37.1 days, the median case duration 0 millisecond, the minimum case duration is 0 millisecond, and the maximum is 3 years 346 days.

The activity – 1\\Tool_compounds was the most frequent activity with 12098 and took 37.55% of all activities. The second wasTARGET1 activity with 11130 and took 34.55% of all activities. These two activities counted 72.1% of overall activities.

### 4.2.2 Interactive dashboards and dynamic visualisations

#### *4.2.2.1 Control flow perspective*

As Figure 47 shows, the activity first batch of Tool_compounds (1\\Tool_compounds) was having the biggest amount of cases – 12098 which were highlighted in dark blue. The activity first batch of TARGET1 (1\\ TARGET1) was the second biggest with 11130 cases which were highlighted in dark blue. The second batch of Tool_compounds (2\\ Tool_compounds) was the sub-path followed by the most various – five routes of activities.

*Figure 47. Full process map of registrations event log 2011-2015 of Disco software*

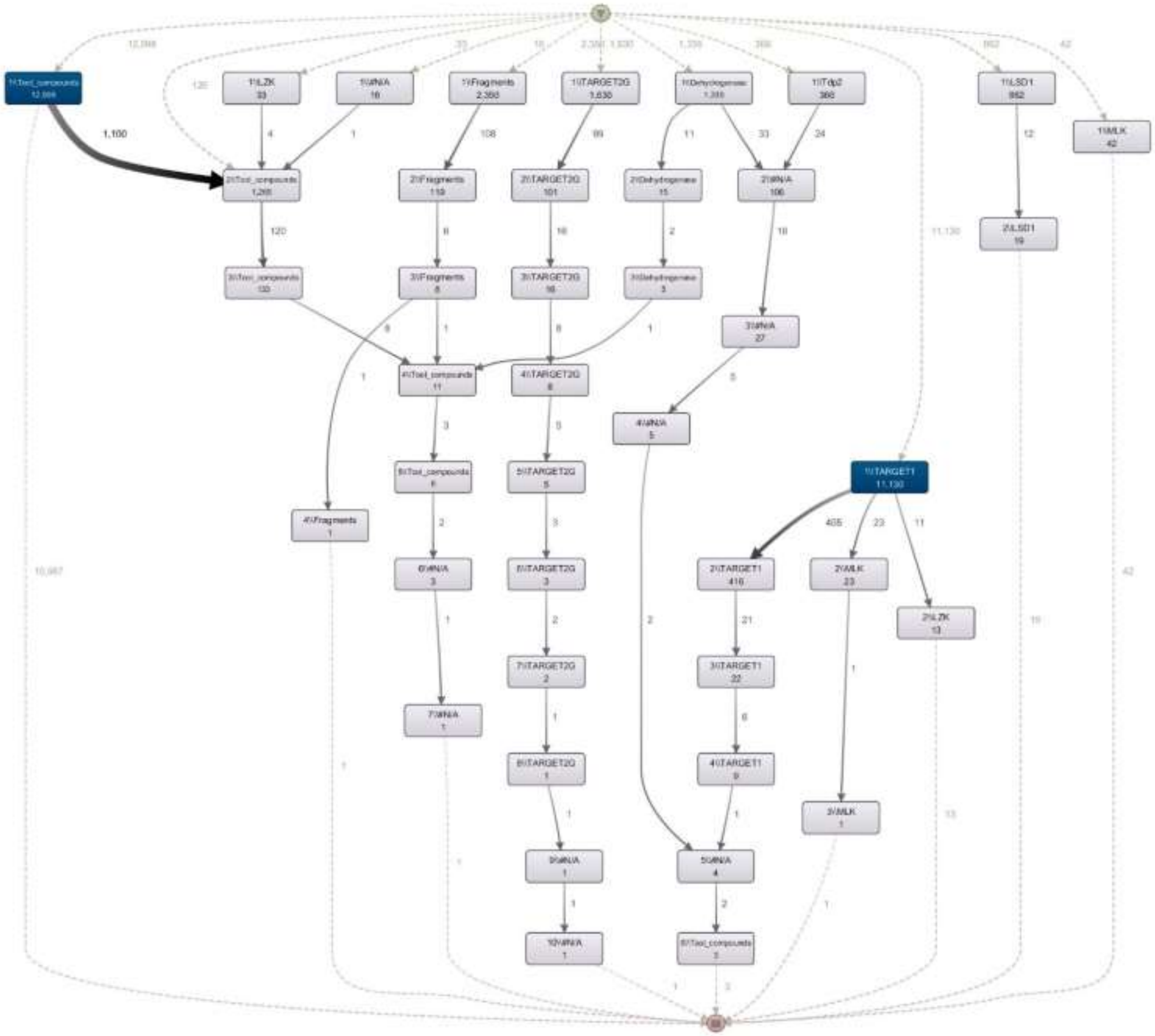

However, the Figure 48 generated by ProM with Inductive Miner-infrequent (IMf) miner (Leemans, Fahland, & van der Aalst, 2016) shows a map different from Disco's. It is because for ProM software. The map shows all cases 30018 were only follow one activity under the setting 100% activities and 0 paths shows. It implies that ProM with Inductive Miner-infrequent (IMf) miner was not a good technique for this event log and model.

*Figure 48. Full process map of registrations event log 2011-2015 of ProM software*

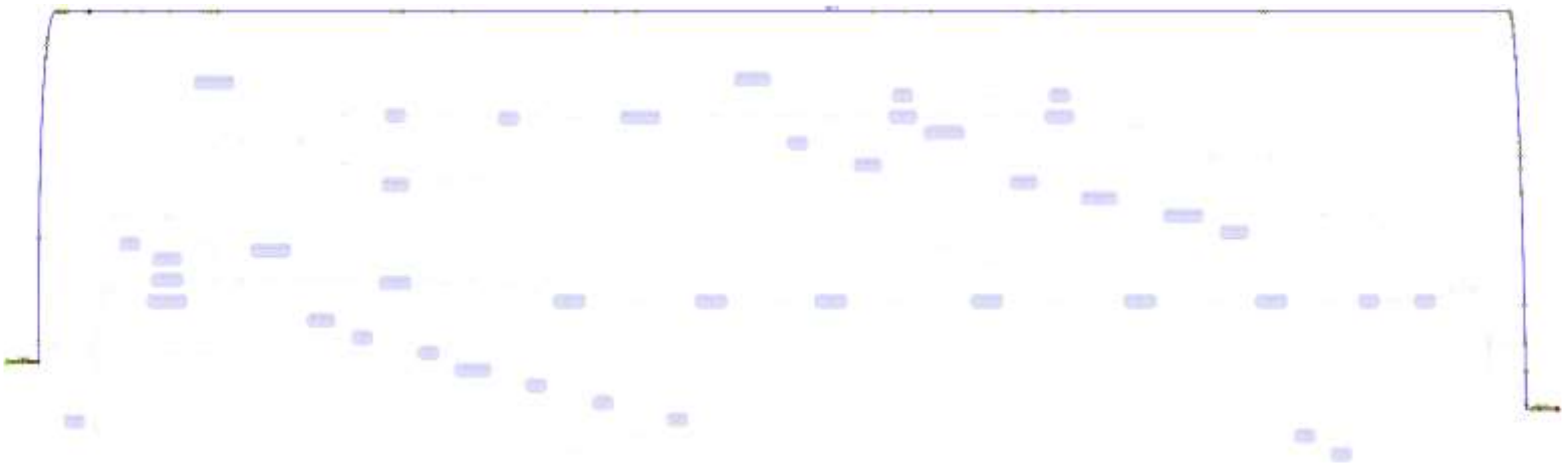

The Figure 49 shows the routes of PROJECT_NAME activities follow their BATCH_NUMBER sequentially.

*Figure 49. Causal activity graph of registrations event log 2011-2015 based on ProM software*

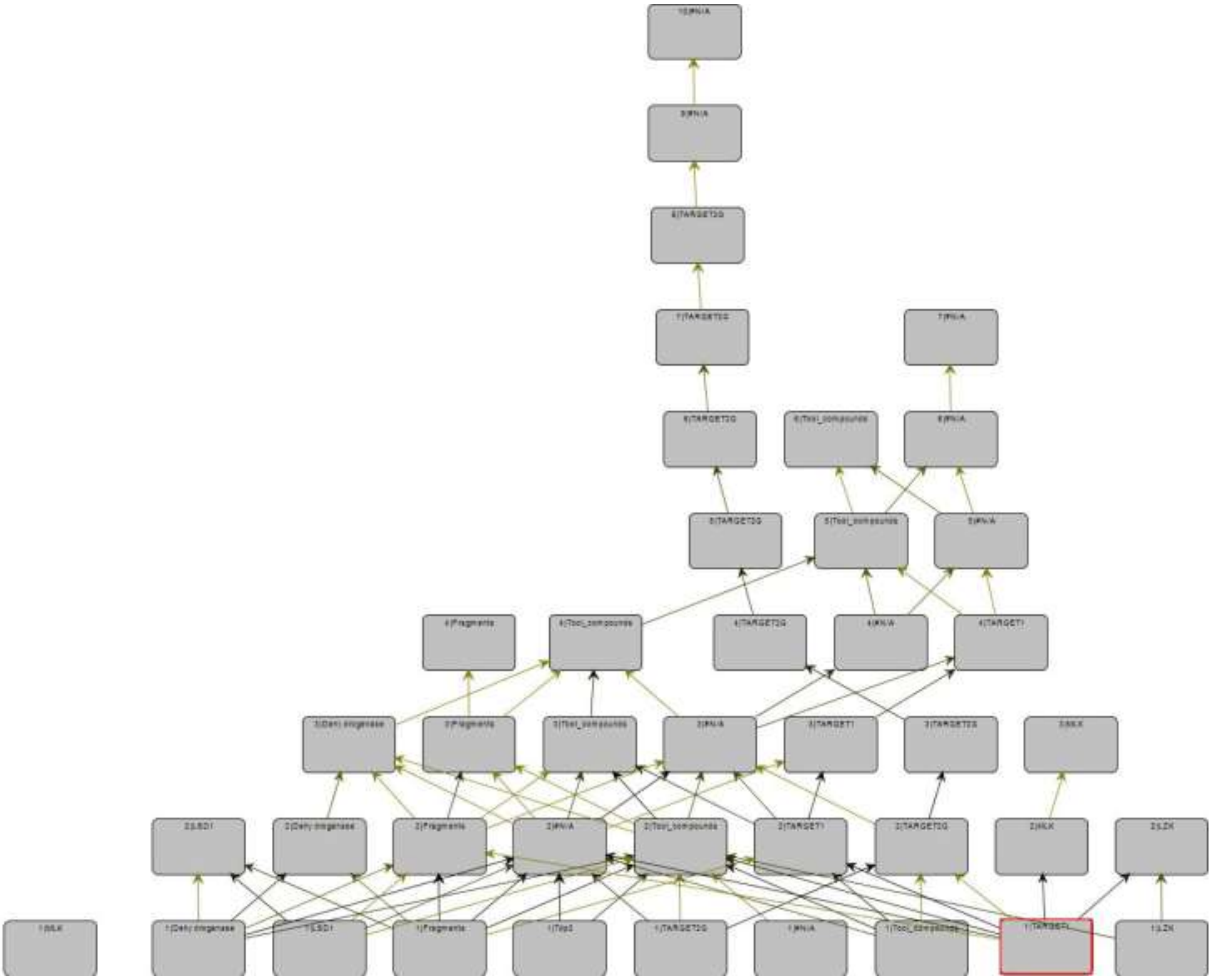

#### *4.2.2.2 Performance perspective*

There was no max repetition in the map of registrations event log 2011-2015 as Figure 50 shows, because of each case go through one activity only.

*Figure 50. Repetition frequency process map of registrations event log 2011-2015 of Disco software*

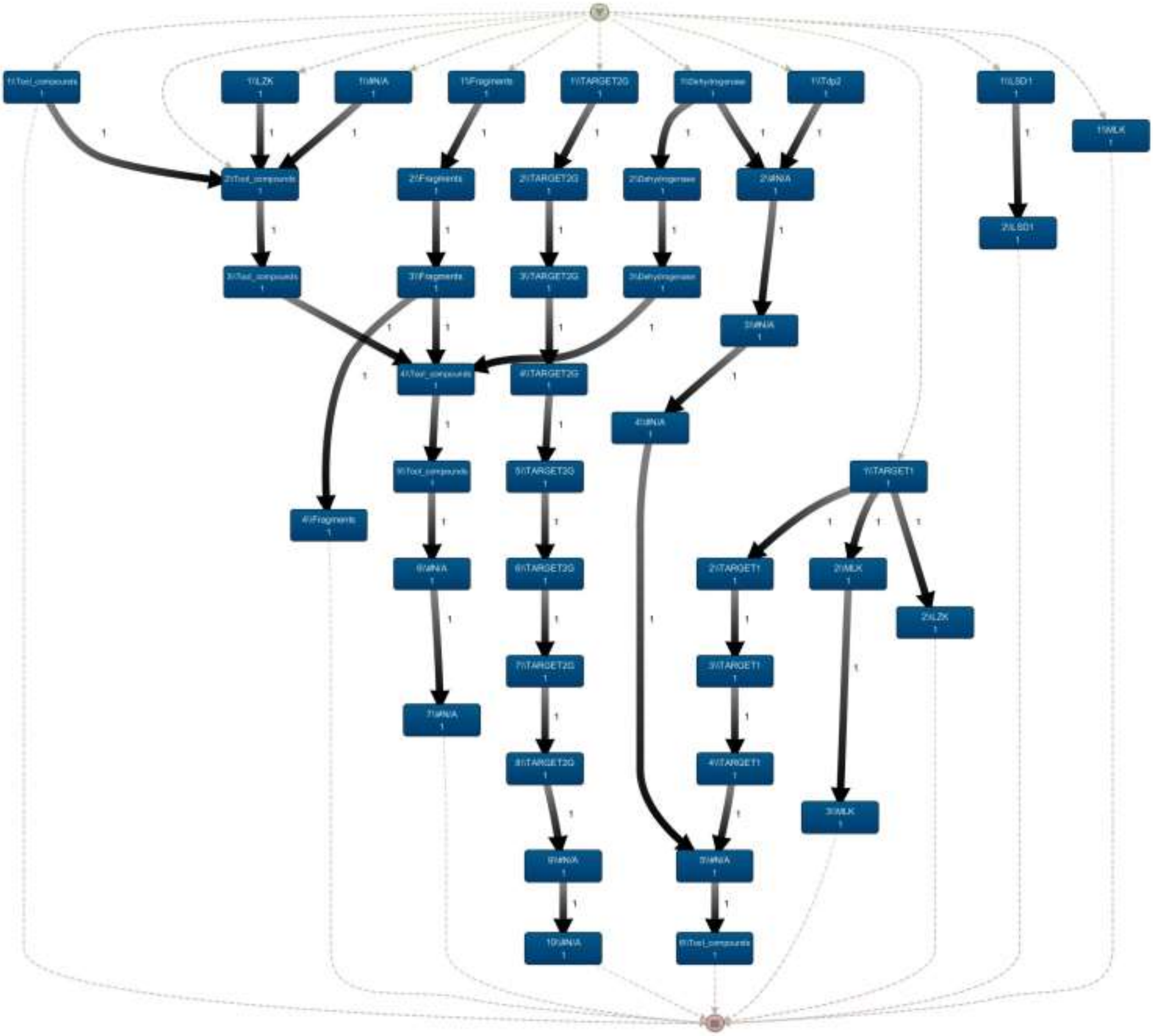

The longest total duration performance was from the first batch of Tool_compounds (1\\ Tool_compounds) to the second batch of Tool_compounds (2\\ Tool_compounds) which takes 2.3 thousand years as Figure 51 shows.

*Figure 51. Total duration performance process map of registrations event log 2011-2015 of Disco software*

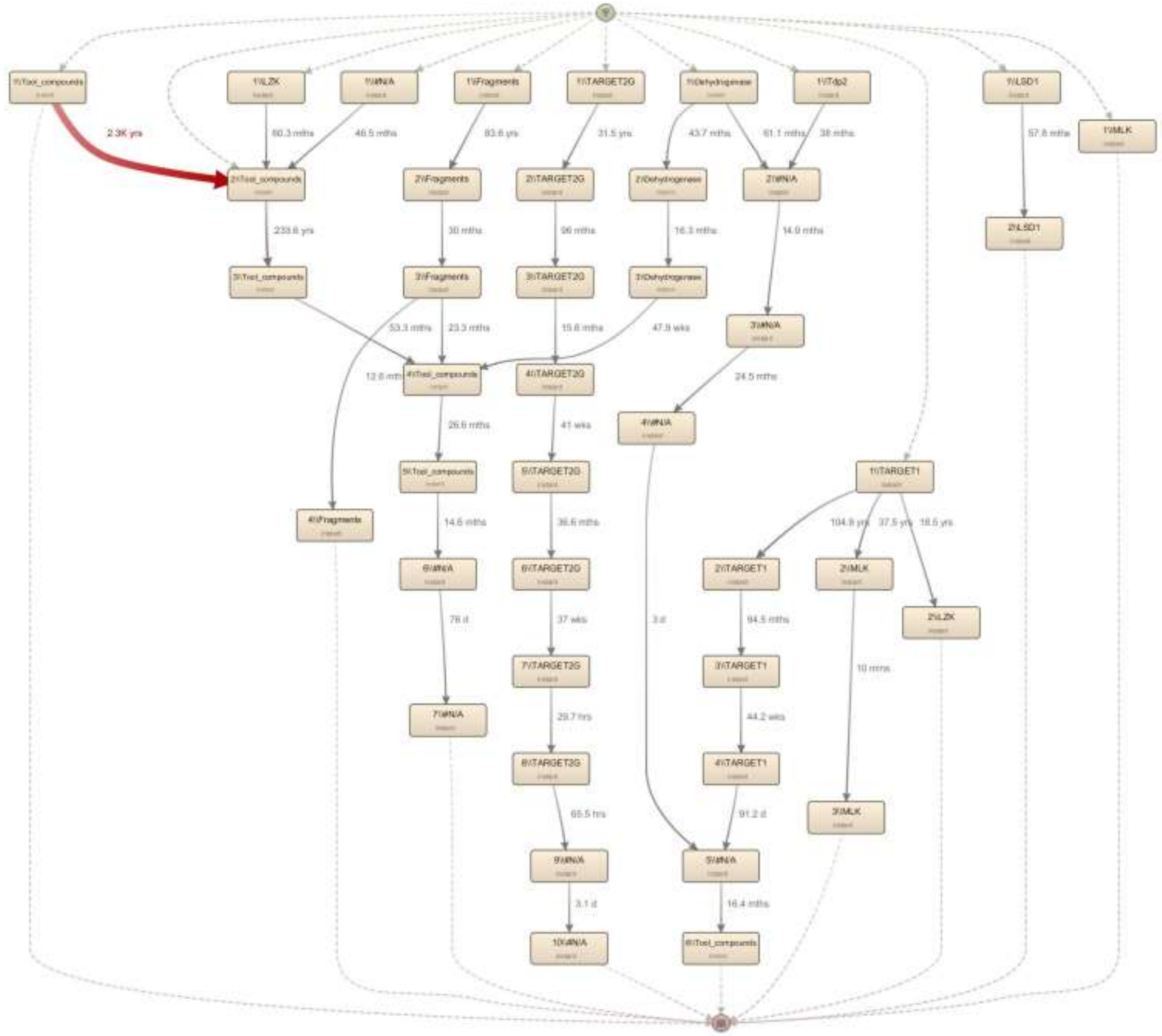

As figure 52 shows, the longest mean duration activities was from 1\\#N/A to 2\\Tool_compounds which took 46.5 months. The second longest duration was from 1\\Tool_compounds to 2\\Tool_compounds which took 25.6 months. Other long duration were including: from 2\\Tool_compounds to 3\\Tool_compounds with 23.4 months; from 3\\Fragments to 4\\Tool_compounds with 23.3 months; from 1\\TARGET1 to 2\\LZK with 20.2 months; from 1\\TARGET1 to 2\\MZK with 19.6 months.

*Figure 52. Mean duration performance process map of registrations event log 2011-2015 of Disco software*

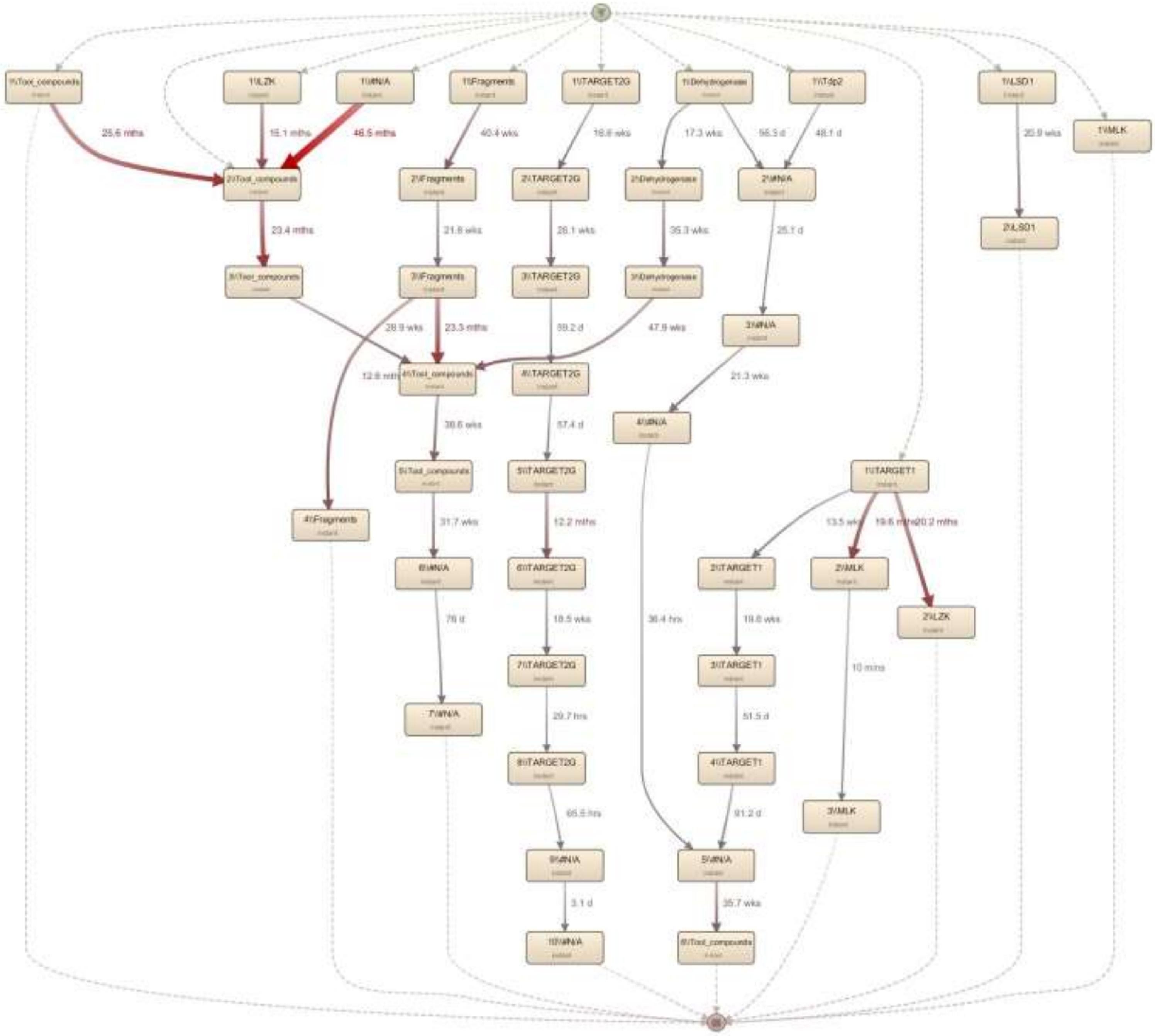

Similar to the event log of experiments, the number of events of registrations was all having a fluctuation which reached the highest peak sharply at the beginning of the year 2012 and 2015 as Figure 53 indicates. The number of events was at the highest point at the beginning of the year 2012 and 2015, then fell down sharply and became normal as usual. However, it is noticed that there were three high number of events in the November of 2011 and 2012, January of 2015. The number of cases was having a similar trend as events, and it maintained a high number of cases from November 2011 to January 2015 continuously as Figure 54 indicates.

*Figure 53.* The number of registrations events over time *2011-2015 of Disco software*

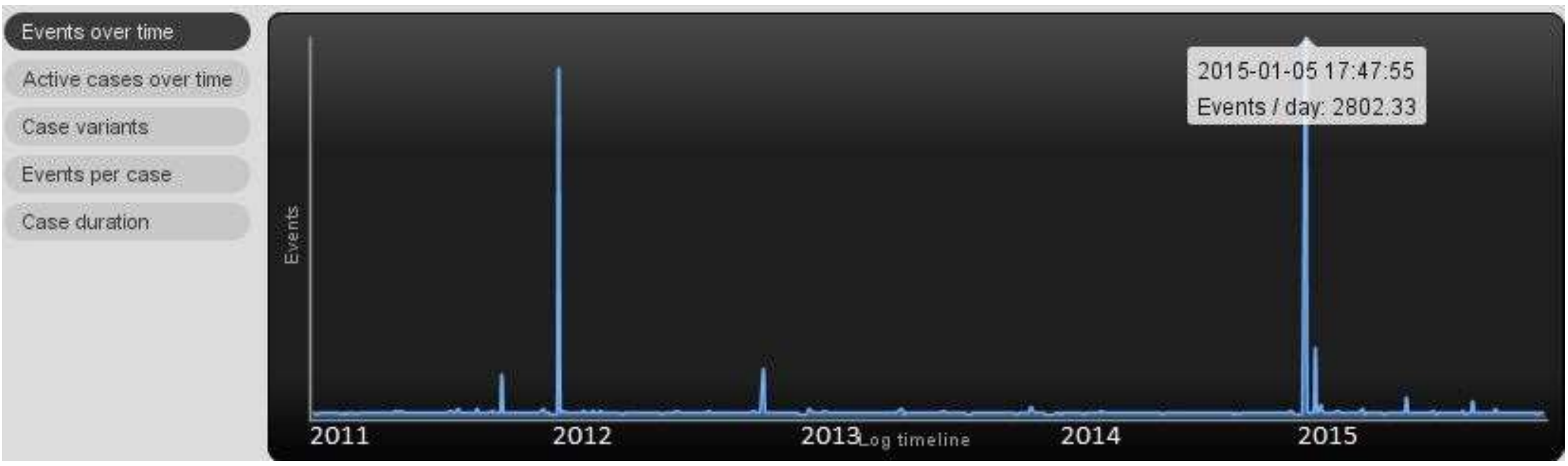

Figure 54. The number of registrations cases over time 2011-2015 of Disco software

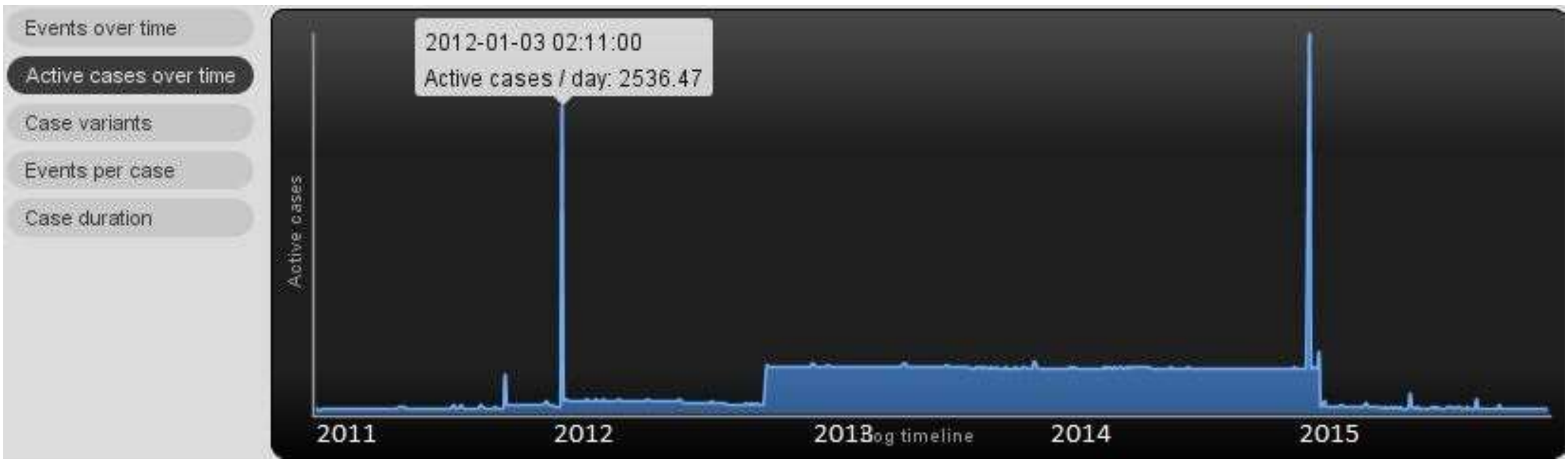

Figure 55 shows the animated dynamic visualised process map of registrations event log at the time 28/04/2013. The animation can be played along with the timeline at the bottom of the dashboard. The spots move along with dashed arrow represented the cases were being performed. The blue boxes represented the activities were being executed, while grey boxes were not. A bigger size of a spot represented a larger amount of cases were performed at a time, such as the big spots went out from 1\\Tool_compounds and 2\\Tool_compounds.

*Figure 55. Animation of process map of registrations event log 2011-2015 of Disco software*

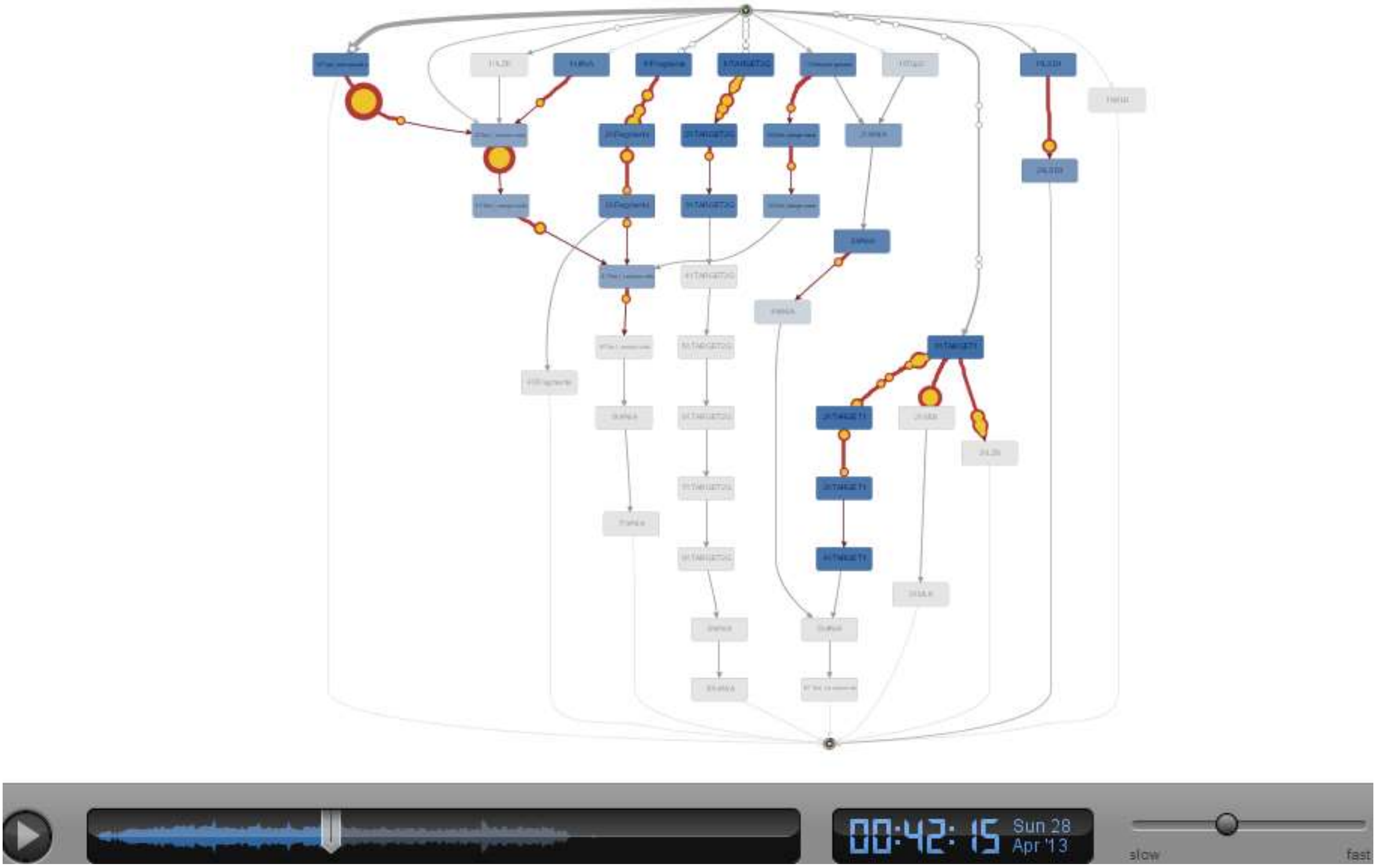

Figure 56 and 57 show the duration (represented by the distance between two dots) spending on each activity and case by staff from 2011 to 2015. Figure 56 implies that the sharply increasing cases (light green - staff_39) at the beginning of 2012 and 2015, did not take a long time to finish. Some extreme cases took as long as 4 and 5 years. While Figure 57 and 58 implies that most of the activities (99.14%) can be completed within the second batch, 92.7% can be completed with the first batch.

*Figure 56. Dotted chart of cases overtime 2011-2015 of ProM software*

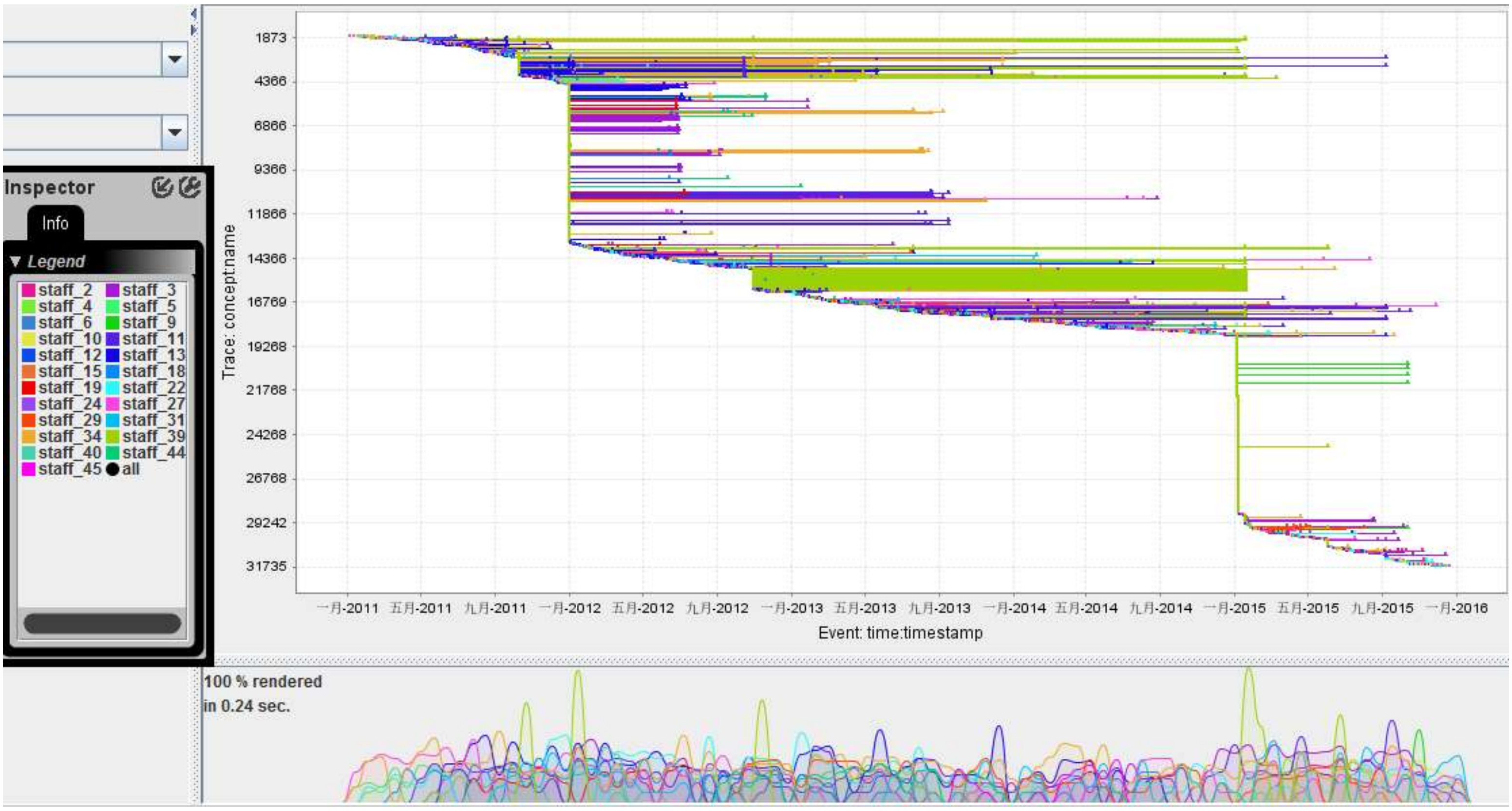

*Figure 57. Dotted chart of activities overtime 2011-2015 of ProM software*

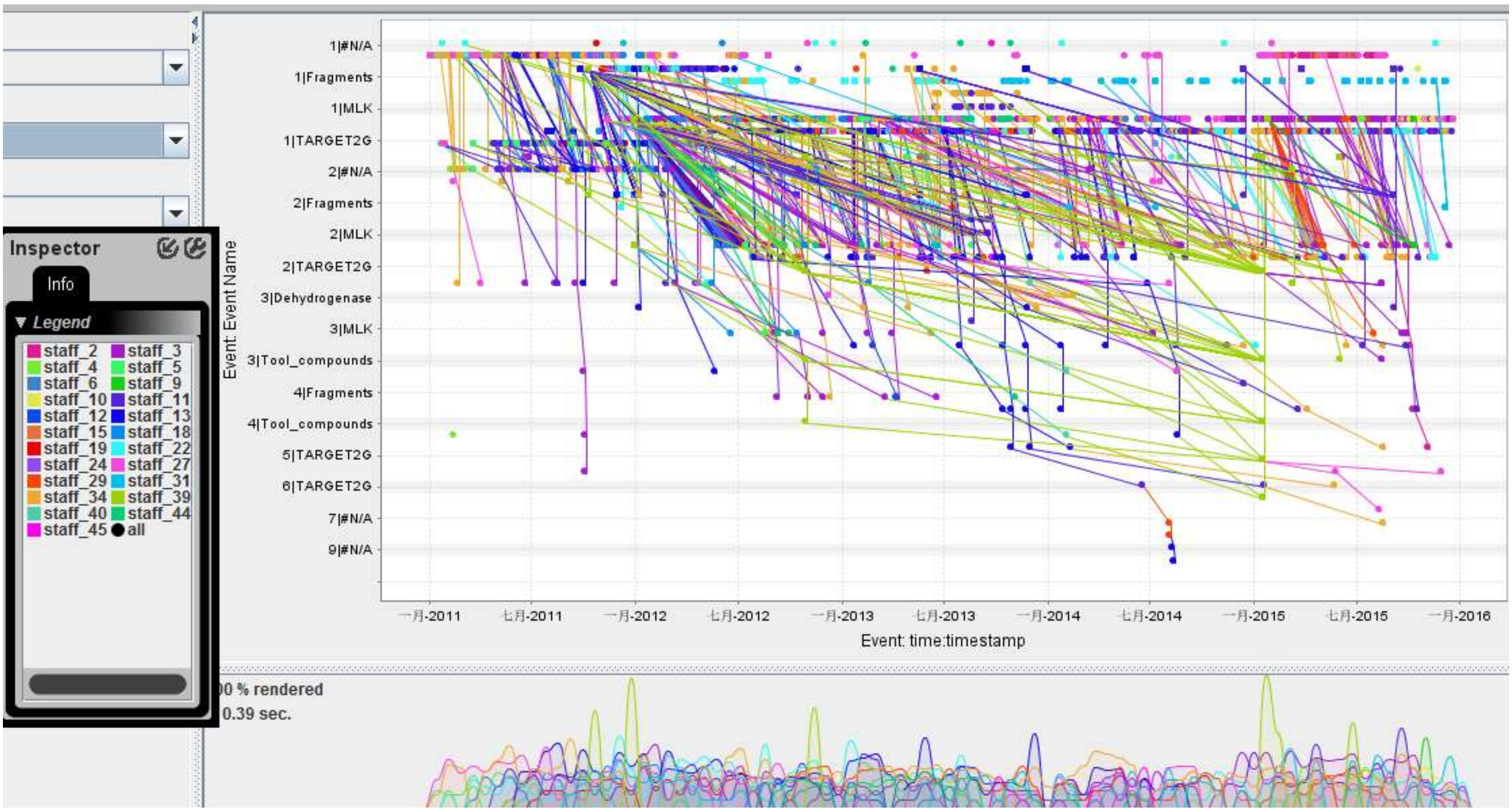

*Figure 58. Descriptive statistics of BATCH_NUMBER 2011-2015 of Disco software*

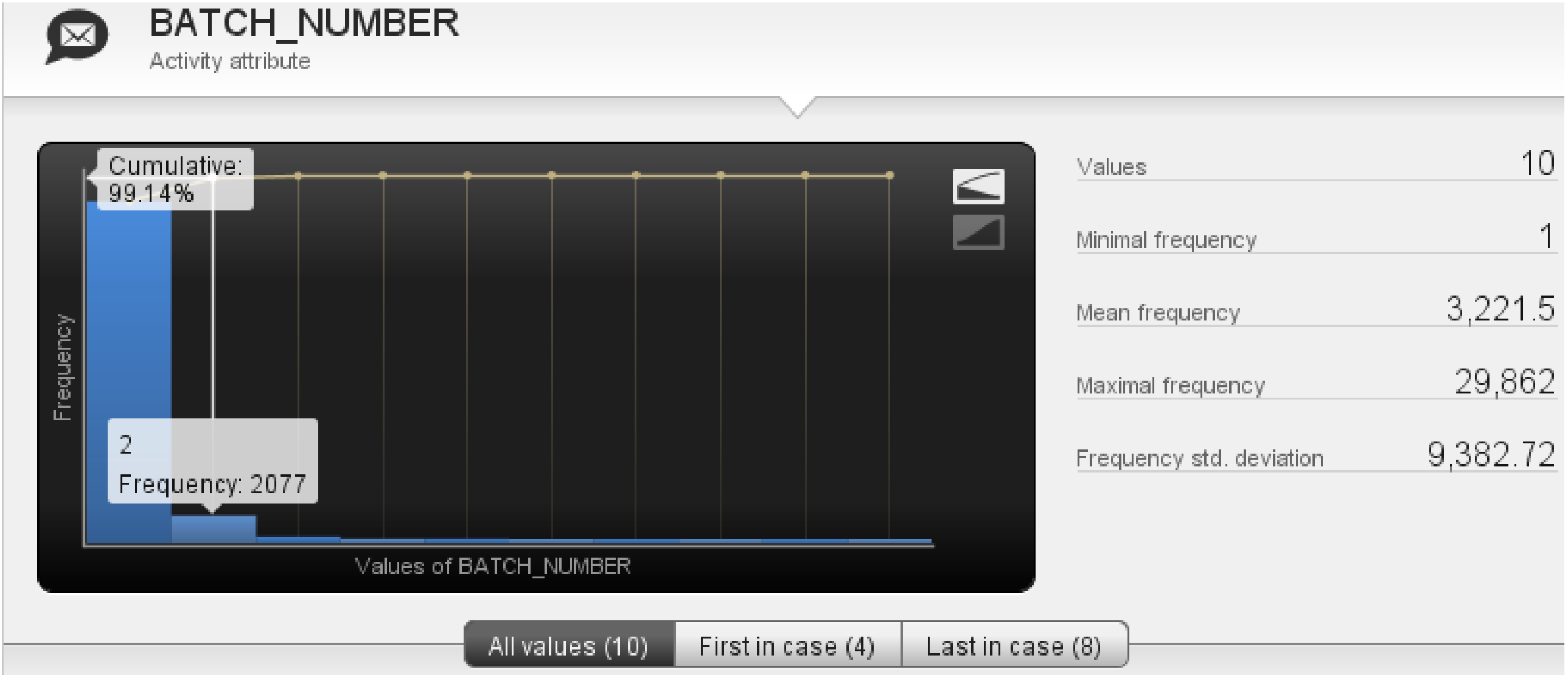

| Value | ▲ Frequency | Relative frequency |
|---|---|---|
| 1 | 29,862 | 92.7 % |
| 2 | 2,077 | 6.45 % |
| 3 | 212 | 0.66 % |
| 4 | 34 | 0.11 % |
| 5 | 15 | 0.05 % |
| 6 | 9 | 0.03 % |
| 7 | 3 | 0.01 % |
| 8 | 1 | 0 % |
| 9 | 1 | 0 % |
| 10 | 1 | 0 % |

#### *4.2.2.3 Organisational perspective*

The figure 59 implies that which activities were mainly executed by whom; who work on the same activities. The PROGRAM_NAME "#N/A" was generated due to missing value, which was mainly because the executors did not input the relevant data field particularly at the time of implementation. Including staff_3, staff_4, staff_5, staff_12, staff_13, staff_18, staff_19, staff_22, staff_27, staff_31, staff_34, staff_39, staff_40, staff_44, staff_45.

*Figure 59. Dotted chart of PROJECT_NAME executed by staff 2011-2015 of ProM software*

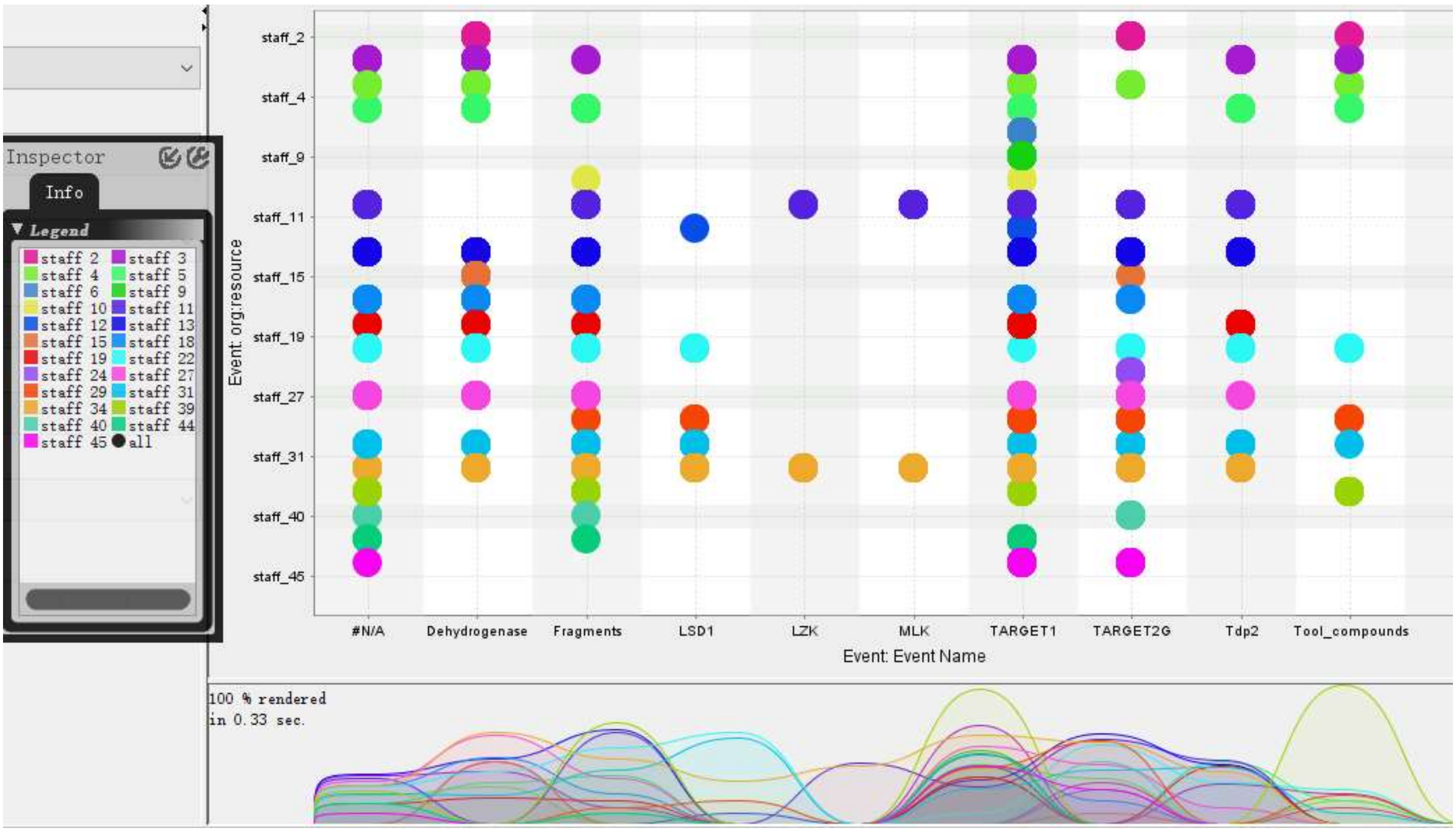

### 4.3 Additional analysis strategy for registrations event log

The different shape of process maps between experiments and registrations event logs are believed because of the different number of event classes per case. The experiments event log has maximum 1 event classes per case, while the registrations have maximum 10 event classes per case as Figure 60 shows. Due to the limited event classes per case of experiments event log, it would be meaningless to apply the additional process mining techniques and algorithms in this section. The number of event classes per case of registrations event log determines that they can use more various process mining techniques and algorithms. This section has demonstrated several additional process mining techniques and algorithms specifically for registrations event log.

*Figure 60. Events and event classes per case – experiments (left) versus registrations (right)*

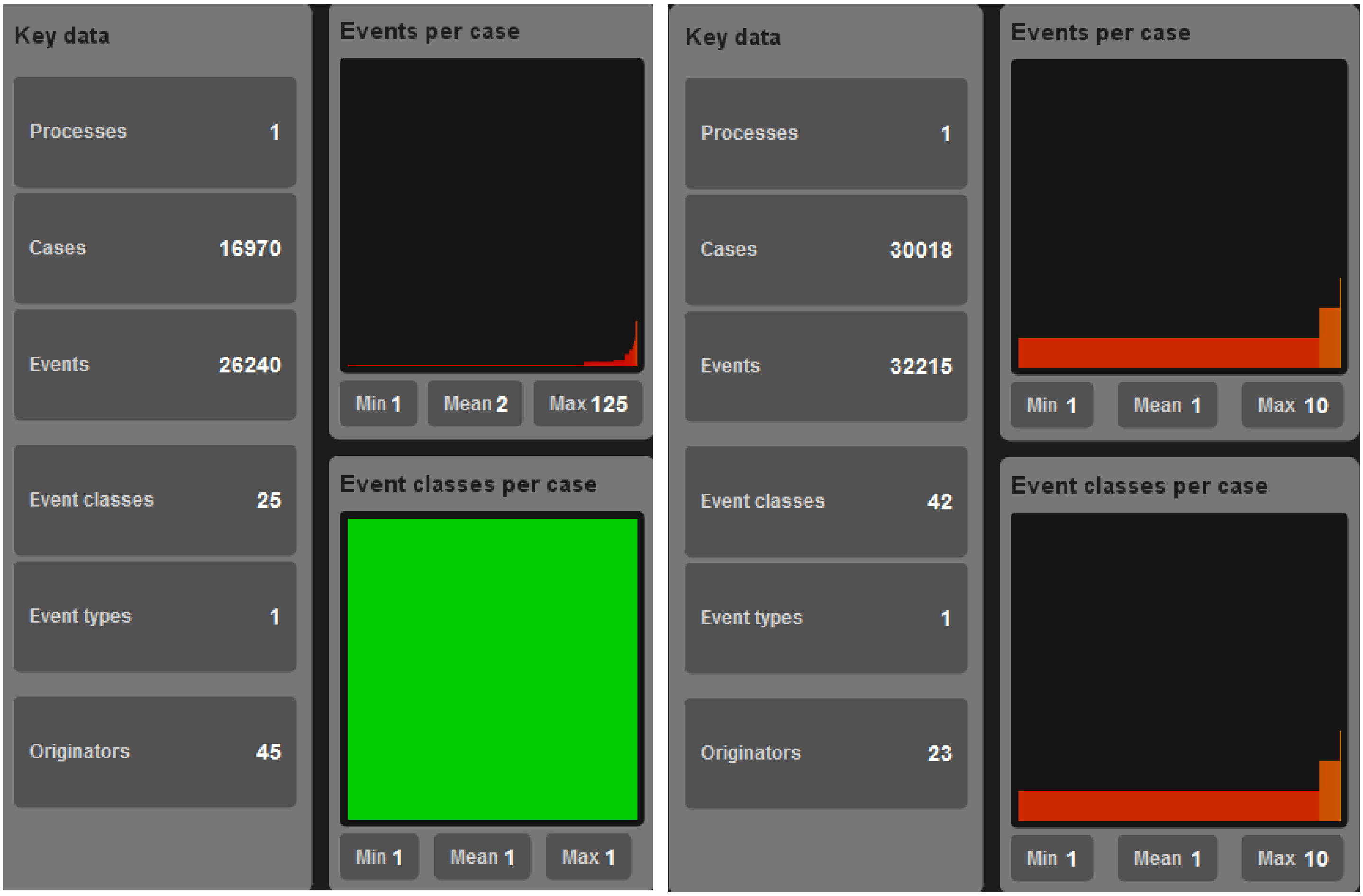

### 4.3.1 More process maps

The Figure 61 Alpha miner Petri net and Figure 62 Graphviz Petri Net provide another view of the process maps.

*Figure 61. Alpha miner Petri net view of registrations event log 2011-2015 of ProM software*

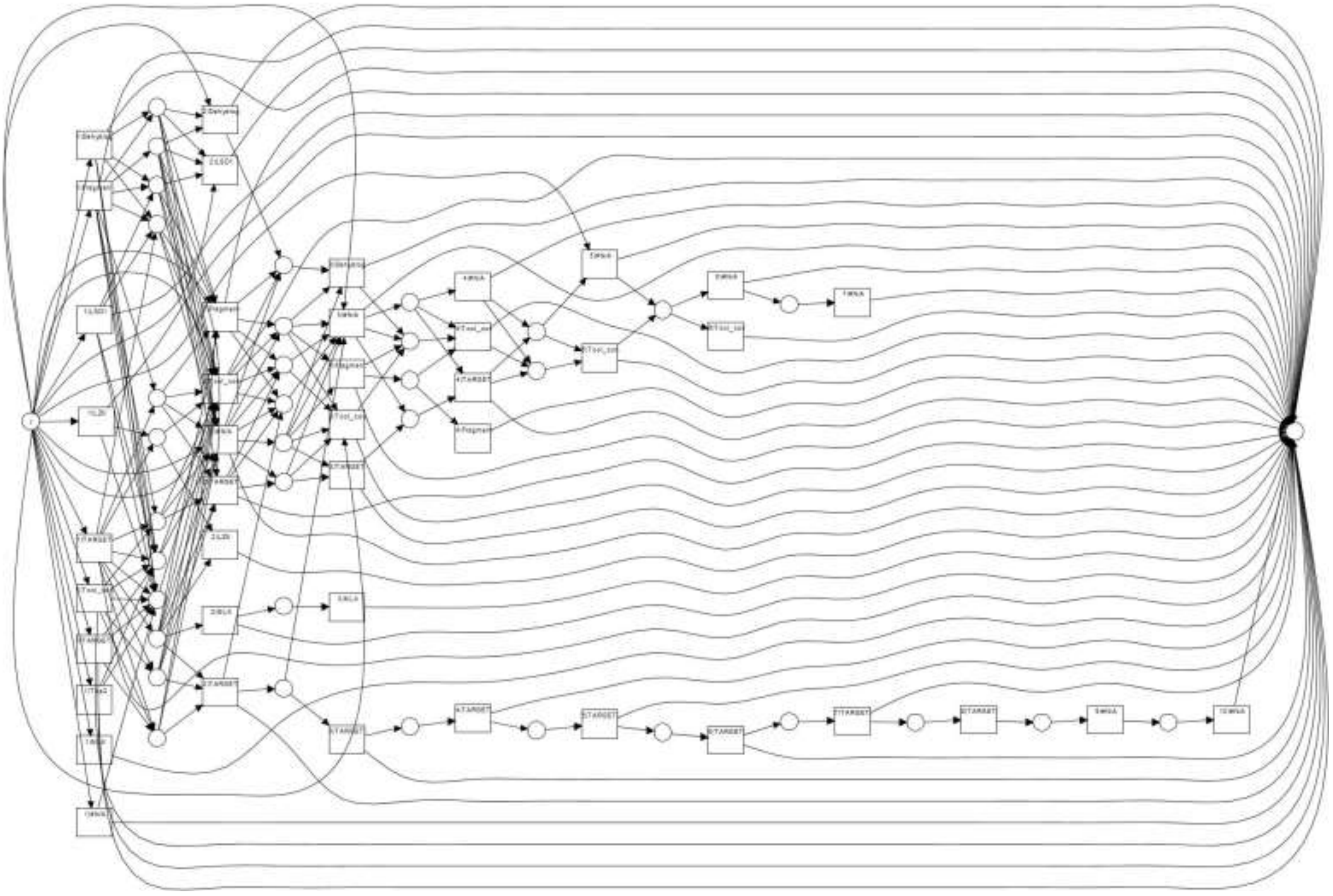

*Figure 62. Graphviz Petri Net Visualisation (Inductive Visual Miner) 2011-2015 of ProM software*

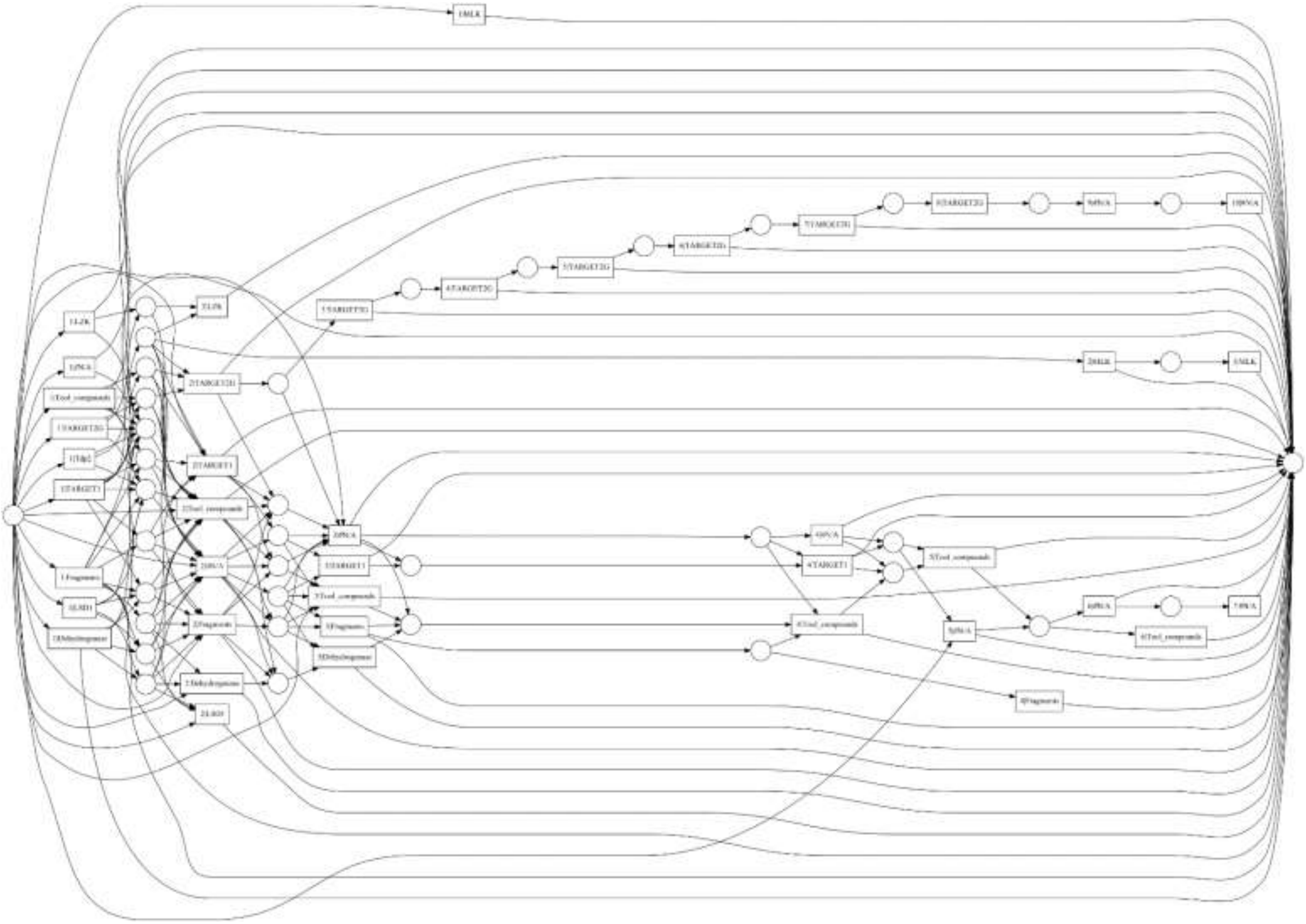

### 4.3.2 Process cluster

The activity cluster array plugin of ProM has classified six higher-level, repetitive clusters for the experiment event log 2011-2015: top left, top right, middle, left, bottom, and bottom right clusters in figure 65. The result follows the correlated dimensions of causal activity matrix in Figure 63 and 64.

*Figure 63. Visualised causal activity matrix using coloured table 2011-2015 of ProM software*

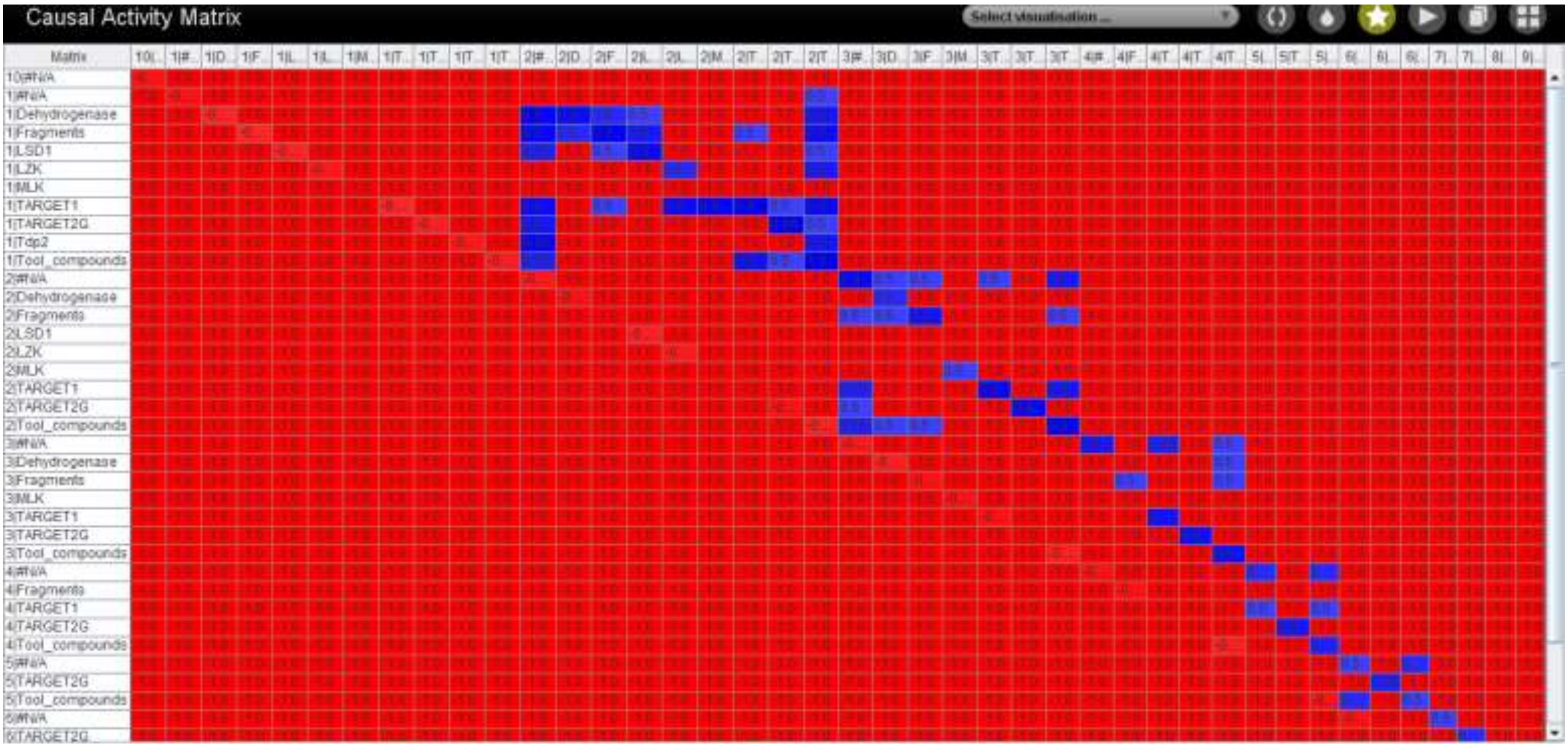

*Figure 64. Visualised causal activity matrix using colour grid 2011-2015 of ProM software*

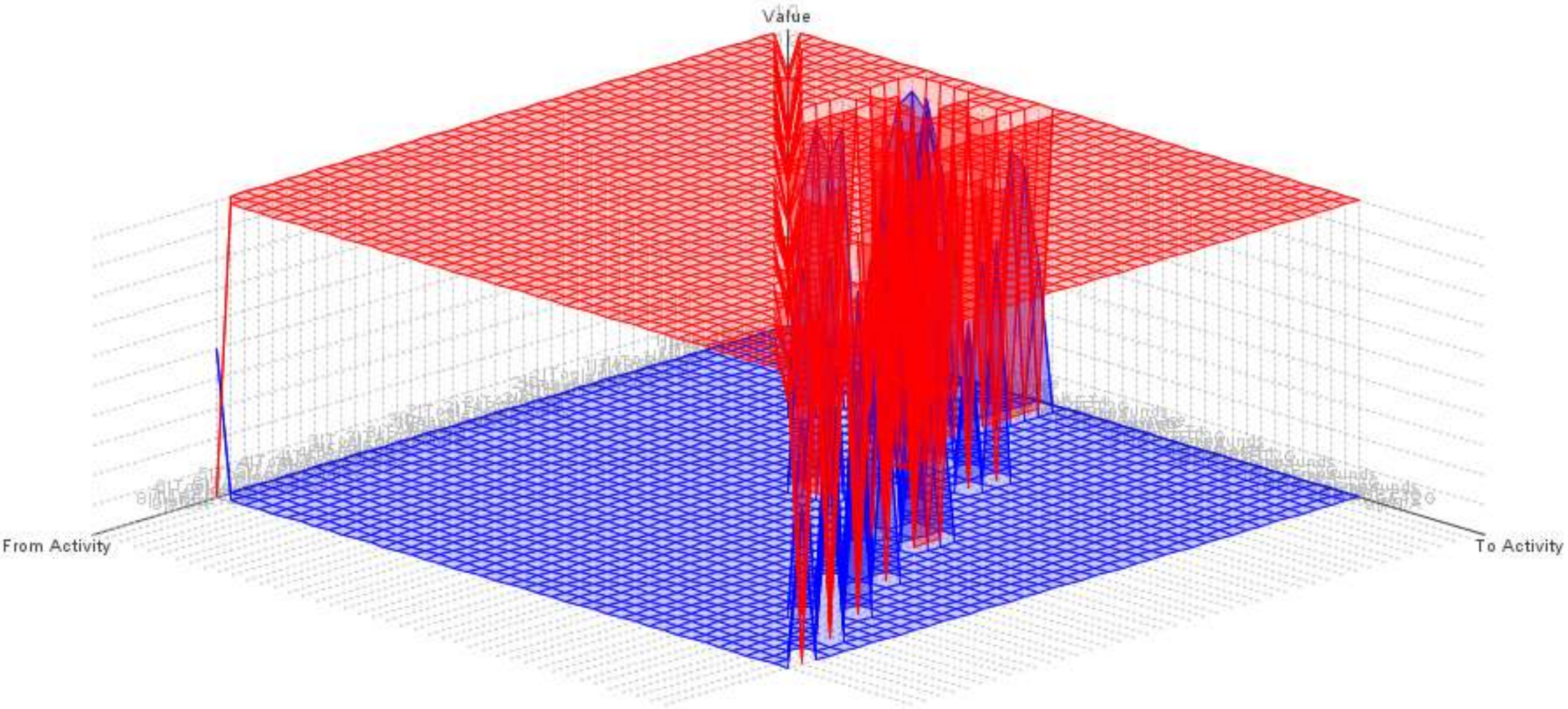

*Figure 65. Activity Cluster array of registrations event log 2011-2015 of ProM software*

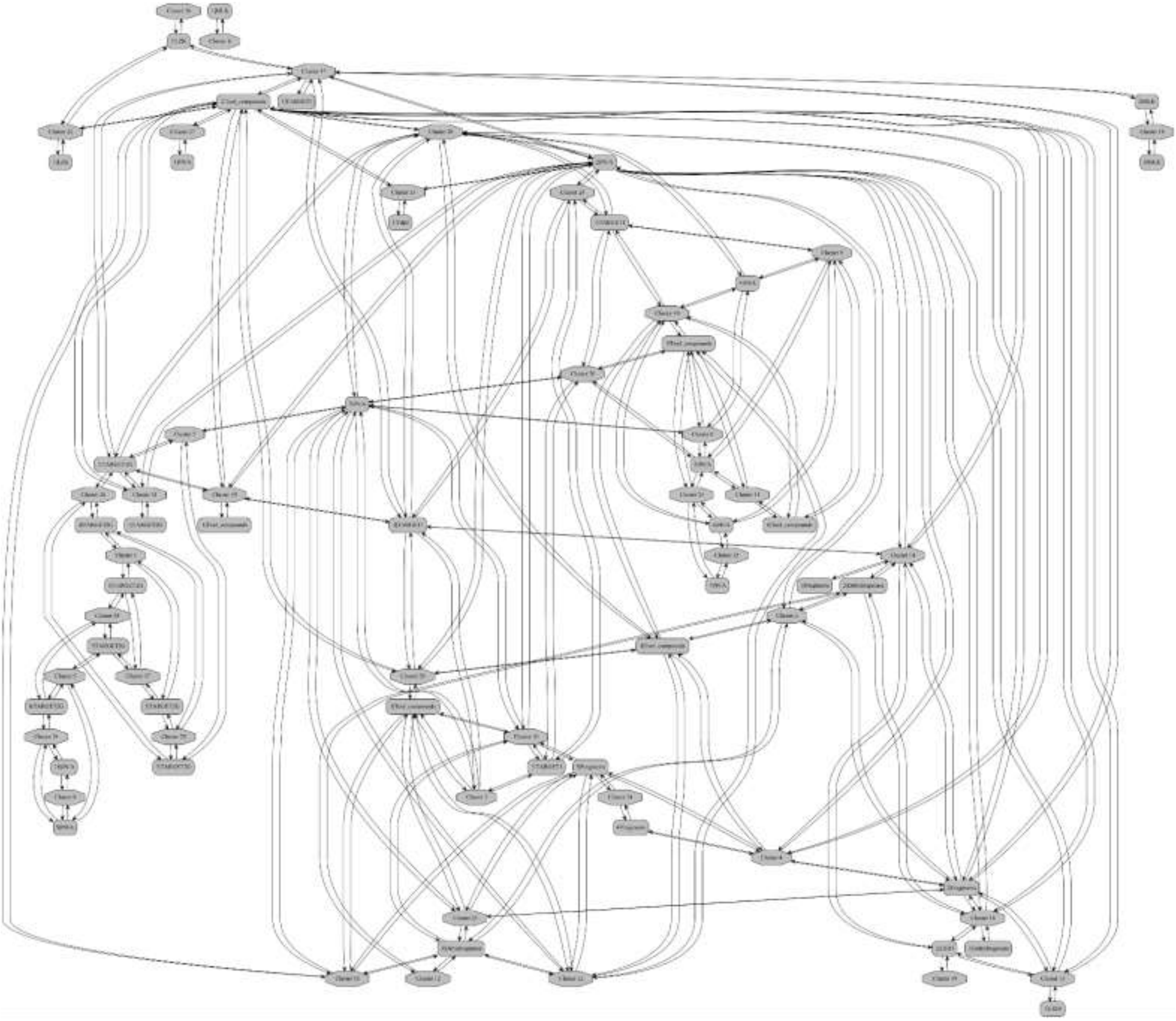

The Figure shows that the activities 1Tool_compounds and 1/TARGET1 were having the highest and second highest levels in frequency significance metric, route significance, and aggregate unary metrics. These imply that 1Tool_compounds and 1/TARGET1 were the most important activities.

*Figure 66. Mine for a Fuzzy Model – Fuzzy unary metrics 2011-2015 of ProM software*

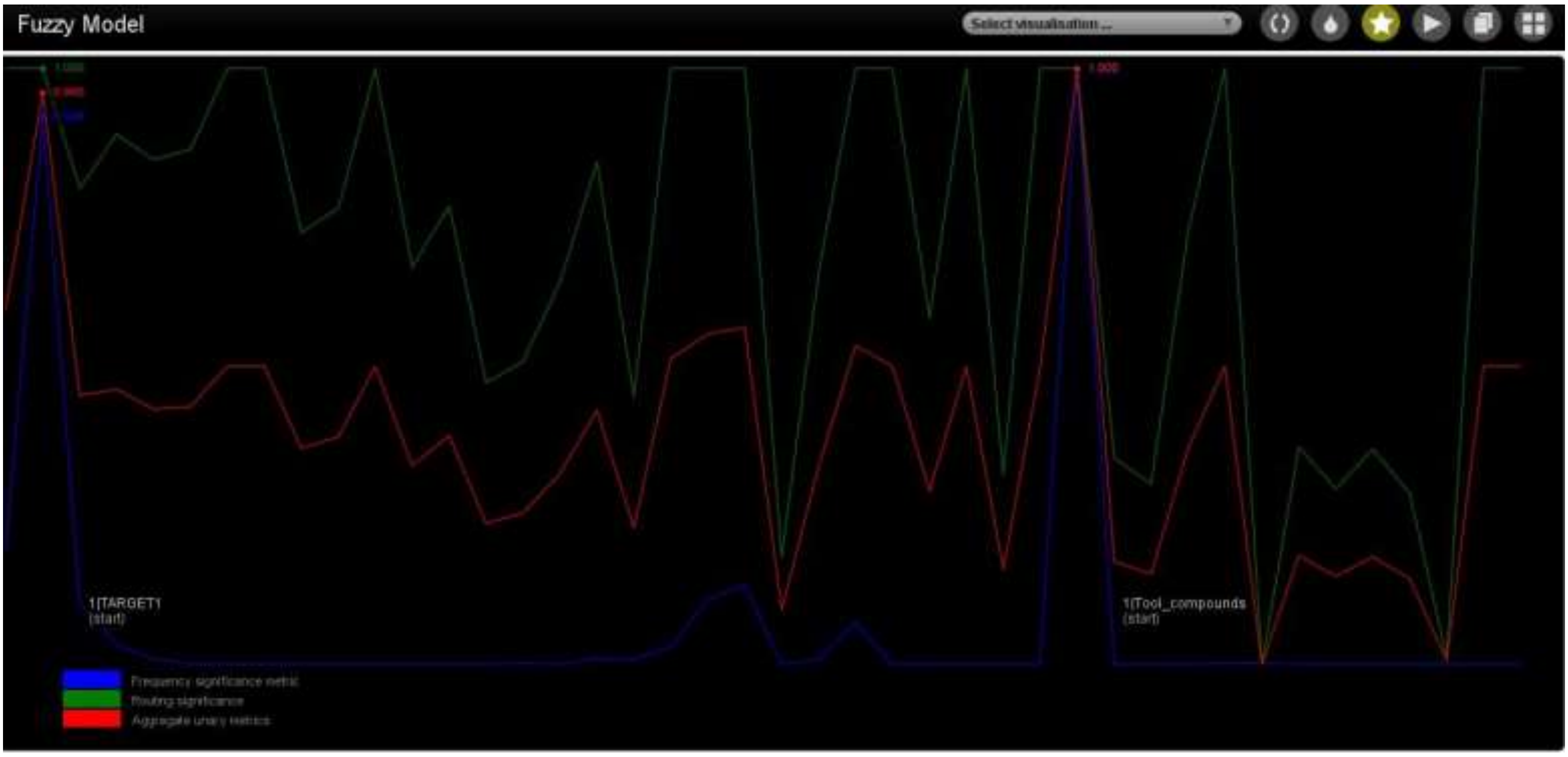

### 4.3.3 Social network

The Social Network pugins of ProM software is specialising answer organisational related questions about the process (ProM Tool, 2010a). Figure 67 shows the social network map of Handover-of-Work between staff. With the layout of ranking view, the correlations of Handover-of-Work between staff were classified five degrees. The ranking view represented from inner more correlations (staff_27, staff_39) with other staff, to outer less (staff_6) or without (staff_15) correlation(s) with other staff. Figure 68 indicates the case 4585 for a handover-of-work sample.

*Figure 67. Mine for a Handover-of-Work Social Network 2011-2015 of ProM software*

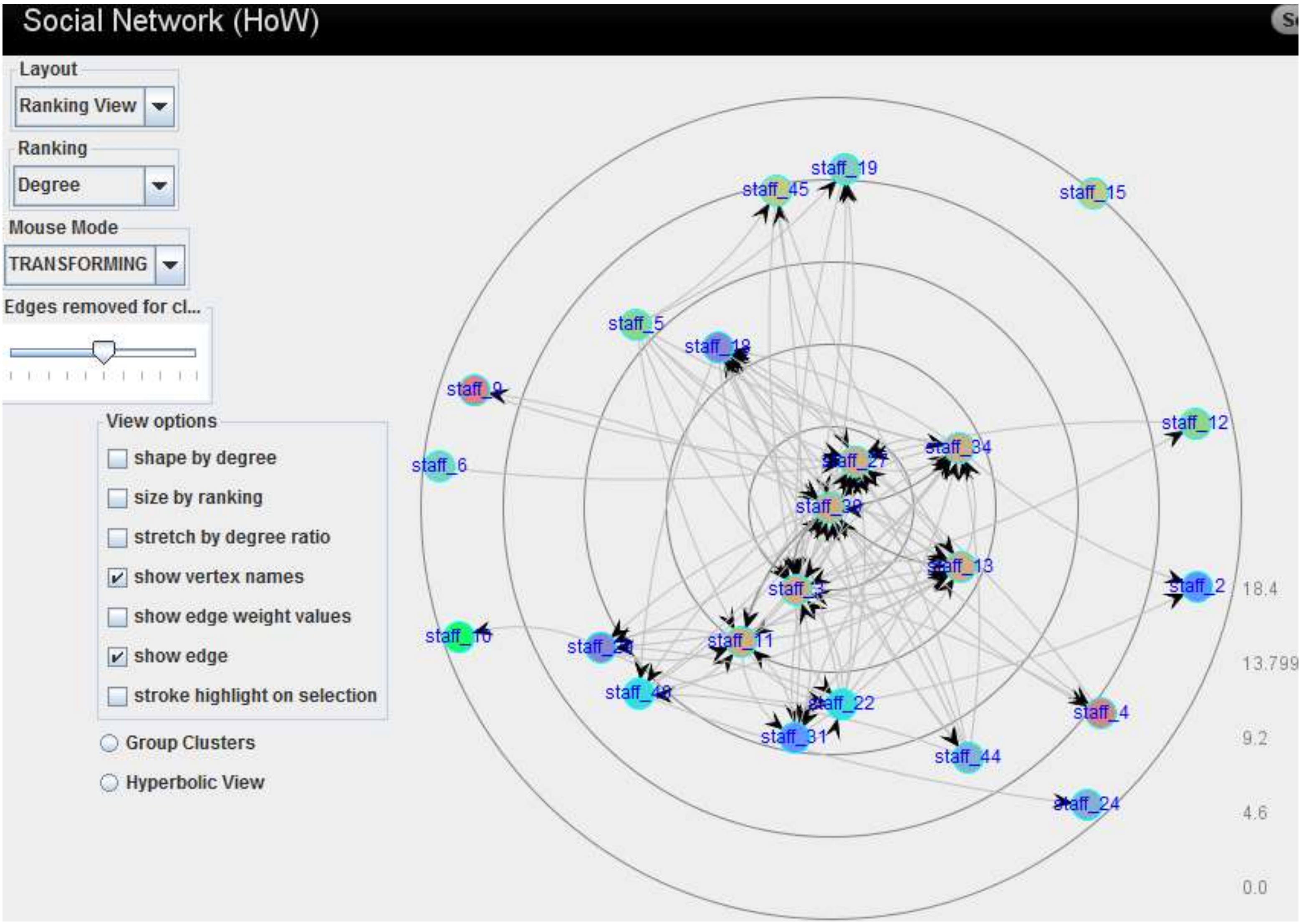

*Figure 68. Mine for Handover-of-Work – sample case 4585 2011-2015 of ProM software*

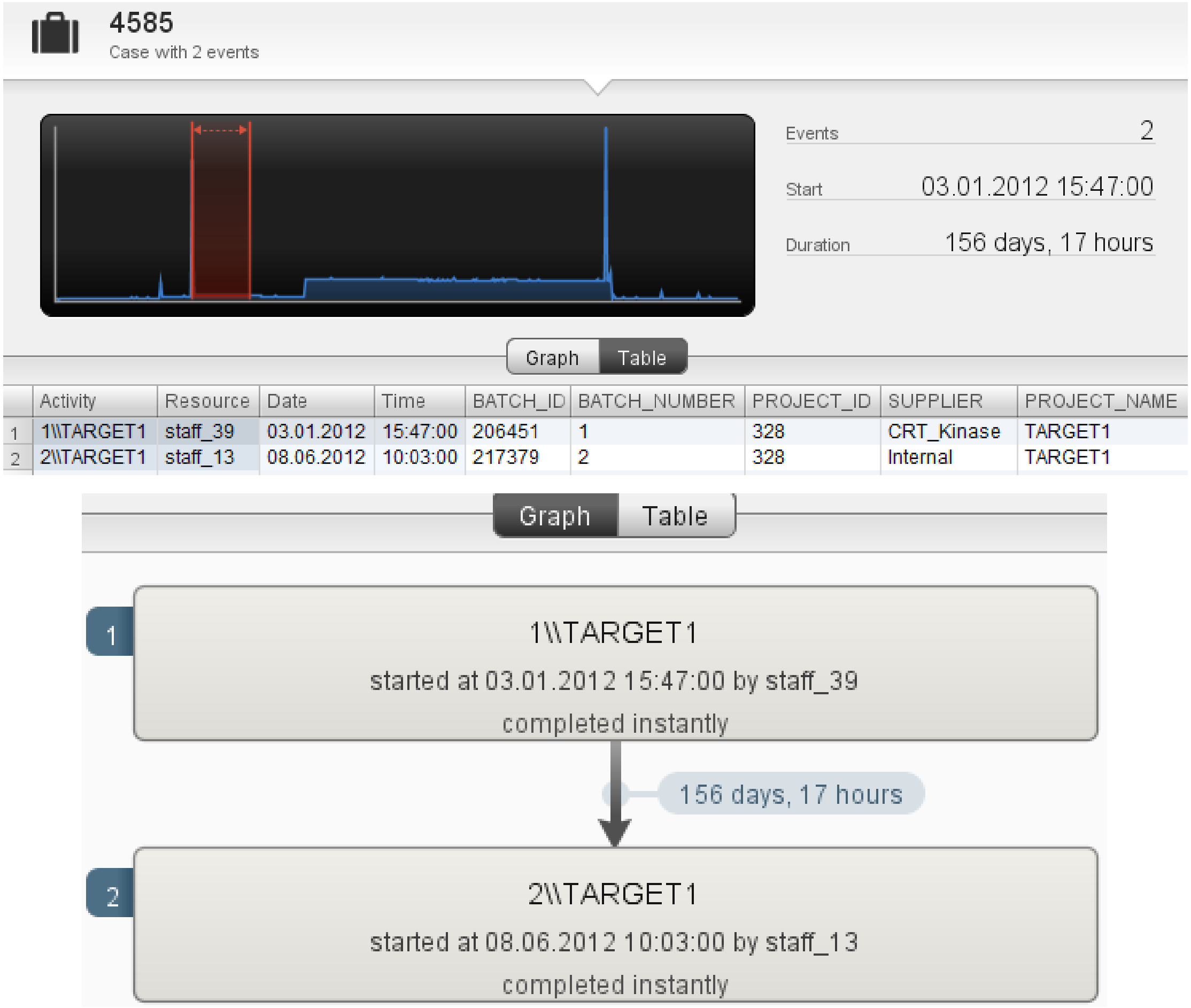

Figure 69 shows the social network map of subcontracting between staff. Based on the KKLayout view, the work was given from a staff to another staff with an arrow towards the end. For instance, both Figure 69 and 70 indicate a case that staff_11 gives a work to staff_13. As Figure 71 draws, the difference between hand-over-work and subcontracting work is that the staff would re-follow up the work for subcontracting, while hand-over-work would not (Van der Aalst, 2004).

*Figure 69. Mine for Subcontracting social network 2011-2015 of ProM software*

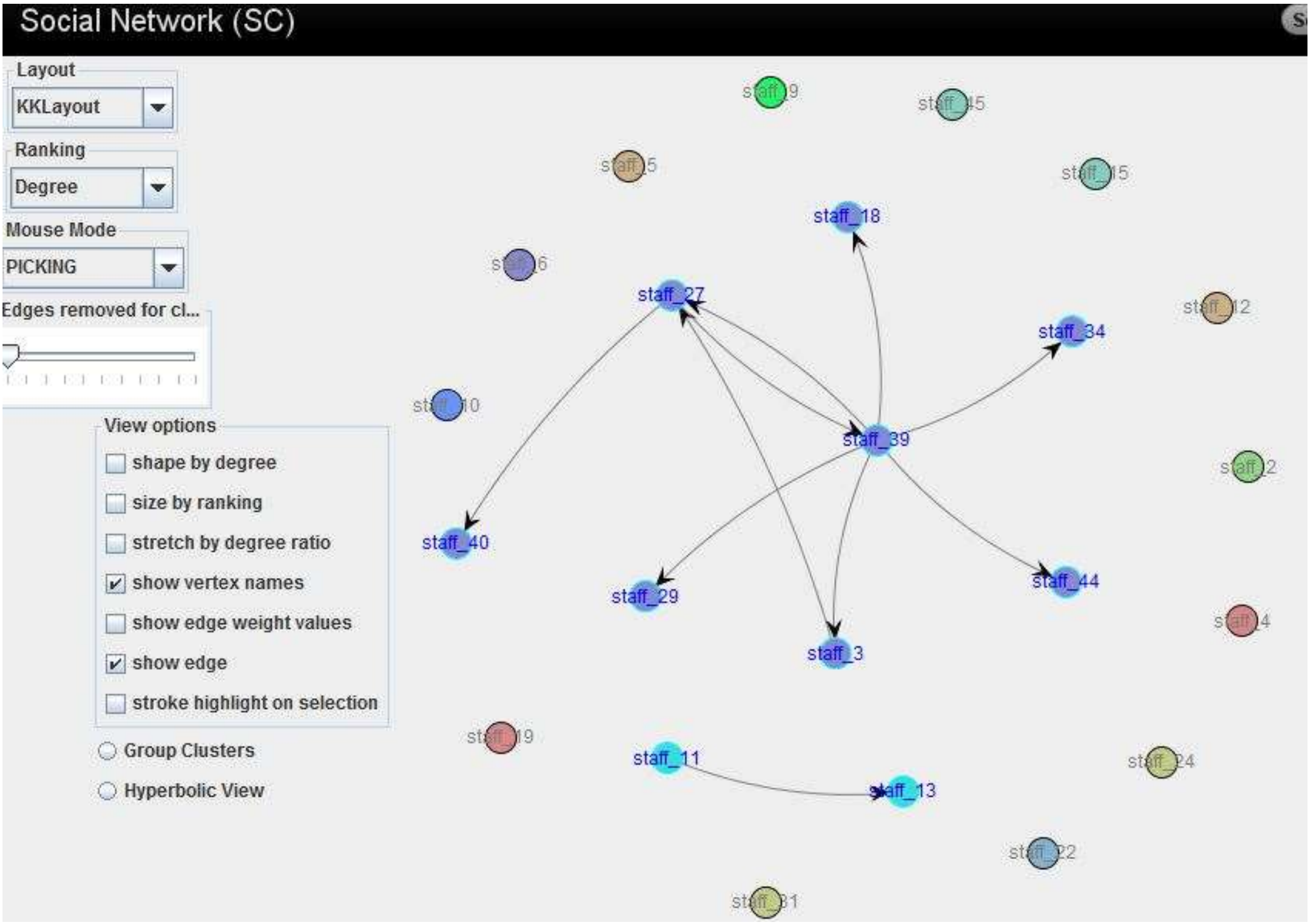

*Figure 70. Mine for Subcontracting – sample case 4032*

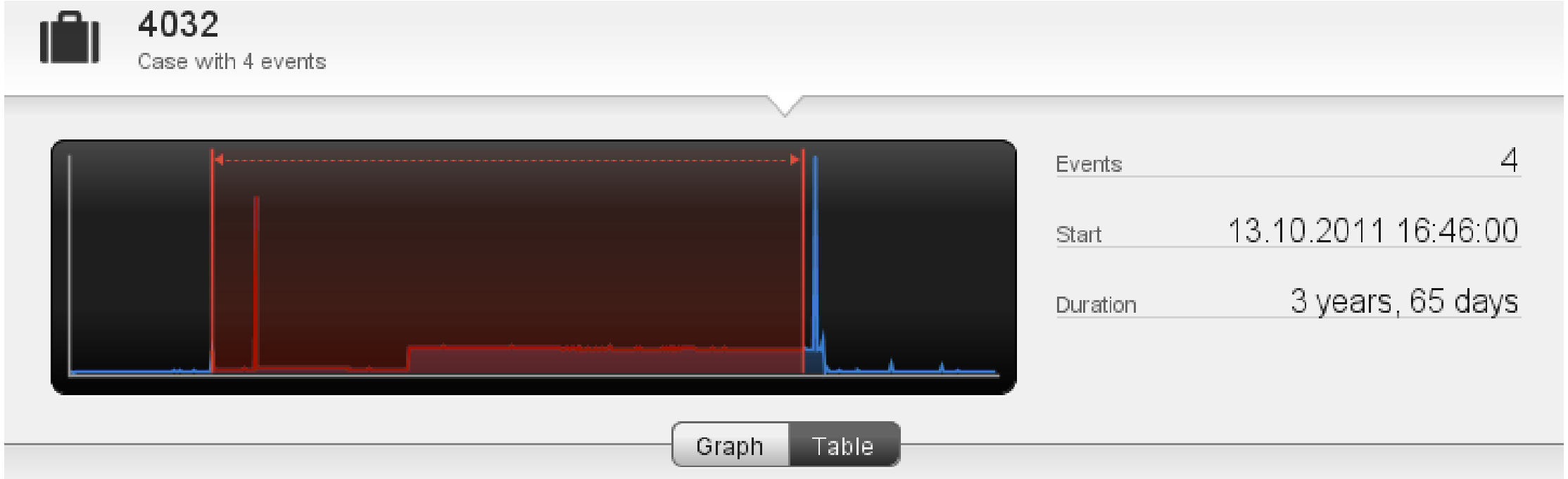

| | Activity | Resource | Date | Time | BATCH_ID | BATCH_NUMBER | PROJECT_ID | SUPPLIER | PROJECT_NAME |
|---|---|---|---|---|---|---|---|---|---|
| 1 | 1\\Fragments | staff_39 | 13.10.2011 | 16:46:00 | 205622 | 1 | 240 | Beatson_Fragment | Fragments |
| 2 | 2\\#N/A | staff_11 | 18.10.2012 | 14:01:00 | 218928 | 2 | NA | UkrOrgSyn | #N/A |
| 3 | 3\\Fragments | staff_13 | 29.11.2013 | 16:09:00 | 224274 | 3 | 240 | External_Fragments | Fragments |
| 4 | 4\\Fragments | staff_11 | 17.12.2014 | 13:31:00 | 227593 | 4 | 240 | UCL | Fragments |

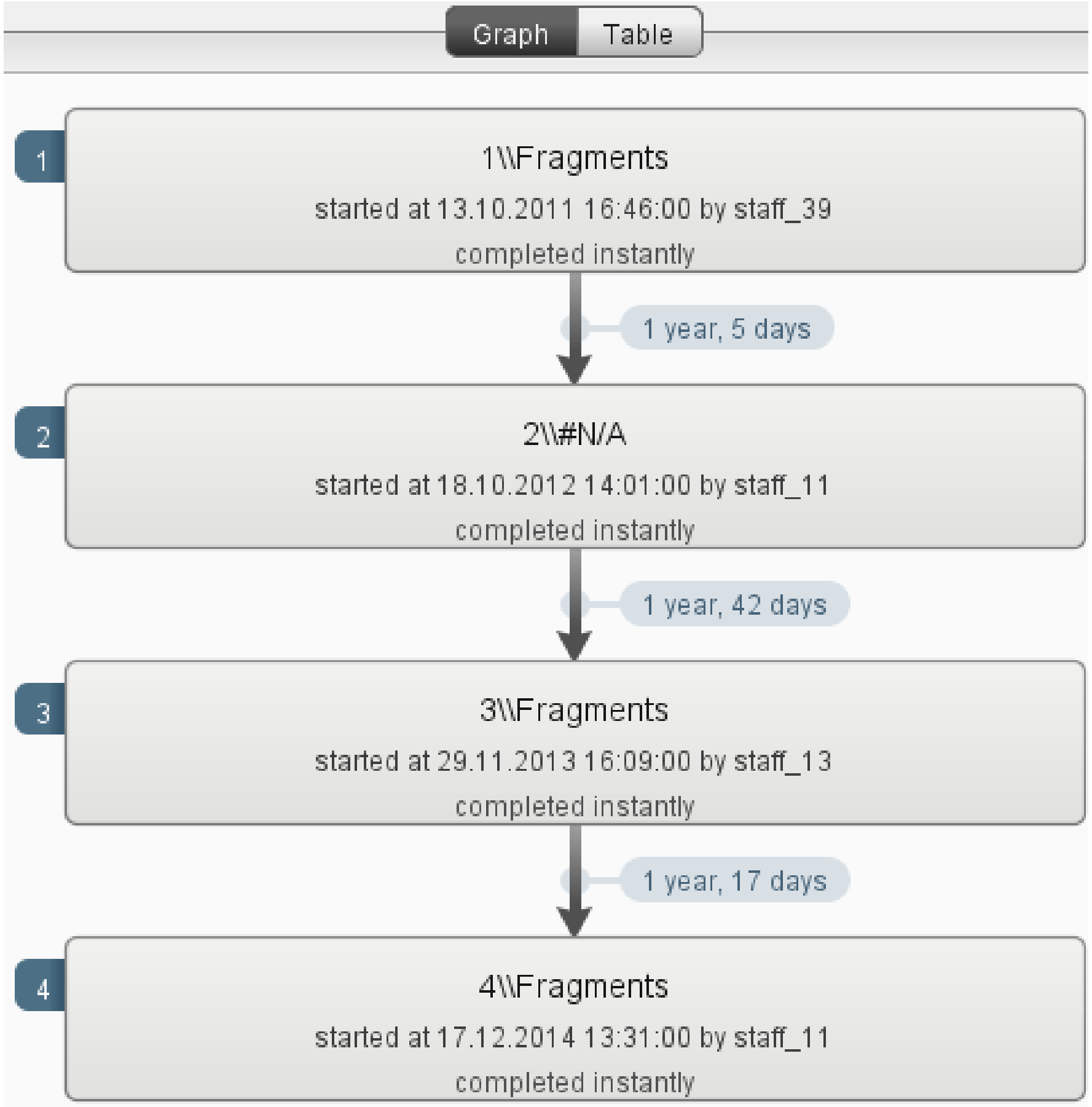

*Figure 71. The diffidence between hand-over-work and subcontracting work (Van der Aalst, 2004)*

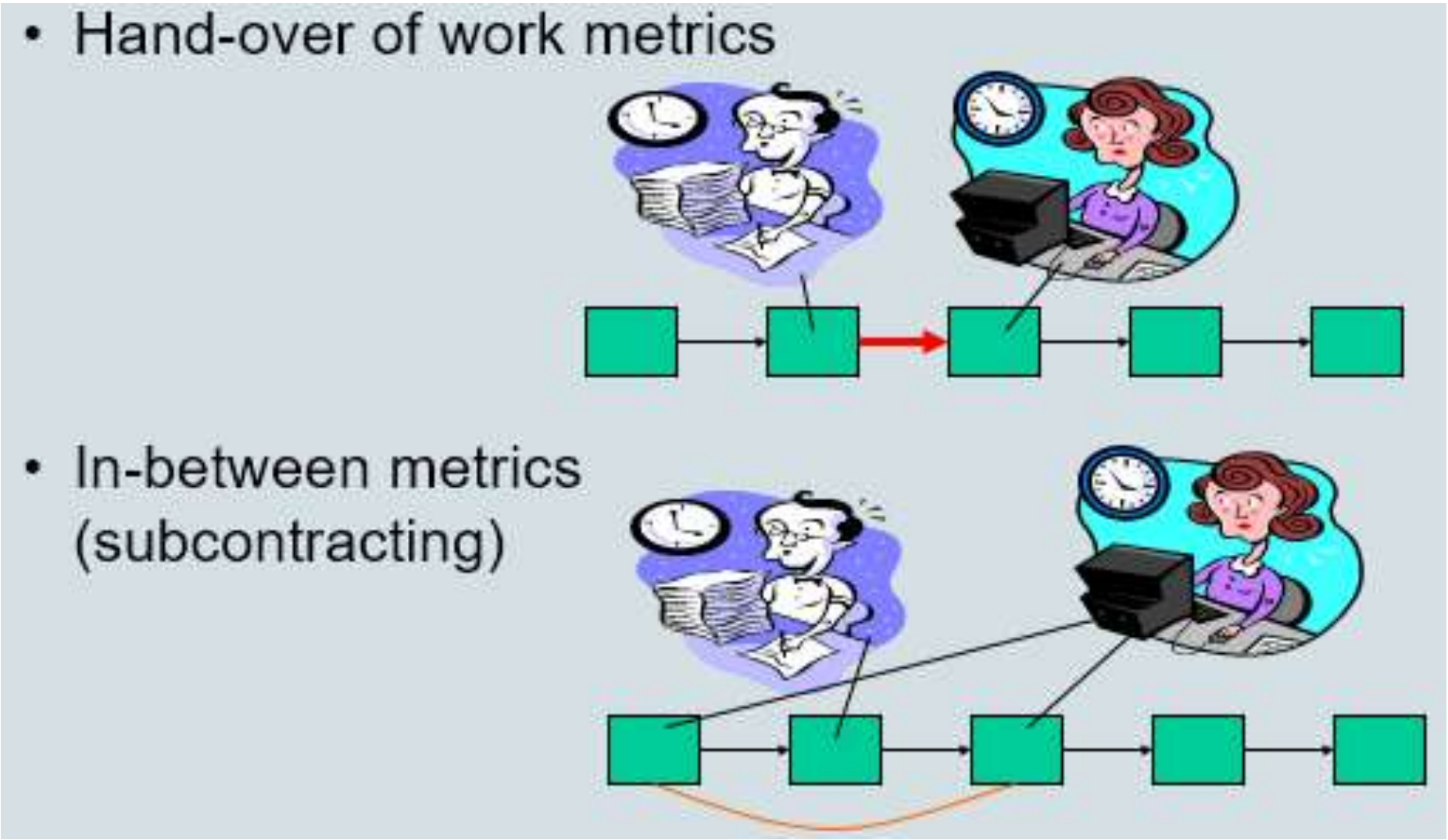

Figure 72 shows the social network map of working-together among staff. Based on the ranking view, the correlations of working-together among staff were classified five degrees, from inner more correlations (staff_27, staff_39) with other staff working together, to outer less (staff_24) or without (staff_15) working together with other staff. Figure 73 indicates the case 16758 for a working-together sample. The staff located in the central of the social network map tended to be the most important to the organisation.

*Figure 72. Mine for working-together social network 2011-2015 of ProM software*

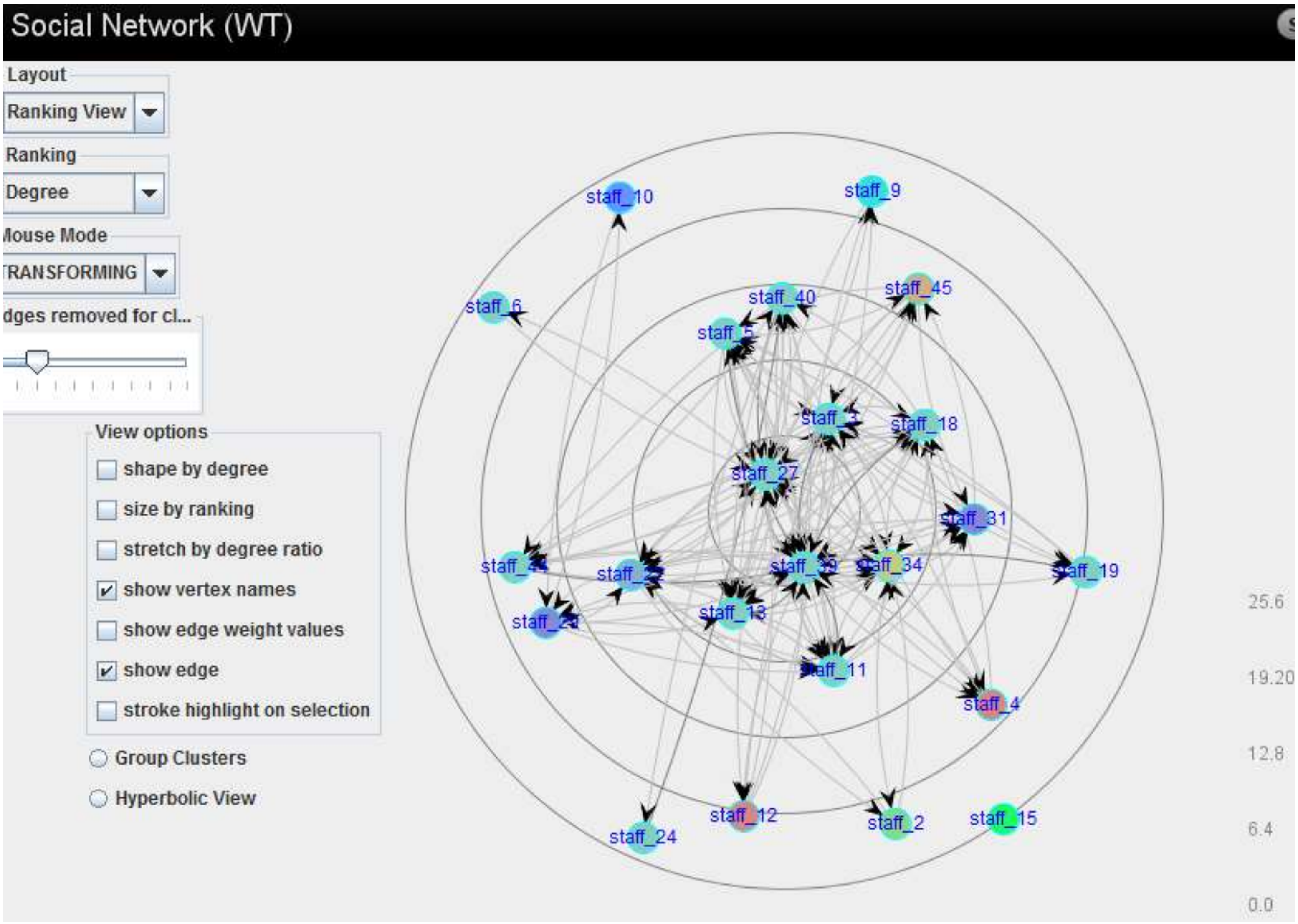

*Figure 73. Mine for working-together – sample case 16758*

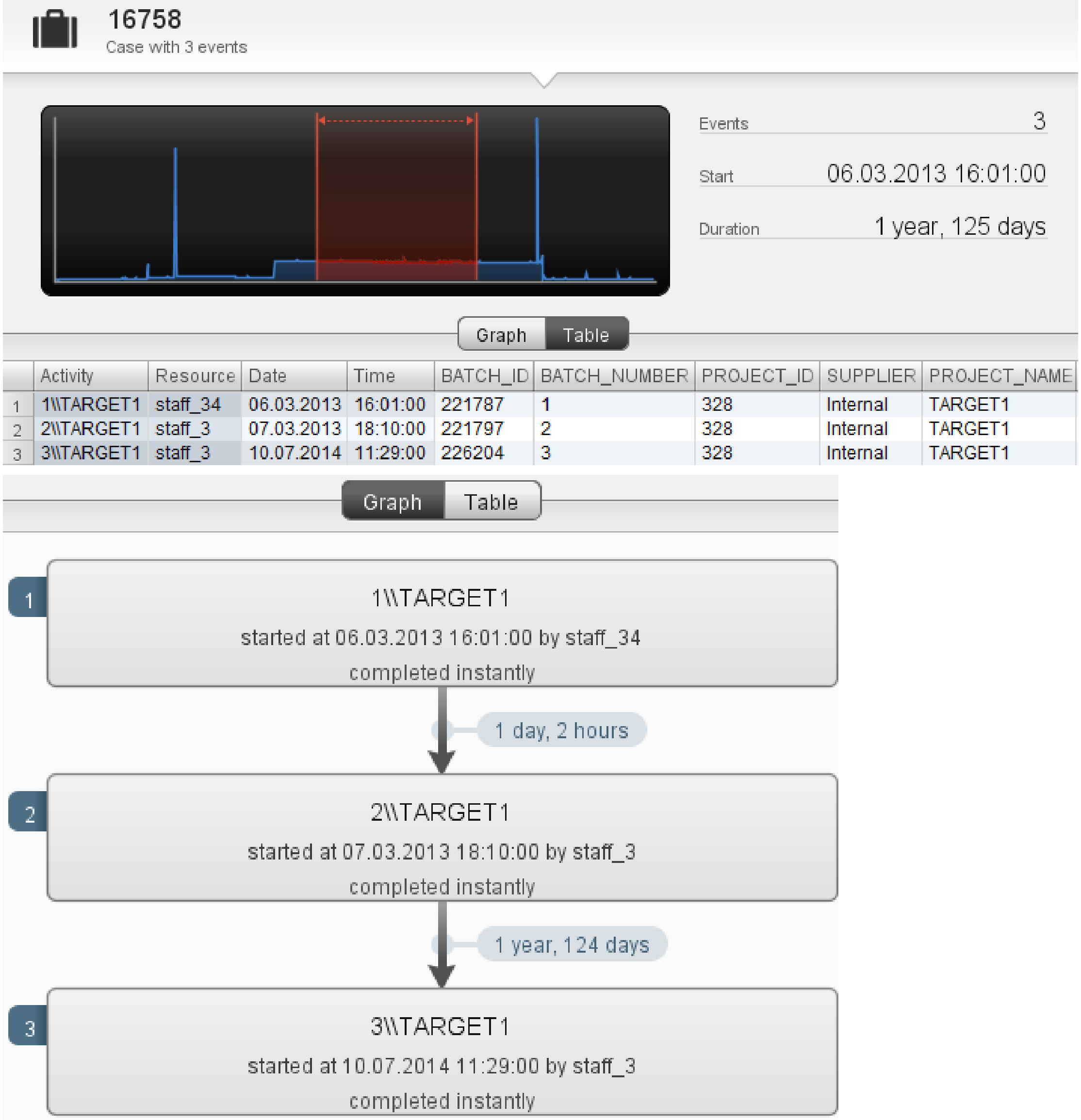

| | Activity | Resource | Date | Time | BATCH_ID | BATCH_NUMBER | PROJECT_ID | SUPPLIER | PROJECT_NAME |
|---|---|---|---|---|---|---|---|---|---|
| 1 | 1\\TARGET1 | staff_34 | 06.03.2013 | 16:01:00 | 221787 | 1 | 328 | Internal | TARGET1 |
| 2 | 2\\TARGET1 | staff_3 | 07.03.2013 | 18:10:00 | 221797 | 2 | 328 | Internal | TARGET1 |
| 3 | 3\\TARGET1 | staff_3 | 10.07.2014 | 11:29:00 | 226204 | 3 | 328 | Internal | TARGET1 |

### 4.3.4 Annual process map conformance checking for 2011-2015

Because the process map of experiments event log only contains one level of the hierarchy. In this section, only the annual process maps of registrations event log have been implemented. From Figure 74 to 78 show the annual process maps from the year 2011 to 2015. There was no unique model can one size fit all, the process map change overtime. It may need to enrich the dataset to give the explanations and evidences of the reason.

*Figure 74. Full process map of registrations event log 2011 of ProM software*

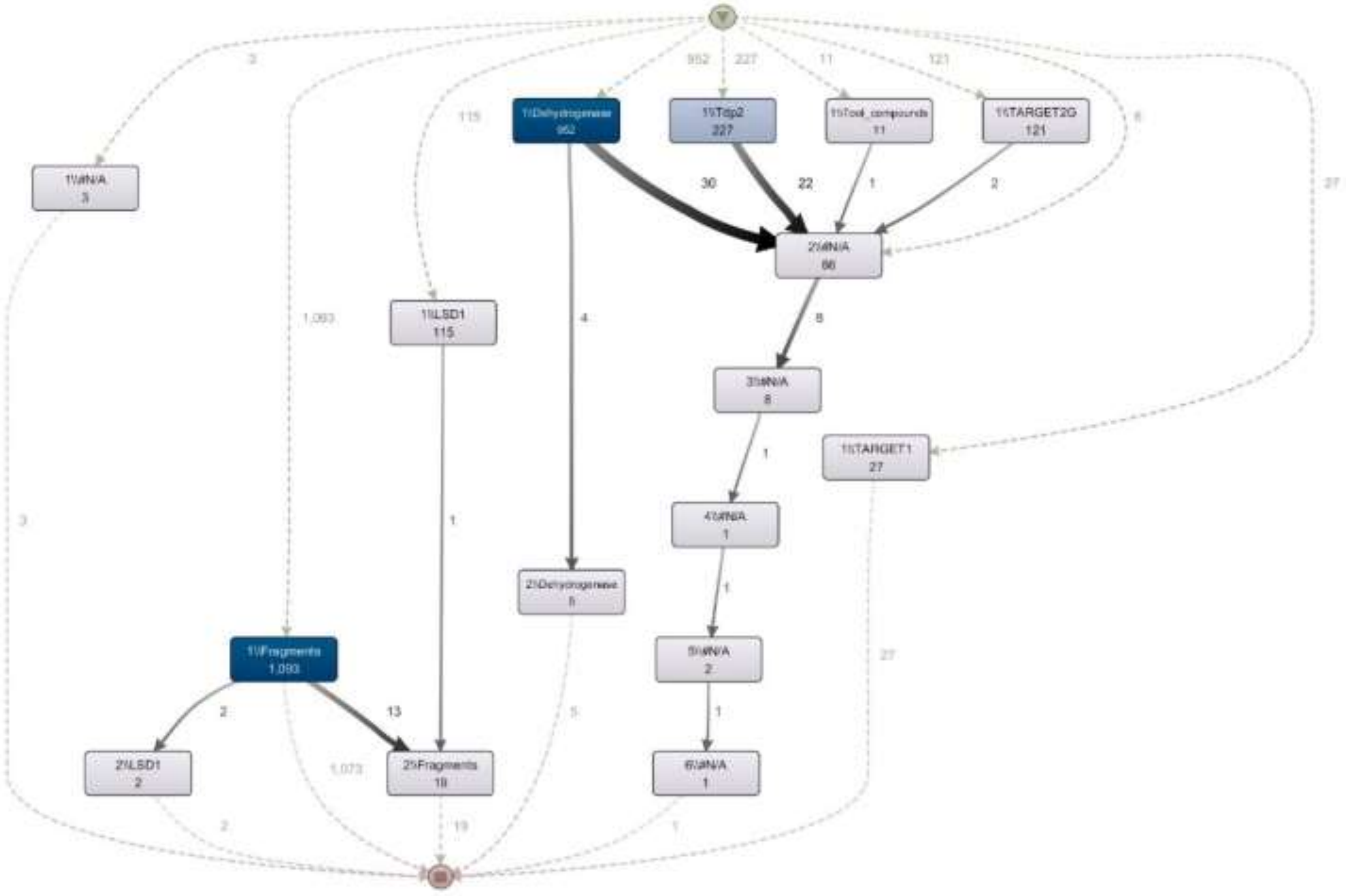

*Figure 75. Full process map of registrations event log 2012 of ProM software*

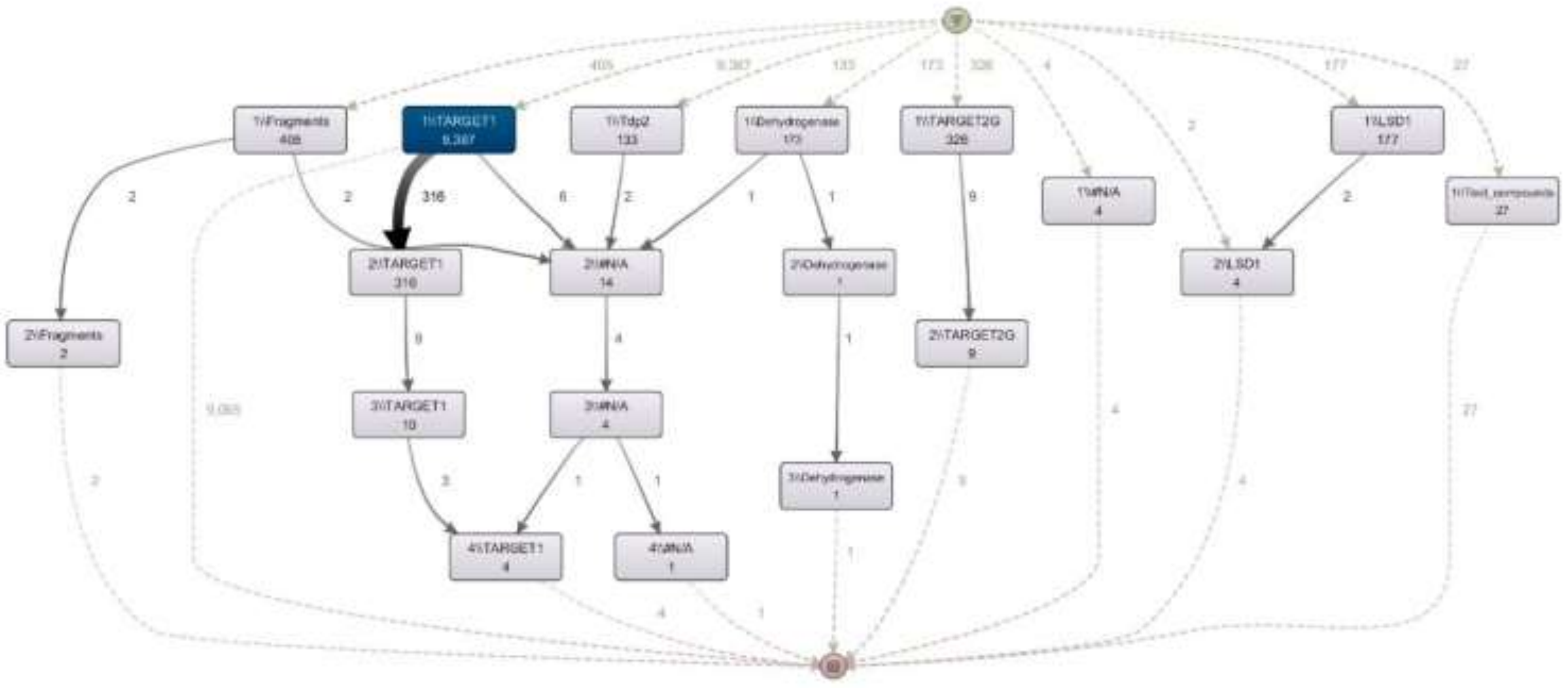

*Figure 76. Full process map of registrations event log 2013 of ProM software*

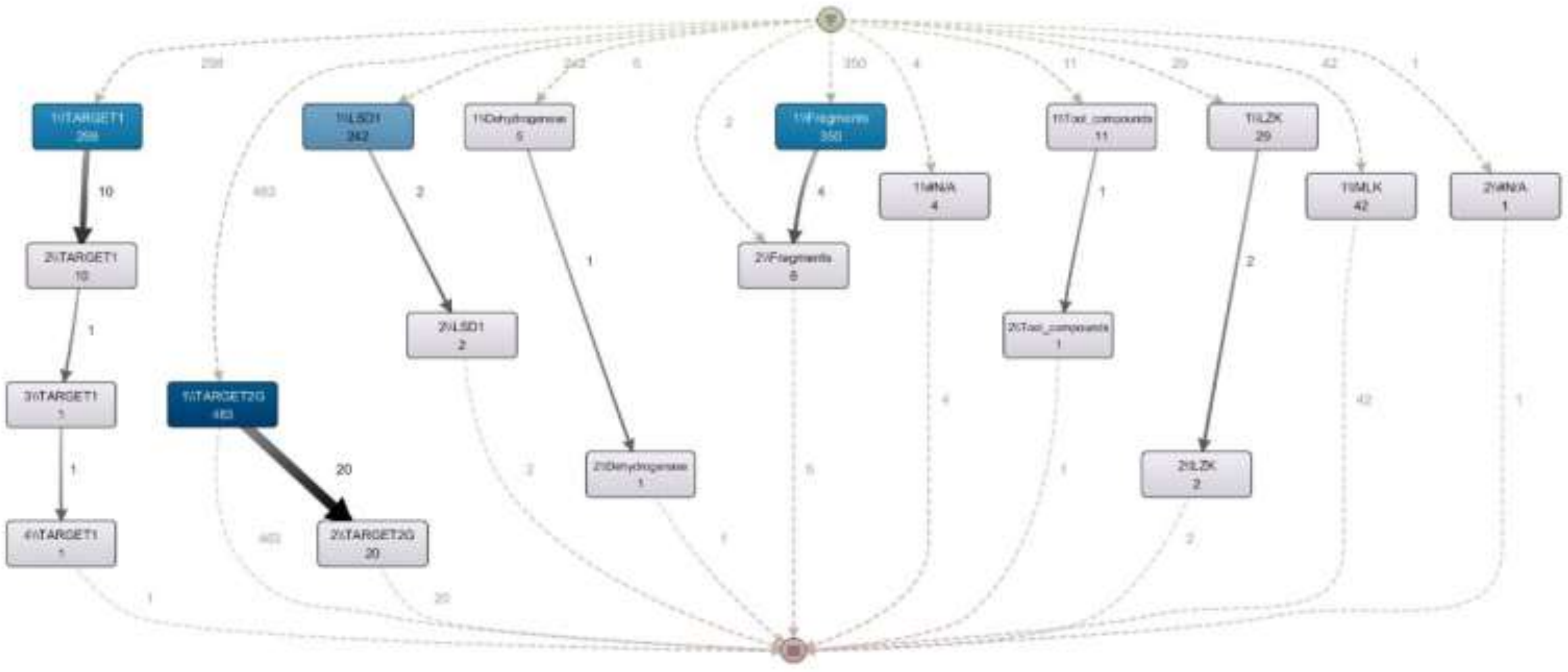

*Figure 77. Full process map of registrations event log 2014 of ProM software*

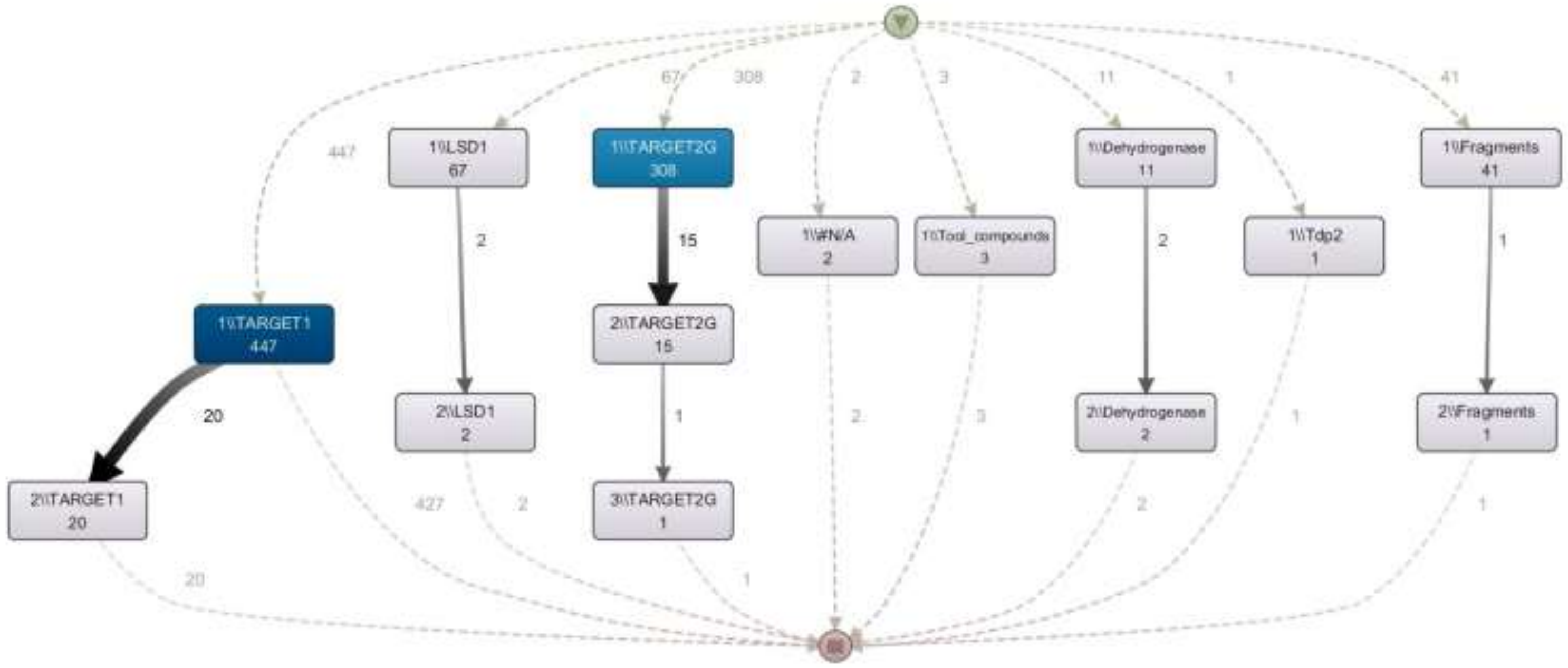

*Figure 78. Full process map of registrations event log 2015 of ProM software*

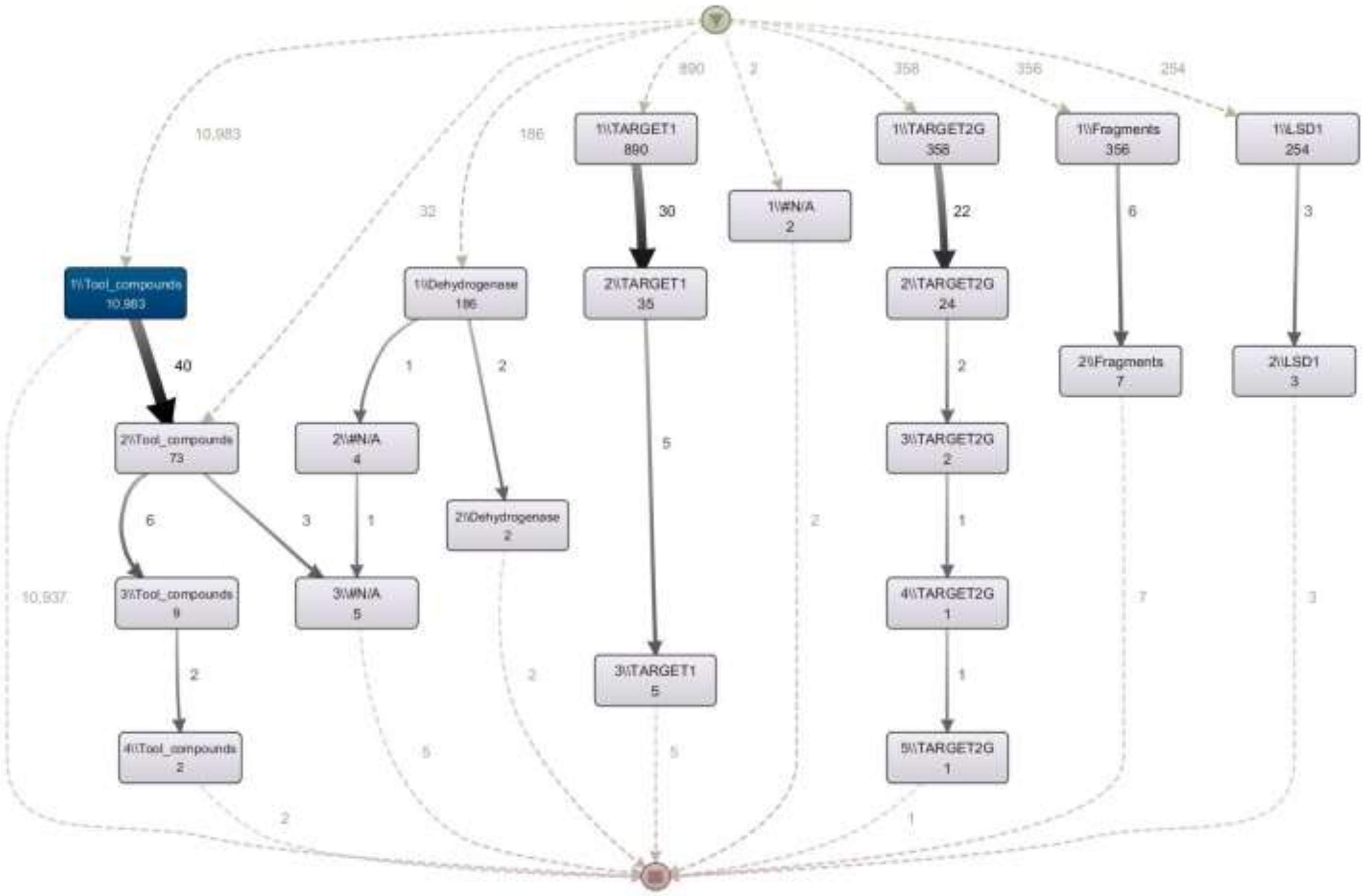

# 5. Chapter Five Discussion

## 5.1 Positioning of process mining types

This paper has conducted the type of process mining research which was named discovery as Table 80 show. The discovery technique generates a model with an event log, but without any a-priori information. It is the fundamental process mining technique. In this paper, two models – experiments and registrations process logs were produced.

In relating to the case of DDU of CRUK, the Figure 79 can explain the positioning of process mining in the organisation as a whole comprehensively. The event logs were experiments and registrations datasets. The software systems were Dotmatics and Oracle which DDU is using. The “world” were including staff, machines, compounds, screenings, protocols, projects, programs, systems, suppliers, organisations, et al.

The “world” were the most complicated element for the implementation of this process model life cycle. It involves so many internal and external characters, in order to collect and integrate all sources of information. For example the data of external systems which need to be uploaded: Cyprotex, Thermal Shift, outsourced chemistry, Mechanism of Inhibition, SPR data.

After the implementation of the process mining techniques, it is considered that it was difficult to link data from a variety of tables and event logs all together. The reasons were identified: firstly, there was no global unique foreign key to link all tables and event logs together, it need to be assigned and redesigned; secondly, the missing data in the event logs obstruct the application of process mining techniques and algorithms; thirdly ,the researcher who performs process mining maybe lack of relative background knowledge. The conformance checking was failed to check because of lack of data. Due to these reasons and time limit, the enhancement has not been implemented. Van

der Aalst et al (2011) claims that significant efforts are required to link events where belong to the same process instance.

*Figure 79. Positioning of the three main type of process mining: discovery, conformance checking, and enhancement (Van der Aalst et al., 2011)*

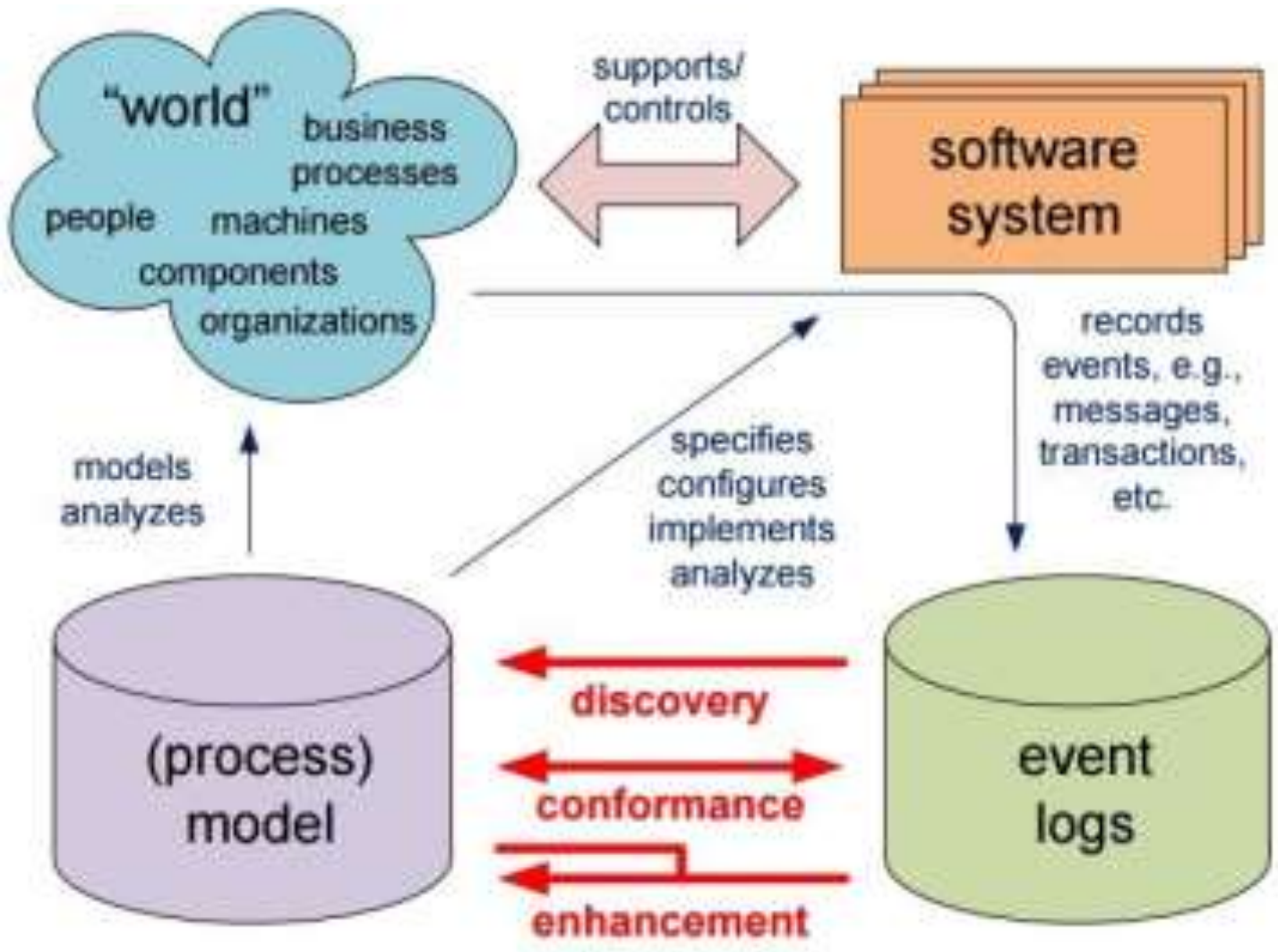

*Figure 80. The three basic types of process mining explained in terms of input and output: discovery, conformance checking, and enhancement (Van der Aalst et al., 2011)*

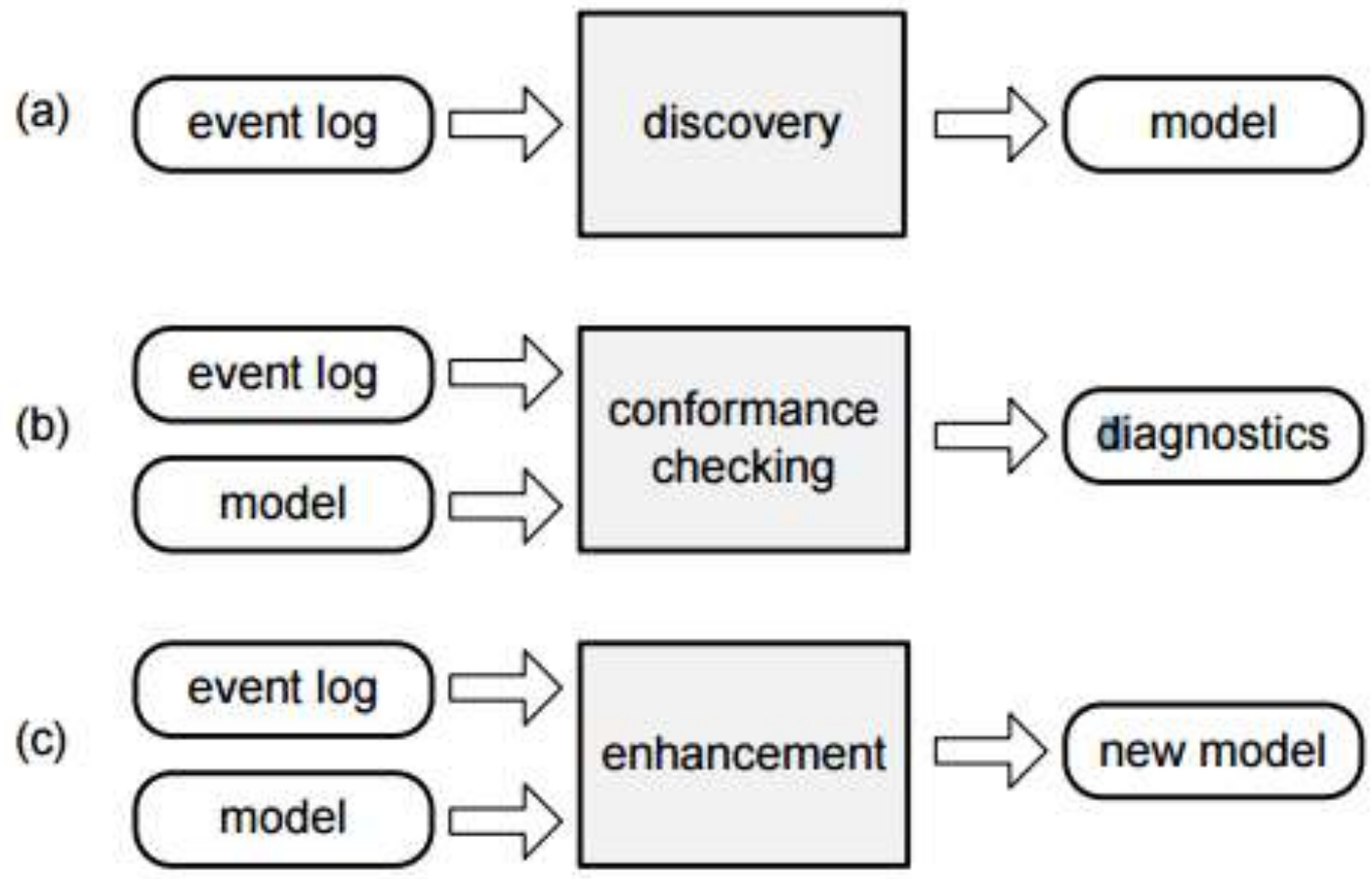

## 5.2 Evaluations

Table 81 indicates that evaluations between tools and event logs. The event log of registrations has better performances in techniques matrixes, clusters, social networks, additional mines/models due to its comprehensive dataset when comparing to event log of experiments.

The tools – Disco and ProM have similar functions in descriptive statistics, process maps, animations, filters, graphic visualisations, conversion, import, export. Disco software was designed for non-experts specifically. Disco has more user friendly graphical interfaces with easy manipulation. Disco's unique interactive dashboards visualise context/content of data interactively, make it easier for the management of organisations.

While ProM is an open source software with more plugins of techniques than Disco like additional process maps, matrixes, clusters, social networks, additional mines/models. These can help process mining experts to research deeper for particular fields and wider ranges.

*Table 81. Evaluations of Tools and Event logs data*

| ***Tools*** | ***Disco*** | | ***ProM*** | |
|---|---|---|---|---|
| ***Techniques\Event logs*** | ***Experiments*** | ***Registrations*** | ***Experiments*** | ***Registrations*** |
| *Descriptive statistics* | Yes | Yes | Yes | Yes |
| *Process maps* | Yes | Yes | Yes | Yes |
| *Additional process maps* | No | No | Yes | Yes |
| *Animations* | Yes | Yes | Yes | Yes |
| *Interactive dashboards* | Yes | Yes | No | No |
| *Filters* | Yes | Yes | Yes | Yes |
| *Graphic visualisations* | Yes | Yes | Yes | Yes |
| *Matrixes* | No | No | Data issues | Yes |
| Clusters | No | No | Data issues | Yes |
| Social networks | No | No | Data issues | Yes |
| Additional mines/models | No | No | Data issues | Yes |
| Conversion | Yes | Yes | Yes | Yes |
| Import | Yes | Yes | Yes | Yes |
| Export | Yes | Yes | Yes | Yes |

## 5.3 Discovery with tools support

In this study, the pre-process procedures and implementations of process mining techniques and algorithms have been executed by Excel, Disco, and ProM as Table 81 shows. No programming skills are required to manipulate these software.

The Excel software was mainly for the procedures of data merging, cleaning, manipulating, and descriptive statistics since the dataset was not large relatively. It is easy to demonstrate and widely used across industries and academics.

The Disco is a closed source software. It has been used in descriptive statistics, interactive dashboards, and dynamic visualisations. The advantage of Disco is its friendly graphical user interface, which makes the interactive dashboards easier to use even for the non-expert in the field of process mining. It does not require much previous relevant background knowledge to explain the meaning of the process maps and figures it generated. It can directly apply to such as CSV file to extract data.

The ProM is an open source software. It has been used in the generations of process maps, dotted charts, matrixes, clusters, metrics, and social network map. It can be applied even more wide range of process mining techniques and algorithms. However, it may be difficult to use for the non-expert in process mining area. Not all of the plug-ins of ProM can be used to all type of process logs. Thus the plug-ins functions need to be selected carefully for different process logs and purposes.

One of the advantages of using different process mining software into same dataset is that any errors/mistakes would soon be identified because of comparisons.

The three types of research questions (Verbeek & Bose, 2010) have been answered through these tools support. The first type research questions were from control flow perspective: how the cases were actually being executed; the mean/minimum/maximum throughput time of cases; the paths/routes take too much time on average; the sub-paths/routes of these paths/routes. The second type research questions were from performance perspective: the identification of the most frequent flow; the average process time for each activity; the time was spent between any two activities in the process model; the number of activities/cases happen on any time slot. The third type research questions were from organisational perspective: the number of staff were involved in a same activity/case; identification of the most important (busy) staff in the process flow; the staff handover the work to whom; the staff subcontract the work to whom, the staff work on the same cases.

### 5.4 Discussion of data types

This paper has conducted two case studies, based on two groups of event logs datasets - experiments and registrations. The different shape of process maps between experiments and registrations event logs are believed due to the different number of event classes per case. The experiments event log has maximum 1 event classes per case, while the registrations have maximum 10 event classes per case as Figure 60 shows. The number of event classes per case of registrations event log determines that they can apply more various process mining techniques and algorithms. In other words, a case with several event classes can have better performance in terms of the application of process mining techniques. The event attribute of registrations event log was based on the integration of data column "BATCH_NUMBER" and "PROJECT_NAME", which enriched the level of event classes. The enrichment of event classes makes the cases can be traced properly.

Because of the limited event classes per case of experiments event log, it would be meaningless to apply the additional process mining techniques and algorithms into experiments event log. As Figure 82 shows, there were complete no connections between staff. This may be due to there was really no connection between staff, but more possible was because of lack of data of event classes per case. However, assign a global unique foreign key to a compound which aligns with different tables can solve this problem. It is common that the database and tables are designed either activity-centric or case-centric, adding a global unique foreign key can help turn the database and tables process-centric. This method can help to identify more event classes per case have gone through, so that it can be better illustrated by process mining techniques.

*Figure 82. Social network map of experiments event log of ProM software. Top left: working-together social network. Top right: subcontracting social network. Bottom left: handover-of-work social network.*

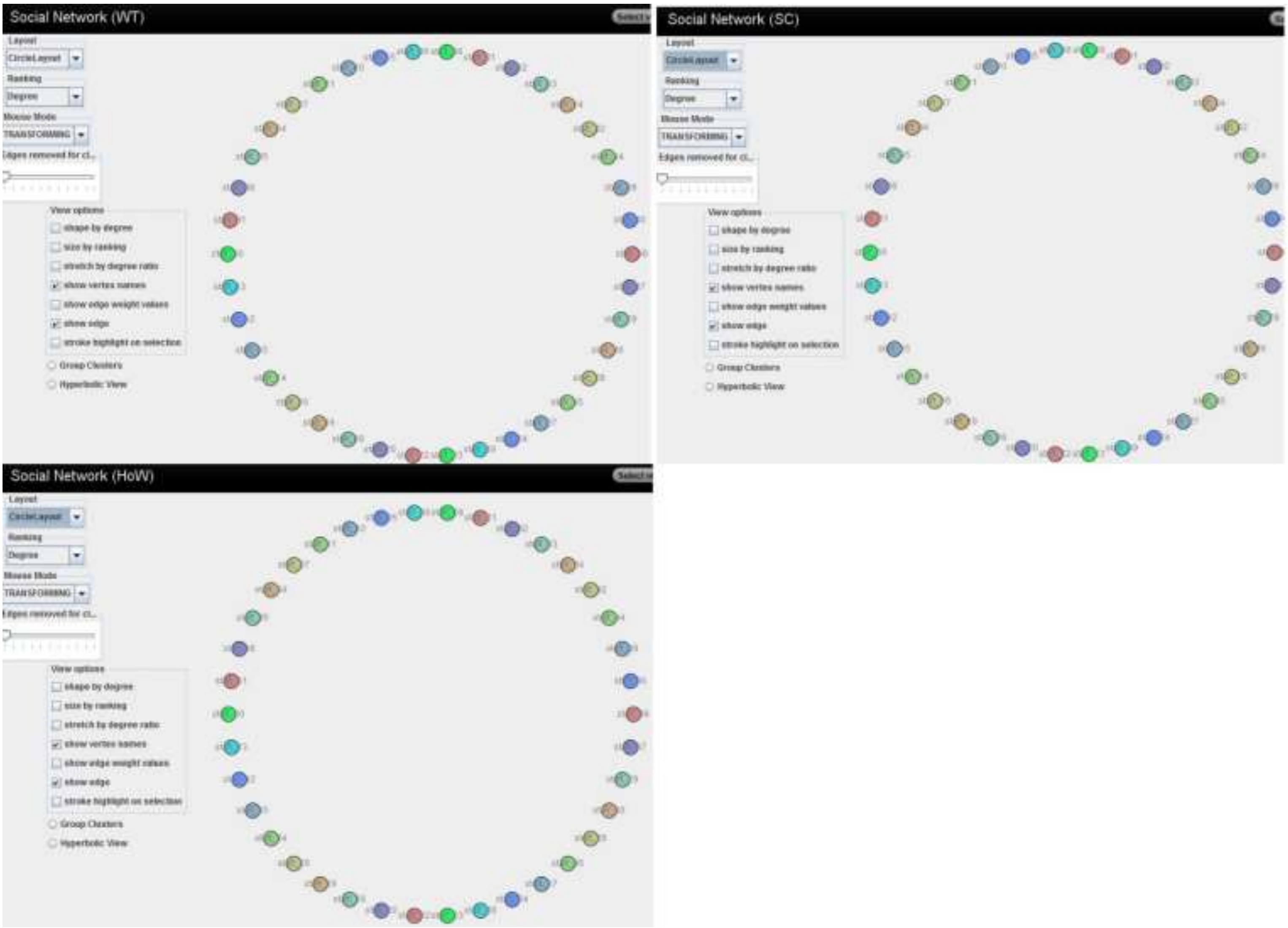

In terms of filters by timeframe, it has to be carefully considered which timeframe should be selected. The options include contained in, intersecting, started in, completed in, and trim to timeframe as Figure 82 shows. This may be slightly effect when the timeframe is long (year, season), but may has a big influence when the timeframe is short (month, week).

*Figure 83. Filters by timeframe function of Disco software*

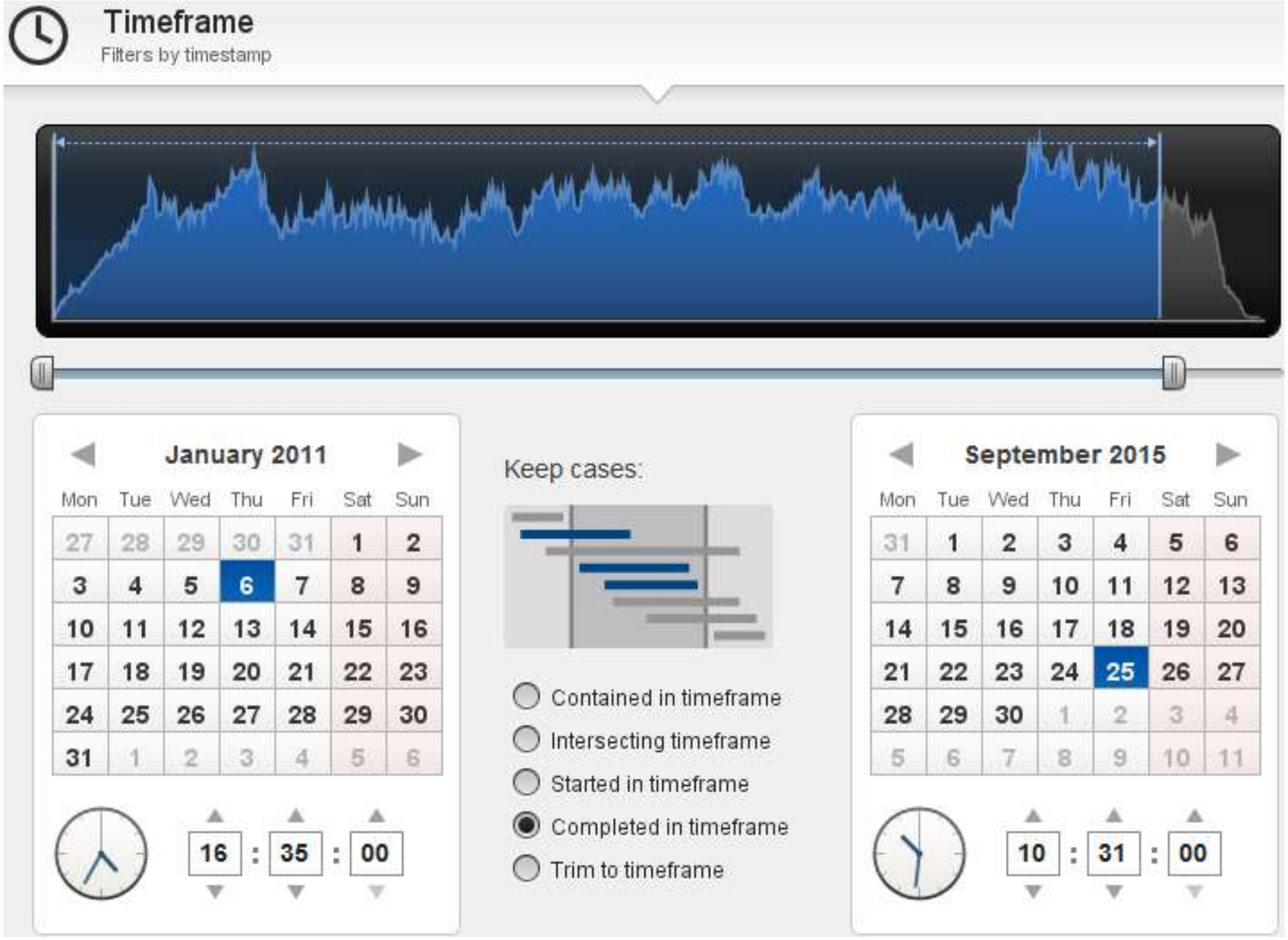

## 5.5 Discussion of data quality

Van der Aalst et al. (2011) proposed three aspects to judge the quality of event data. Events should be trustworthy. In this case, the content can be assumed to be correct and match reality, event logs should be complete. In this case, there were missing data. Some of the recorded events have not well-defined semantics. Moreover, the event data should be safe in terms of privacy and security concerns were addressed when recording the events. For example, originators should aware that events have been recorded and the way they were used. In this case, the names of staff should be anonymised before hand out to process mining researchers.

Based on the event logs data and sources provided by DDU of CRUK, the characterisation of event log data have been classified as between level three and four as Table 84 shows (Van der Aalst et al., 2011). The events were recorded in an ERP system - Dotmatics software which was having both manual (such as key entry missing and typo error) and automatic approaches in a systematic and reliable manner. The events need to be extracted from a variety of tables. The process semantics, instances (cases) and activities were not supported in an explicit manner, which needs to be explained by relevant experts in details.

*Table 84. Maturity levels for event logs (Van der Aalst et al., 2011)*

| Level | Characterization |
|---|---|
| ⋆⋆⋆⋆⋆ | Highest level: the event log is of excellent quality (i.e., trustworthy and complete) and events are well-defined. Events are recorded in an automatic, systematic, reliable, and safe manner. Privacy and security considerations are addressed adequately. Moreover, the events recorded (and all of their attributes) have clear semantics. This implies the existence of one or more ontologies. Events and their attributes point to this ontology. *Example:* semantically annotated logs of BPM systems. |
| ⋆⋆⋆⋆ | Events are recorded automatically and in a systematic and reliable manner, i.e., logs are trustworthy and complete. Unlike the systems operating at level ⋆⋆⋆, notions such as process instance (case) and activity are supported in an explicit manner. *Example:* the events logs of traditional BPM/workflow systems. |
| ⋆⋆⋆ | Events are recorded automatically, but no systematic approach is followed to record events. However, unlike logs at level ⋆⋆, there is some level of guarantee that the events recorded match reality (i.e., the event log is trustworthy but not necessarily complete). Consider, for example, the events recorded by an ERP system. Although events need to be extracted from a variety of tables, the information can be assumed to be correct (e.g., it is safe to assume that a payment recorded by the ERP actually exists and vice versa). *Examples:* tables in ERP systems, events logs of CRM systems, transaction logs of messaging systems, event logs of high-tech systems, etc. |
| ⋆⋆ | Events are recorded automatically, i.e., as a by-product of some information system. Coverage varies, i.e., no systematic approach is followed to decide which events are recorded. Moreover, it is possible to bypass the information system. Hence, events may be missing or not recorded properly. *Examples:* event logs of document and product management systems, error logs of embedded systems, worksheets of service engineers, etc. |
| ⋆ | Lowest level: event logs are of poor quality. Recorded events may not correspond to reality and events may be missing. Event logs for which events are recorded by hand typically have such characteristics. *Examples:* trails left in paper documents routed through the organization ("yellow notes"), paper-based medical records, etc. |

## 5.6 Guiding principles of process mining

Van Der Aalst (2012) has summarised six guiding principles:
The guiding principle one is the event data should be treated as a priority (Van Der Aalst, 2012). The events should be trustworthy, it should be safe to assume that the recorded events actually happened and that the attributes of events are correct. Such as staff A uses staff B's log in to execute event should not happen. Event logs should be complete, that is, given a particular scope, no events may be missing. Any recorded event should have well-defined semantics. Moreover, the event data should be safe in the sense that privacy and security concerns are addressed when recording the event log. Such as the event data should not contain any personal private information, passwords.

The guiding principle two is the extraction of event logs data should be questions driven (Van Der Aalst, 2012). Without concrete questions, the thousands of tables in the database of an ERP system like Oracle do not know where to start to extract meaningful event data. The event data being extracted should support concurrency.

The guiding principle three is choice and other basic control-flow constructs (Van Der Aalst, 2012). Basic workflow patterns should be supported by process mining techniques. These include all mainstream languages (e.g., BPMN, EPCs, Petri nets, BPEL, and UML activity diagrams) are a sequence, parallel route (AND-splits/joins), choice (XOR-splits/joins), and loops.

The guiding principle four is events should be related to model elements (Van Der Aalst, 2012). Once the model has been built, the events should be related to models elements. Especially conformance check and enhancement are dependent on the links between models and event logs. The event logs should be able to be simulated in the models and to replay the events. The simulation can be used to identify the mismatch between event logs and models. The simulation also can improve the model with extra information extracted from the event logs. For example the identification of bottlenecks with timestamps in the event logs.

The guiding principle five is models should be treated as purposeful abstractions of reality (Van Der Aalst, 2012). Models derived from event logs provide the views of behaviour and reality. There may be multiple views available for an event log.

The guiding principle six is process mining should be a continuous process (Van Der Aalst, 2012). The process may be changed overtime and seasonally. A fixed model is not enough to illustrate the event logs with flexibility. A dynamic process mining model with the regular update is necessary. Users and analyst also advised reviewing the dynamic model regularly.

## 5.7 Process mining Challenges

There are eleven challenges proposed by Van der Aalst et al. (2011):

The first challenge is finding, merging, and cleaning event data (Van der Aalst et al., 2011). Integration of distributed data over a variety of sources, incomplete event data, an event log with outliers, logs may contain events at the different level of granularity (for example timestamp ranging from day to second).

The second challenge is dealing with complex event logs having diverse characteristics (Van der Aalst et al., 2011). Different characteristics can be contained in the event logs. Some event logs can be extremely large making them difficult to deal with while other event logs are too small to make reliable conclusions. It is suggested that low-level events can be aggregated into high-level events. At the same time, a trial-and-error approach needs to be used to see whether the event log is fit for process mining. A fast feasibility test is crucial to discover potential process mining issues before going into detail.

The third challenge is creating representative benchmarks (Van der Aalst et al., 2011). The benchmarks of classical data mining techniques are available which have inspired researchers to improve the performance of their techniques. But some good process mining benchmarks is yet to be decided for this technology. For example, the metric for measuring the quality of process mining results; the criteria to judge the quality of process mining result, the consensus on the creation of synthetic datasets to capture specific characteristics.

The fourth challenge is dealing with concept drift (Van der Aalst et al., 2011). Concept drift refers to the process is changing while being analysed. The relationship of two activities may be concurrent at the beginning, but changed to sequential at the end. Only few research state and understanding such concept drifts. But it is important for the management of processes.

The fifth challenge is improving the representational bias used for process discovery (Van der Aalst et al., 2011). A specific language (such as Petri nets) is a process discovery technique which used to produce the model. There are limitations come along with the implicit assumptions of a target language. It is necessary to separate the visualisation result from a different language to avoid representational bias.

The sixth challenge is to find models that balancing between all four competing quality dimensions fitness, simplicity, precision, and generalisation (Van der Aalst et al., 2011). A model with good fitness allows the traces in the log can be replayed by the model from beginning to end. A model with good simplicity can explain the behaviour seen in the log. A model with precision cannot allow “too much” behaviour seen in the log. A model with good generalisation is not to restrict behaviour to only the examples seen in the log.

The seventh challenge is cross-organisational mining (Van der Aalst et al., 2011). There are increasingly various use cases of event logs available for analysis across multiple organisations who work together or have common process instances (e.g., outsourcing, supply-chain integration, and cloud computing).

The eighth challenge is providing operational support (Van der Aalst et al., 2011). Process mining does not only work on off-line historical data analysis but also can be used as online real-time analysis for operational support.

The ninth challenge is combining process mining with other types of analysis from different fields to extract more insights from event data (Van der Aalst et al., 2011). These analysis approaches including such as operation management, optimisation techniques, mathematical models, data mining, simulation, visual analytics.

The tenth challenge is improving usability for non-experts (Van der Aalst et al., 2011). The user-friendly interfaces that automatically set parameters and suggest suitable types of analysis can help non-expert to utilise process mining techniques and algorithms easily without sophisticated implementations.

The eleventh challenge is improving understandability for non-experts (Van der Aalst et al., 2011). The generation of an easy understanding conclusion can help decision makers who are non-experts. The trustworthiness and limitations of the result should be clearly stated (Van Der Aalst, 2012).

# 6. Chapter Six Conclusion

In conclusion, this paper has extracted knowledge from the event logs data which recorded during drug discovery process. A trial-and-error approach has been performed for the integration of process mining and drug discovery process.

It has been highlighted that the data are the most important and priority elements during the implementation of the application. Such as lack of information on event classes per case makes the data become isolated, thus lead to a poor performance to apply process mining techniques. Whereas the abundant of event log information can help researchers fully apply the process mining techniques thereby generate useful discovery.

Due to the time limit, only one of the three basic types of process mining, discovery type has been performed in this research successfully. The conformance was failed to check because of not enough data. For future prospects, the other two types – conformance checking and enhancement should be considered to perform for diagnostics and making a new model.

There are three recommendations. The first recommendation is that it need to understand more on the target organisational behaviour and organisational business process. For the conformance checking, it requires comparing the process flow in reality and process flow in the model for diagnostics purpose. Furthermore, in terms of enhancement, it needs to build a new model and compare with the ideal process flow. The second recommendation is that the process mining is not once and for all but is a regular operation management process. The content and context of the drug discovery process are changing overtime. It is suggested that building a fast process mining manipulating standard for organisational process improvement is necessary. The third recommendation is researchers have to be aware that event logs data are the most important and priority elements in process mining. There may be many database and tables are used by an organisation. The researchers have to understand the event logs, extract event logs data according to their research questions. These recommendations may all require the process mining researchers working closer with the management of the target organisation.